\documentclass[preprint]{aastex631}

\usepackage{amsmath} 
\usepackage{mwe,tikz}\usepackage[percent]{overpic}

\begin{document}

 \title{Large Interferometer For Exoplanets (\emph{LIFE}): XII. The Detectability of Capstone Biosignatures in the Mid-Infrared - Sniffing Exoplanetary Laughing Gas and Methylated Halogens }

\author[0000-0001-6138-8633]{Daniel Angerhausen}
\affiliation{ETH Zurich, Institute for Particle Physics \& Astrophysics, Wolfgang-Pauli-Str. 27, 8093 Zurich, Switzerland}
\affiliation{ National Center of Competence in Research PlanetS, Gesellschaftsstrasse 6, 3012 Bern, Switzerland}

\author[0000-0001-9771-7953]{Daria Pidhorodetska}
\affiliation{Department of Earth and Planetary Sciences, University of California, Riverside, CA, USA}

\author[0000-0003-1906-5093]{Michaela Leung}
\affiliation{Department of Earth and Planetary Sciences, University of California, Riverside, CA, USA}

\author[0009-0003-1247-8378]{Janina Hansen}
\affiliation{ETH Zurich, Institute for Particle Physics \& Astrophysics, Wolfgang-Pauli-Str. 27, 8093 Zurich, Switzerland}\affiliation{ National Center of Competence in Research PlanetS, Gesellschaftsstrasse 6, 3012 Bern, Switzerland}

\author[0000-0002-0006-1175]{Eleonora Alei}
\affiliation{ETH Zurich, Institute for Particle Physics \& Astrophysics, Wolfgang-Pauli-Str. 27, 8093 Zurich, Switzerland}\affiliation{ National Center of Competence in Research PlanetS, Gesellschaftsstrasse 6, 3012 Bern, Switzerland}
\affiliation{NPP Fellow, NASA Goddard Space Flight Center, Greenbelt, MD 20771, USA}

\author[0000-0002-5476-2663]{Felix Dannert}
\affiliation{ETH Zurich, Institute for Particle Physics \& Astrophysics, Wolfgang-Pauli-Str. 27, 8093 Zurich, Switzerland}\affiliation{ National Center of Competence in Research PlanetS, Gesellschaftsstrasse 6, 3012 Bern, Switzerland}

\author[0000-0003-2769-0438]{Jens Kammerer}
\affiliation{European Southern Observatory, Karl-Schwarzschild-Straße 2, 85748 Garching, Germany}
\affiliation{Space Telescope Science Institute, 3700 San Martin Drive, Baltimore, MD 21218, USA}

\author[0000-0003-3829-7412]{Sascha P. Quanz}
\affiliation{ETH Zurich, Institute for Particle Physics \& Astrophysics, Wolfgang-Pauli-Str. 27, 8093 Zurich, Switzerland}\affiliation{ National Center of Competence in Research PlanetS, Gesellschaftsstrasse 6, 3012 Bern, Switzerland}\affiliation{ETH Zurich, Department of Earth Sciences, Sonneggstrasse 5, 8092 Zurich, Switzerland}
\author[0000-0002-2949-2163]{Edward W. Schwieterman}
\affiliation{Department of Earth and Planetary Sciences, University of California, Riverside, CA, USA}
\affiliation{Blue Marble Space Institute of Science, Seattle, WA, 98154, USA}


\collaboration{20}{and the \textit{LIFE} initiative}
\affiliation{https://life-space-mission.com/}



\begin{abstract}
This study aims to identify exemplary science cases for observing N$_2$O, CH$_3$Cl, and CH$_3$Br in exoplanet atmospheres at abundances consistent with biogenic production using a space-based mid-infrared nulling interferometric observatory, such as the  \emph{LIFE} (Large Interferometer For Exoplanets) mission concept.
We use a set of scenarios derived from chemical kinetics models that simulate the atmospheric response of varied levels of biogenic production of N$_2$O, CH$_3$Cl and CH$_3$Br in O$_2$-rich terrestrial planet atmospheres to produce forward models for our LIFE\textsc{sim} observation simulator software. 
In addition we demonstrate the connection to retrievals for selected cases. We use the results to derive observation times needed for the detection of these scenarios and apply them to define science requirements for the mission.  
Our analysis shows that in order to detect relevant abundances with a mission like \emph{LIFE} in it's current baseline setup, we require:
   (i) only a few days of observation time for certain very near-by ``Golden Target" scenarios, which also motivate future studies of ``spectral-temporal" observations 
   (ii) $\sim$10 days in  certain standard scenarios such as temperate, terrestrial planets around M star hosts at 5 pc, 
   (iii) $\sim$50 - 100 days in the most challenging but still feasible cases, such as an Earth twin at 5 pc. A few cases for very low fluxes around specific host stars are not detectable.
 In summary, abundances of these capstone biosignatures are detectable at plausible biological production fluxes for most cases examined and for a significant number of potential targets. 

\end{abstract}



\section{Introduction} \label{sec:intro}

Nitrous oxide (N$_2$O) and  methylated halogens (e.g., CH$_{3}$Cl, CH$_{3}$Br) are part of the group of molecules that have been proposed as  potential biosignatures\footnote{For a more general and detailed introduction about biosignatures we refer to \cite{2002AsBio...2..153D, 2016AsBio..16..465S, 2018AsBio..18..663S, 2018SciA....4.5747K} and references therein.} for detecting life in exoplanetary contexts  \cite[see e.g.,][]{2022ApJ...937..109S, 2022ApJ...938....6L}. A large inferred flux of these biogenic gases is most consistent with a productive global photosynthetic biosphere, which could be revealed in tandem with O$_3$.  None of these three gases are produced in chemical equilibrium, which reduces the false positives potential when compared to other potential biosignatures that do have equilibrium sources (e.g., CH$_4$, PH$_3$, NH$_3$). N$_2$O is produced by certain microbes during nitrification and denitrification processes, which are common in many environments on Earth. Similarly, CH$_3$X gases are produced by marine microbes and terrestrial plants. The short photochemical lifetimes of these gases require large production rates to sustain significant (or detectable) abundances. The detection of nitrous oxide or  methylated halogens in the atmosphere of a terrestrial, temperate exoplanet could therefore indicate the presence of fluxes that can only be explained by a global biosphere of (microbial) life producing it. In the next decades our community will start to to investigate the atmospheric properties for a statistically significant, comprehensive and consistent, large observational set of rocky, warm exoplanets \cite[see e.g.,][]{2018AsBio..18..739F}. First steps, albeit with a focus on larger planets, in this direction will be made with selected or proposed ground-based projects   \cite[see e.g.,][]{2013ApJ...764..182S, 2017A&A...599A..16L, 2018SPIE10702E..6NB} and space-based missions such as the JWST \citep{Greene2016,morley2017}, Roman \citep{kasdin2020} or the Atmospheric Remote-sensing Infrared Exoplanet Large-survey \citep[\textit{Ariel}, ][]{2018ExA....46..135T}.
The NASA Habitable Worlds Observatory (HWO) is a proposed flagship-class mission that aims to study planets beyond our solar system and search for signs of habitability and life probing the exoplanets in reflected stellar light. The HWO will utilize a combination of spectroscopic and imaging techniques to explore the atmospheres and surfaces of temperate and rocky exoplanets \citep{2021pdaa.book.....N} \footnote{HWO is based on the studies for the Large UV/Optical/IR Surveyor \citep[\textit{LUVOIR,}][]{2019arXiv191206219T} , and the Habitable Exoplanet Observatory \citep[\textit{HabEx},][]{2020arXiv200106683G}}.

An alternative approach to the coronagrapy-based HWO concept is to cancel out the stellar contribution and separate the thermal emission of the planet using a nulling interferometer. The Large Interferometer for Exoplanets (\emph{LIFE}) is a European-lead international project with the goal to consolidate various efforts and define a road map that leads to the launch of such a space-based mid-infrared nulling interferometer \citep{2018SPIE10701E..1IQ, 2019arXiv190801316Q, LIFE1}. This mission will be designed with the capability to investigate the atmospheric properties of a large sample of (primarily) terrestrial exoplanets. \textit{LIFE} directly addresses the scientific theme of detecting and characterizing temperate exoplanets in the MIR, which was recommended with \textit{'highest scientific priority'} by ESA’s Voyage 2050 Senior Committee report \footnote{https://www.cosmos.esa.int/web/voyage-2050} as a candidate topic for a future L-class mission in the ESA Science Programme. \citet{LIFE1, LIFE2} and \citet{LIFE6} discussed the detection yield of a mission like \textit{LIFE}. 
  \citet{LIFE3} and \citet{LIFE5} critically checked the ability of the proposed mission to characterize an Earth-twin at 10 pc and Earth analogs over its geologic history. \citet{LIFE4, LIFE7}  compared various technical implementations of the interferometric setup, while \citet{LIFE11} explored phase-space synthesis decomposition (PSSD) to potentially improve the sensitivity of a mission like \textit{LIFE}. \citet{LIFE9} analysed the detectability of a Venus twin exoplanet and in particular how clouds impact the spectral retrieval process.  The observability of exocomets with \textit{LIFE} was discussed in  \citet{2023A&A...671A.114J}. \citet{LIFE10} studied how \textit{LIFE }can detect of currently already known exoplanets and how it can leverage synergies HWO. Lastly \citet{LIFE8} reported on the detectability of phosphine (PH$_3$) in different exoplanetary scenarios with \textit{LIFE}.  
 Here we explore the detectability of nitrous oxide  and  methylated halogens with a mid-infrared space nulling interferometer observatory like \textit{LIFE}. In section \ref{sec:models} we introduce the exoplanetary atmospheric chemistry models and software used to simulate the \textit{LIFE} observations. Section \ref{sec:results}  summarizes the results of our feasibility study of various observational scenarios. In section \ref{sec:disc} we discuss these results and their implications and conclude with our main findings.

\section{Models and Simulations}\label{sec:models}

\subsection{Photochemical Modelling} 
In this study, we focus on warm, terrestrial, ``Earth-like'' planets. To maintain consistency, simplify reproducibility and to focus solely on examining the impact of changing surface molecular fluxes for the respective gases, we assume an Earth-like bulk atmosphere (78$\%$ N$_{2}$, 21$\%$ O$_{2}$ by volume) along with the Earth's temperature-pressure profile, which has a globally averaged surface temperature of 288 K. We also assume a planetary radius and surface gravity identical to that of Earth (R=6371 km; g=9.8 m s$^{-1}$). We provide further details in the subsections below and the original sources for these atmospheric simulations \citep{2022ApJ...937..109S, 2022ApJ...938....6L}.
\subsubsection{Nitrous Oxide Atmospheres}

N$_{2}$O is produced by life as an intermediate product of biological denitrification, the transformation of NO$_{3}^{-}$ to N$_{2}$ gas through multiple steps, which is an essential component of the biological nitrogen cycle on Earth. N$_{2}$O is also produced by other microbial metabolisms including the direct oxidation of ammonia by certain bacteria and archaea \citep{Prosser2012}. The modern biological production rate of N$_{2}$O is $\approx$0.4 teramole per year ($10^{12}$ moles per year; Tmol/yr), which includes both marine and terrestrial sources \citep{Tian2020} and results in a modern atmospheric N$_{2}$O concentration of $\approx$330 ppb (parts-per-billion). However, the biological N$_{2}$O flux may have varied greatly through geological time. For example, the H$_{2}$S-rich Proterozoic (2.5 to 0.541 Ga) oceans would have sharply limited the availability of copper, an essential component of the nitrous oxide reductase enzyme that facilitates the last step in the denitrification cycle from N$_{2}$O to N$_{2}$ \citep{Buick2007}. In the case of severe limitation of this enzyme (or its failure to evolve on an inhabited exoplanet), the N$_{2}$O flux to the atmosphere would be limited instead by the denitrification flux of the biosphere, which is substantially greater than the modern N$_{2}$O flux. Photolysis and reactions with radical species transforms atmospheric N$_{2}$O into N$_{2}$ and O$_{2}$ abiotically with an average lifetime of $\approx$120 years on Earth \citep{Prather2015}, though this would vary for other host stars and atmospheric compositions (e.g., O$_{2}$ concentrations). There are no substantial surface sinks for N$_{2}$O once it is released into the atmosphere. 

\citet{2022ApJ...937..109S} used a biogeochemical model and a photochemical model to circumscribe the plausible envelope of N$_{2}$O concentrations on Earth-like exoplanets with microbial biospheres. We refer to that work for a more expansive discussion of the biological and abiotic sources and sinks of N$_{2}$O. For the purposes of this work, we are interested what range of N$_{2}$O production fluxes can result in detectable N$_{2}$O signatures in the directly imaged thermal infrared spectra of terrestrial exoplanets with the understanding that detectability at lower production rates enhances the probability such a signature may be found. 

For the cases presented here, we chose those results with Earth’s modern O$_2$ concentration (21\%) and N$_2$O fluxes corresponding to biological production levels of (i) 1 teramole per year (equivalently, 3.7$\times$10$^{9}$ molecules cm$^{-2}$ s$^{-1}$), (ii) 10 Tmol/yr (3.7$\times$10$^{10}$ molecules cm$^{-2}$ s$^{-1}$), and (iii) 100 Tmol/yr (3.7$\times$10$^{11}$ molecules cm$^{-2}$ s$^{-1}$). These values correspond to scenarios where (i) 5-10\% of Earth’s denitrification flux is released as N$_2$O, (ii) 50\%-100\% of Earth’s total denitrification flux is released as N$_2$O, and (iii) 50-100\% of the denitrification flux of an Earth-like planet with twice the nutrient availability of Earth’s oceans is released as N$_2$O. These flux levels were obtained via biogeochemical modeling as described in Section 2 of \citet{2022ApJ...937..109S}. As points of comparison, the 100 Tmol/yr case corresponds to an N$_{2}$O flux about three times the modern CH$_{4}$ flux and is similar to the PH$_{3}$ flux required to generate a detectable PH$_{3}$ signature on a CO$_{2}$-rich planet orbiting an M dwarf star \citep[][see their Table 3]{Sousa-Silva2020}.

The chemical profiles for our N$_2$O detectability analyses were sourced from \cite{2022ApJ...937..109S}, which describes their generation in detail. Briefly, these abundance profiles were calculated with the photochemical model component of the \textit{Atmos} code \citep{2016AsBio..16..873A}\footnote{https://github.com/VirtualPlanetaryLaboratory/atmos}. The photochemical model is based originally on the work of \citet{1979JGR....84.3097K}  but has undergone numerous subsequent modifications and upgrades \citep{2001JGR...10623267P,2006Gbio....4..271Z,2017ApJ...836...49A, 2018ApJ...867...76L}. The model template used here includes 50 species and 238 photochemical reactions, including all major species in Earth’s modern atmosphere and the dominant photochemical destruction channels for N$_{2}$O. This version is appropriate for modeling atmospheres with substantial free oxygen concentrations (pO$_2$ $\geq$ 1\% present atmospheric level; PAL) and was previously used to model possible Earth-like atmospheric scenarios for Proxima Centauri b \citep{2018AsBio..18..133M}. The simulated atmosphere is divided into 200 layers of 0.5 km thickness with a maximum altitude of 100 km. The model incorporated recommended H$_2$O cross-sections and reaction rate corrections from \cite{2020ApJ...896..148R}. We assumed the modern Earth’s temperature-pressure profile for all simulations, to isolate the unique impact of different N$_2$O fluxes (abundances). The chemical mixing ratio profiles were otherwise calculated self-consistently with respect to varied stellar spectra and N$_2$O flux levels. These chemical profiles were generated assuming host star spectra identical to those of the Sun \citep{2004AdSpR..34..256T}, the K6V star HD  85512 \citep{2016ApJ...820...89F, 2016ApJ...824..102L}, the M5.5 Ve star Proxima Centauri \citep{2018ApJ...867...71L, 2020ApJ...895....5P, 2014AJ....148...64S}, and the ultracool dwarf (M8V) star  TRAPPIST-1 \citep{2019ApJ...871..235P}. See Appendix \ref{sec:app_fluxabund} and Table \ref{table:N2OFLUXABUND} for the ground-level N$_{2}$O mixing ratios that correspond to each flux and host star combination. 

We briefly note here that all photochemical simulations were done with 1D models using key assumptions (such as average solar zenith angle) most appropriate for rapidly rotating planets, and thus do not take into account circulation impacts from synchronous rotation, which would be likely for habitable zone planets orbiting M dwarf stars. However, \citet{2018ApJ...868L...6C} showed that the differences between full 3D and 1D predictions for abundant biosignature gases like N$_{2}$O  and CH$_{4}$ differ by only $\sim$20$\%$, which is relatively small given the uncertainty from other intrinsic and extrinsic factors such as bulk atmospheric composition, biosignature gas production rate, and distance from the host star. \citet{2022ApJ...937..109S} quantitatively examine the sensitivity of N$_{2}$O flux-abundance relationships to other 1D model assumptions (such as vertical mixing parameterizations and background N$_{2}$ pressure) in their Section 5.5, finding small or modest effects, and we direct the reader there for more details. We emphasize the universe of potential atmospheric compositions and planetary parameters for habitable worlds is much greater than assessed here.

\subsubsection{Methylated Halogen Atmospheres}

The chemical profiles of Earth-twin planets with various levels of biological CH$_3$Cl and CH$_3$Br production were sourced from \cite{2022ApJ...938....6L} who determined the abundance using the photochemical model part of the \textit{Atmos} code \citep{2016AsBio..16..873A}. In contrast to the version used for the N$_2$O calculations described above, this version of the photochemical model incorporated additional reactions to account for the inclusion of Cl and Br chemistry. There are a total of 89 unique kinetically active chemical species and 413 reactions in this version. To compute CH$_{3}$X flux-abundance calculations, we adopted the same stellar spectra as described above for HD 85512 (K6V), Proxima Centauri (M5.5 Ve), and TRAPPIST-1 (M8V). We additionally use the M3.5 Ve star AD Leo with stellar spectra from the Atmos library, which was ultimately sourced from \citep{Segura2005}. We do not consider G dwarf (Sun-like) or F dwarf host stars because the atmospheric accumulation of CH$_{3}$ gases is spectrally insignificant for these scenarios even at the largest plausible fluxes \citep{2022ApJ...938....6L}.

CH$_3$Cl and CH$_3$Br fluxes were independently and jointly co-varied according to factors of 1, 10, 100, and 1000 times the modern flux of each species. In other words, we varied: CH$_{3}$Cl while other gas fluxes are held constant (cases labeled ``CH$_{3}$Cl"), CH$_{3}$Br while other gas fluxes are held constant (cases labeled ``CH$_3$Br"), and both  CH$_{3}$Cl+CH$_{3}$Br together, by the same factor based on their original Earth-like fluxes (cases labeled ``CH$_{3}$X"). For more details see Appendix \ref{sec:app_fluxabund}, Table \ref{table:CH3XFLUXABUND}. Importantly, even the highest flux scenarios are equal to or less than flux of these species present in productive local environments \citep{2022ApJ...938....6L}. Since the methylation process is often an adaptation used for environmental detoxification \citep{Jia2013-nn}, the gas fluxes produced can be directly related to the abundance of metals and metalloids in the local environment. A higher supply of substrates (Br, Cl, etc.) could generate higher fluxes. Since the highest fluxes are produced at wetlands, salt marshes, and other marine/terrestrial cross over ecosystems \citep{Tait1995-pn, Yang2022-lj}, a planet with greater productivity may simply require a greater percentage of these environments or accelerated weathering processes that concentrate halogens and other heavy ions \citep{Fuge1988-dp}. Highly productive salt marsh environments cover only .00017 \% of the Earth's surface, so our most productive case would only require 1.7\% total surface coverage, equivalent to 5\% of vegetation present today \citep{Murray2022-ck, Zhu2016-za}. The actual coverage necessary may be higher or lower depending on the uniformity of production across salt marsh environments. Alternative evolutionary paths such as widespread radiation of the methylation adaptation could yield higher global productivity. Furthermore, the greatest atmospheric loss process is through reaction with OH, which sizably decreases in abundance for later star types \citep{Segura2005}. For targets such as TRAPPIST-1, this bestows a considerable advantage in sustaining atmospheric buildup and generating potentially detectable atmospheric features. For further discussion of methylated biosignature candidate production and accumulation, see \cite{2022ApJ...938....6L}.

\subsection{Spectral Simulations} 
\subsubsection{PSG}

\cite{2022ApJ...937..109S} used the Planetary Spectrum Generator (PSG) to calculate the radiance (thermal emission) spectra for the Earth spectra with varying N$_2$O fluxes. PSG is a highly flexible and publicly available radiative tool used to simulate the remote spectral observables of planetary objects across a full range of viewing geometries and distances   \citep{2018JQSRT.217...86V,2022fpsg.book.....V}. PSG uses the HITRAN 2020 database for its input infrared opacities \citep{2022JQSRT.27707949G}. Planetary spectra were produced from 5-20 $\mu$m and are averaged 50\% clear sky and 50\% cloud scenarios. The clouds consisted of equal fractions liquid water cloud and cirrus cloud particles. 

\subsubsection{SMART}

\cite{2022ApJ...938....6L} used the Spectral Mapping and Radiative Transfer code \citep[SMART; ][]{1996JGR...101.4595M} to simulate emission spectra. This model has an extensive history of use for both solar system and exoplanet objects and has previously been used to simulate biosignatures such as O$_2$ and organic sulfur compounds \citep{2011AsBio..11..393R, 2014JGRE..119.1860A, 2018AsBio..18..630M, 2011AsBio..11..419D}. As input, SMART uses the Line-by-Line Absorption Coefficients code (LBLABC) which are calculated based on the HITRAN 2020 linelists \citep{2022JQSRT.27707949G}.  For this application, 50\% cloudy spectra with 25\% cirrus and 25\% stratocumulus clouds were modeled.

\subsection{LIFE\textsc{sim} observation and population simulations}

LIFE\textsc{sim} is a state of the art simulator software specifically designed for the Large Interferometer For Exoplanets. As a tool it provides accurate and reliable simulations of the interferometric measurement processes involved in the operation of \textit{LIFE}. With LIFE\textsc{sim}, users can explore the capabilities of \textit{LIFE} and gain insights into the parameter space accessible to \textit{LIFE} for the detection and characterization of exoplanets.

In this work we used two features of LIFE\textsc{sim}:
\begin{itemize}

\item a module based on the population synthesis tool P-Pop \citep{Kammerer&Quanz2018} that simulates random exoplanetary systems around an input catalog of target stars based on current exoplanet statistics, e.g., from transit or radial velocity surveys. LIFEsim can then compute the detectability of each individual simulated planet and derive an optimized observing sequence for the stellar sample. This can be used to compute detection yields in the blind (target) search \textit{detection phase} of \textit{LIFE}  \citep[see][]{LIFE1, LIFE2, LIFE6}. In section \ref{sec:yields} we used this to calculate the number of potential targets available for the analysis discussed here.  The assumed mission time available for the blind search phase (2.5 yrs, 20\% overheads) is optimally distributed among the synthetic planetary systems to optimize the number of planets detected in the HZ.
 \item a simulator for individual spectroscopic observations in 
\textit{LIFE}'s \textit{characterization phase} which is used for the feasibility study in the following sections following the same approach as \cite{LIFE8}.
    
\end{itemize}

LIFE\textsc{sim} incorporates astrophysical noise sources such as stellar leakage and thermal emission from local zodiacal and exo-zodiacal dust. While only fundamental photon noise is considered in this study, the software is also flexible to accommodate instrumental noise terms in the future. LIFE\textsc{sim} offers an accessible way to predict the expected SNR of future observations based on various instrument and target parameters \cite[see][for more details]{LIFE2} .

  \begin{table}[ht]
\caption{Overview of simulation parameters used in LIFE\textsc{sim}. These are the same standard values as e.g. used in \cite{LIFE1} or \cite{LIFE8}}              
\label{table:1}      
\centering                                      
\begin{tabular}{l l}          
\hline\hline                        
Parameter & Value \\    
\hline                                   
    Quantum efficiency & 0.7\\      
    Throughput & 0.05\\
    Minimum Wavelength & 4 $\mu$m       \\
    Maximum Wavelength & 18.5 $\mu$m   \\
    Spectral Resolution & 50      \\
    Interferometric Baseline & 10-100 m \\
    Apertures Diameter & 2m \\
    Exozodi & 3x local zodi \\
\hline
\end{tabular}\label{tab:lifesim}
\end{table}

Following the same approach as in \cite{LIFE8} we used LIFE\textsc{sim} to produce synthetic observations of the outlined exoplanet cases with different flux levels of the discussed species and also without them being present in their atmospheres. For the presented output spectra, LIFE\textsc{sim} is configured with the current \textit{LIFE} ``baseline" setup with four apertures of 2~m diameter each, a broadband wavelength range of 4-18.5  $\mu$m, a throughput of 5 percent,  and a spectral resolution of $R = 50$. While the detection phase simulation sampled over a distribution of exozodi levels, we assume an exo-zodi level of 3 times the local zodi density for the characterization observations \cite[following the results from the HOSTS survey for the expected median level of emission in][]{Ertel2020} and a nulling baseline setup between 10 and 100 m \footnote{Here we assume that the planets are known from other surveys or detected in the \textit{LIFE} detection phase, so that we can optimize the baselines for each case.} (see Table \ref{tab:lifesim}). In section \ref{sec:tec_req} we present a short analysis of the impact that changes in this setup have for the dectectability of selected cases.

\section{Results} \label{sec:results}

\subsection{Predicted Distance Distribution of HZ Planets around M and FGK-type stars detectable with LIFE}\label{sec:yields}

\begin{figure}[ht]
\begin{minipage}[\textheight]{0.5\textwidth}
\centering
  \includegraphics[width=\textwidth]{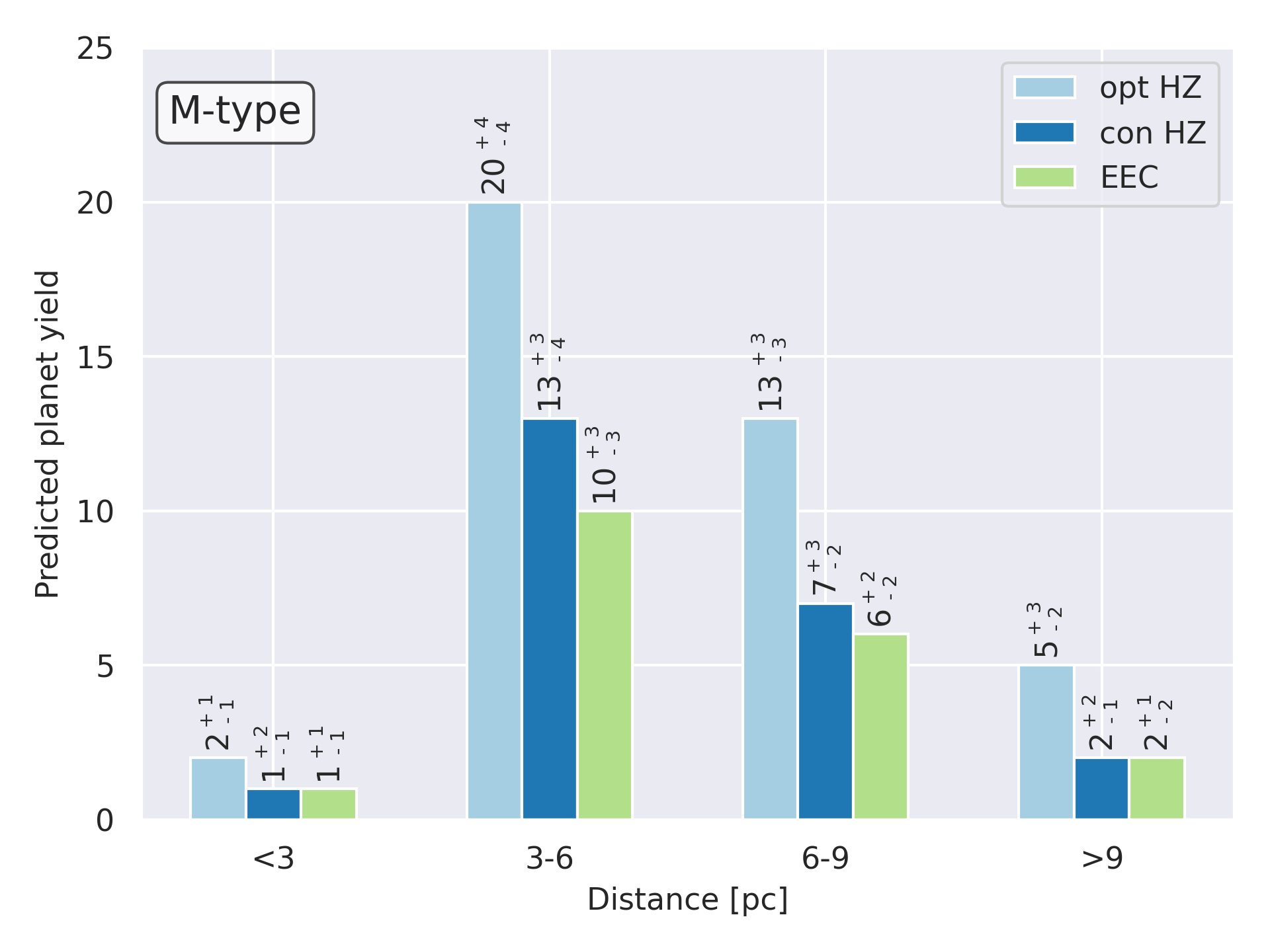}
  \centering
\end{minipage}
\hfill
\begin{minipage}[\textheight]{0.5\textwidth}
\centering
  \includegraphics[width=\textwidth]{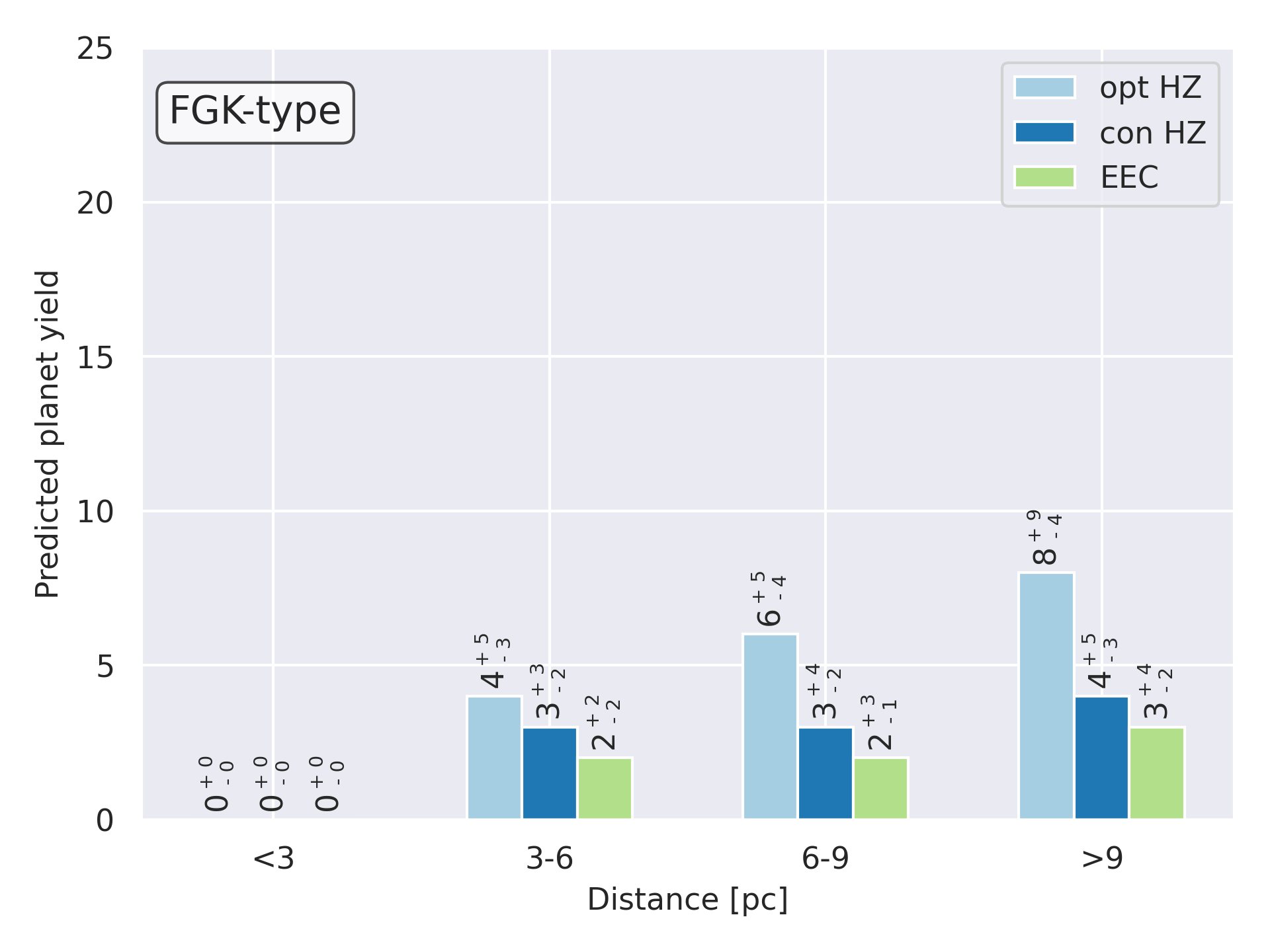}
  \centering
 \end{minipage}
\caption{Distance distribution of HZ planet populations around M and FGK-type stars detectable with \textit{LIFE} in the current baseline setup (see Table \ref{tab:lifesim}). Left: Detection yield predictions for planets around M-type stars with an underlying AFGKM exoplanet population following occurrence rate estimates from NASA's ExoPaG SAG13 \citep{Kopparapu2018} and \citet{D&C15}. Right: Detection yield predictions around FGK-type stars with an underlying FGK exoplanet population based on occurence rates from \citet{Bryson2021}. We differentiate between three HZ definitions: optimistic (opt) HZ, conservative (con) HZ and exo-Earth candidates (EECs). The bars show the predicted median planet yields and 1-$\mathrm{\sigma}$ uncertainties derived from the Monte Carlo simulations.}
\label{fig:HZplanet_distro}
\end{figure}

As a first step in our study we identify how many planets we would encounter in our survey for this analysis and at what typical distances from the solar system their host stars appear. Therefore we used the LIFE\textsc{sim} population part to estimate the distance distribution of habitable zone (HZ) planets with radii between 0.5 and 1.5 $R_\Earth$ around M and FGK-type stars within 20\,pc of the Sun that are detectable with \textit{LIFE}. For the sake of comparability to previous studies, our yield predictions were based on planet populations generated for \citet{LIFE1} and \citet{LIFE6} respectively. 

Hence, we note that the number of planets around M and FGK-type stars were calculated in two separate simulations, (1) optimizing on maximum yield of HZ planets in general using a synthetic planet populations around a sample of  AFGKM-type and (2) optimizing on number of HZ planets around FGK stars only using a subsample of FGK-type host stars within our target database \citep{LIFE1}.

These populations were derived from two sets of rocky planet occurrence rate estimates: (1) Similar to \citet{LIFE1}, occurrence rates from NASA's ExoPaG SAG13 \citep{Kopparapu2018} were used for AFGK-type stars with effective temperatures between 3940-9700\,K and results from \citet{D&C15} were applied to M-type stars with effective temperatures between 2440-3800\,K. (2) As in \citet{LIFE6}, exoplanet occurrence rate estimates from \citet{Bryson2021}, based on \textit{Kepler} DR25 planet catalog data \citep{Thompson_2018} and Gaia-based stellar properties \citep{Gaia2018}, were used to compute separate planet yield predictions around a subsample of FGK-type stars with effective temperatures between 3940-7220\,K. In comparison to estimates from NASA's ExoPaG SAG13 \citep{Kopparapu2018}, planet occurrence rates in \citet{Bryson2021} cover the entire parameter range of the optimistic HZ around early-type stars. Here, we used the \textit{model 1 hab2 stars high bound} scenario from \citet{Bryson2021}. We focused on model 1 because it considers more parameters to model the underlying planet population, including the stellar effective temperature. Moreover, the hab2 stars sample is based on a larger number of planet candidates and a wider range of stellar effective temperatures. Finally, high bound is considering a more optimistic extrapolation of the Kepler occurrence rates beyond orbital periods of 500~days for which completeness characterization is not available in \textit{Kepler} DR25 \citep{Thompson_2018}, in which the completeness beyond 500~days is assumed to be zero. For further details about the planet occurrence rate model, we refer the reader to \citet{Bryson2021}. Based on these occurrence rates, (1) 500 and (2) 1000 synthetic planet populations were generated around each star in our target sample \citep{LIFE1} with the population synthesis tool P-Pop \citep{Kammerer&Quanz2018}. We associated all planetary systems with circular orbits which were uniformly distributed on the sphere. A multi-planet stability criterion was assumed to have a negligible impact on the derived yield estimates. For further details see \citet{LIFE6}. By averaging over a multitude of Monte Carlo realizations, we marginalized over different population properties such as number of planets per simulated universe, exozodiacal dust levels as well as orbital parameters and with this are able account for uncertainties in the underlying occurrence rate model. 

For retrieving the integration time required to detect each planet in the synthetic planet populations, we simulated photon noise from all major astrophysical noise sources with the observation simulation tool LIFE\textsc{sim} \citep{LIFE2}. The observation time was distributed such that the number of planets in the HZ around their respective host stars was optimized. Since instrumental noise is not considered in the current implementation of LIFE\textsc{sim} (Dannert et al., in prep), we assumed a conservatively high signal-to-noise ratio (SNR) of 7 integrated over the full wavelength range, partly to also leave some margin for currently unaccounted instrumental errors. We assigned each planetary system an exozodiacal dust level randomly drawn from the nominal exozodi level distribution with a median of $\sim$ 3\,zodi according to results from the HOSTS survey \citep{Ertel2018, Ertel2020}. Since \citet{LIFE6} previously showed the planet yield to be largely independent of the interferometric baseline length optimization scheme, we chose a baseline length optimization for the center of the optimistic HZ. We assumed a reference implementation of \textit{LIFE} mission parameters corresponding to an aperture size of $D$\,=\,2\,m, throughput of 5\% and a wavelength coverage of $\lambda$\,=\,4-18.5\,$\mathrm{\mu m}$. Other adopted mission specifications correspond to the descriptions in \citet{LIFE1}. We compared predicted planet yields for three different HZ definitions: planets located within the optimistic HZ, within the conservative HZ and exo-Earth candidates (EEC). For incident stellar flux and planetary radii limits of the respective HZ models see \citet{Kopparapu2014} and \citet{Stark2019}. The derived distance distributions of detectable HZ planets around M and FGK-type stars are shown in Figure \ref{fig:HZplanet_distro}.  

In summary we find that we will have a typical sample of 15-20 HZ planets at distances less than 6 pc. 
This motivates and justifies the choice of our examples at 5 pc. These would be the typical distance for HZ planets around late type stars and the case for a few best targets around earlier types.

\subsection{Detectability of N$_2$O}

We modelled the detectability of N$_2$O for a set of observation times of 10 days, 50 days and 100 days using the \textit{LIFE} baseline setup shown in Table \ref{tab:lifesim}. Based on our experience from previous LIFE studies \citep[e.g.][]{LIFE3, LIFE5, LIFE8}  the 10 days case represents a relatively short (potentially preliminary) characterization observation that can be conducted for a larger sample of targets, 50 days represent a more thorough observation and 100 days will be the deepest characterization observation for only the most interesting/promising targets for \textit{LIFE}. In section \ref{sec:tec_req} we also vary the integration times and explore setups beyond the \textit{LIFE} baseline. It is worthwhile to mention that these numbers may correspond to a significant fraction of the orbital periods of the considered planets, and in some cases they even clearly exceed them. Here we assume that the planet signal is stationary in LIFE\textsc{sim} and that we can add the signal of multiple visits without significantly increasing the noise contributions.

For all simulated observations we calculated two metrics for the detectability of the discussed cases: the maximum difference in a single spectral line (usually in the absorption band of the feature in question) in units of the respective sensitivity in that channel and a bandpass-integrated SNR defined as 
\begin{equation*}
SNR = \sqrt{\Sigma_{i=1}^n (\frac{\Delta y_i}{\sigma (y_i)})^2 }
\end{equation*} 

where $\Delta y_i$
is the difference between the two spectra (one spectrum containing the species and the other that does not) in
each of the n spectral bins and $\sigma (y_i)$
is the \textit{LIFE} sensitivity the respective bin.  This assumes independence between the bins and was used in a similar manner in e.g. \cite{2019AJ....158...27L} or \cite{Bixel_2021}. As a rule of thumb we assume a case is observable (i.e. the feature distiguishable) if the band-integrated SNR is  above $5-10$ and the single line sigma above 3 (more on the justification of these number in section \ref{sec:retr_test}, where we compare those to some exemplary full retrievals).

\begin{figure*}[htp]

\centering
\includegraphics[width=.45\textwidth]{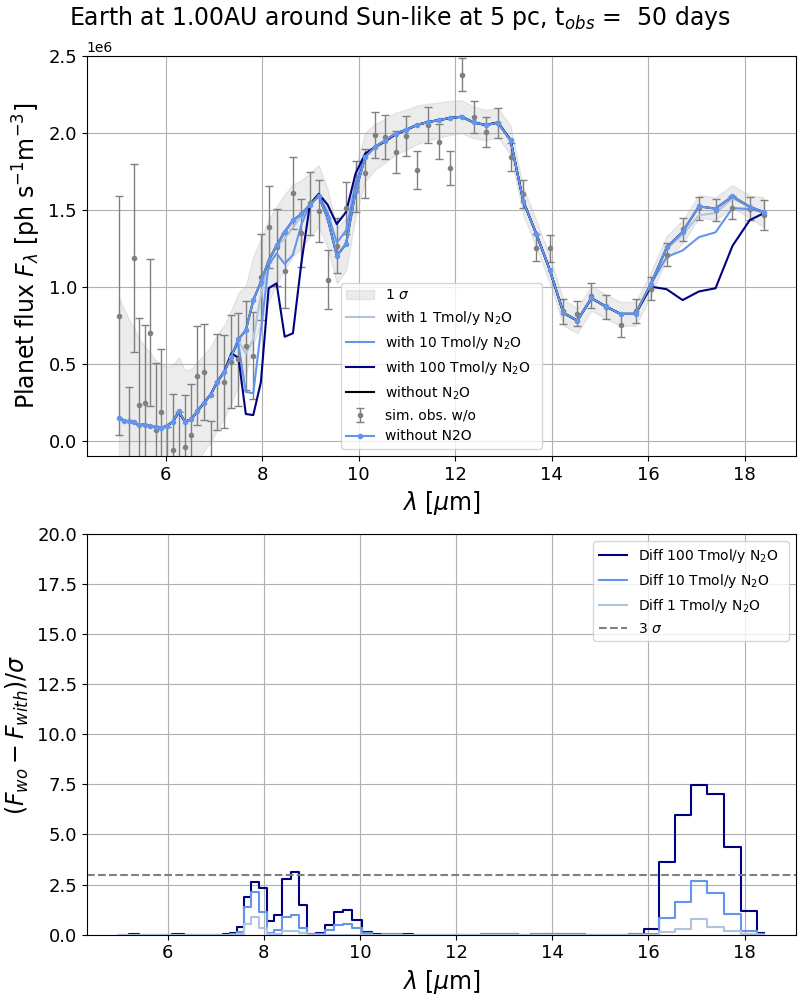}\hfill
\includegraphics[width=.45\textwidth]{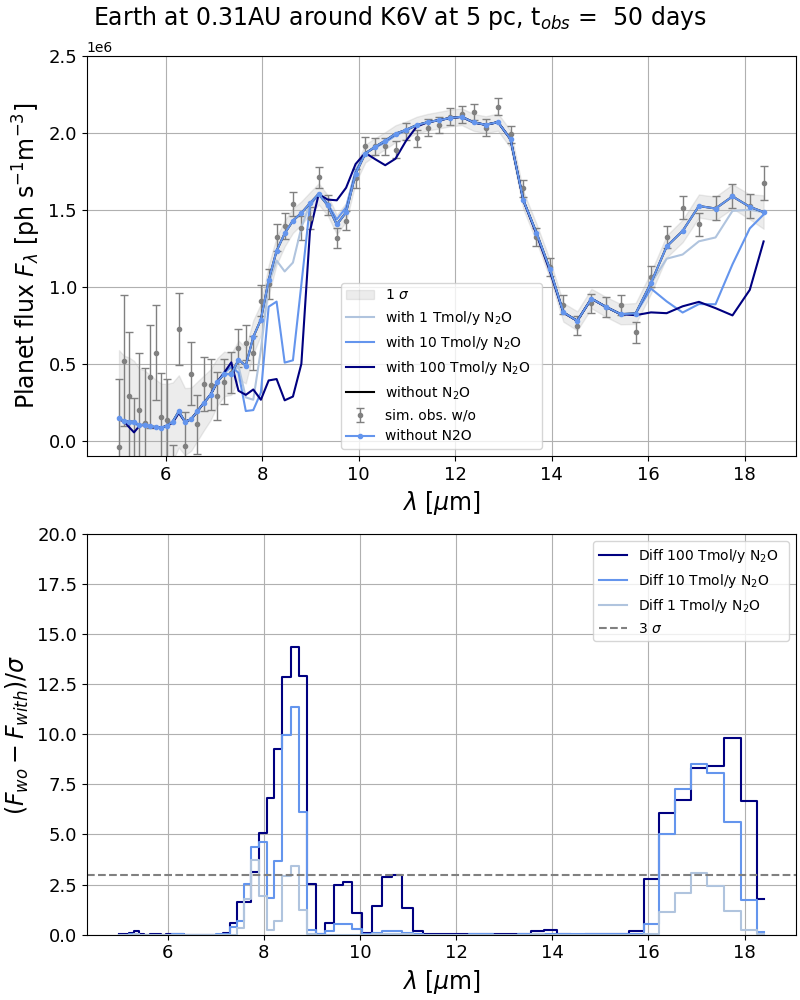}\hfill

\includegraphics[width=.45\textwidth]
{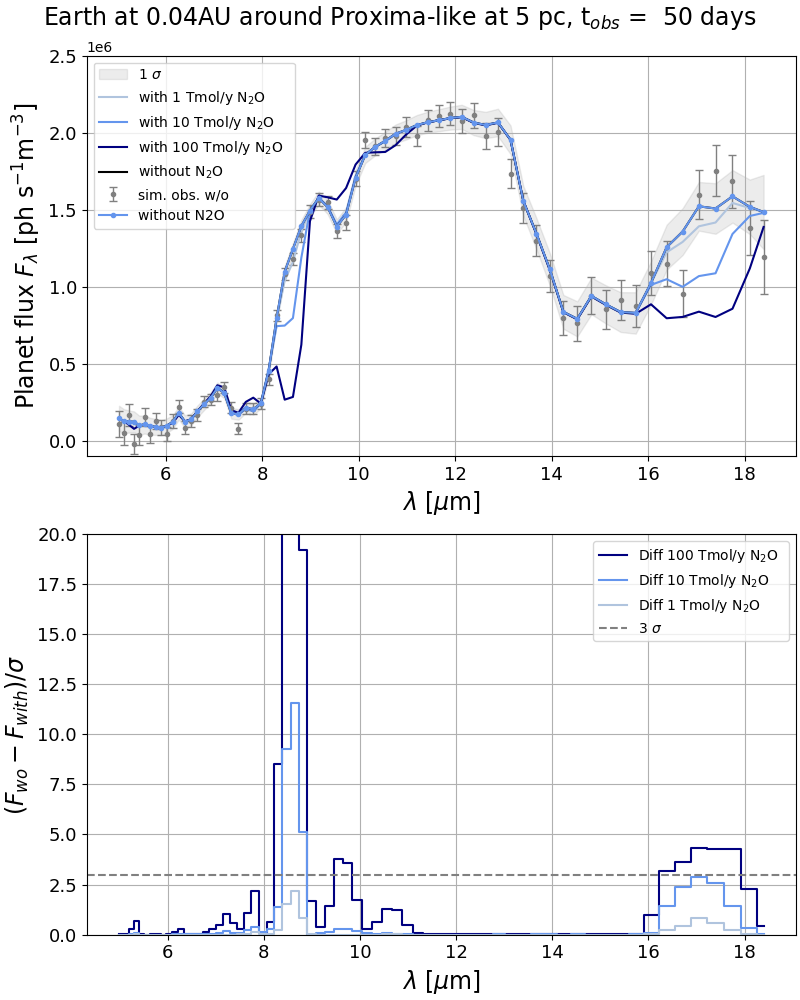}\hfill
\includegraphics[width=.45\textwidth]{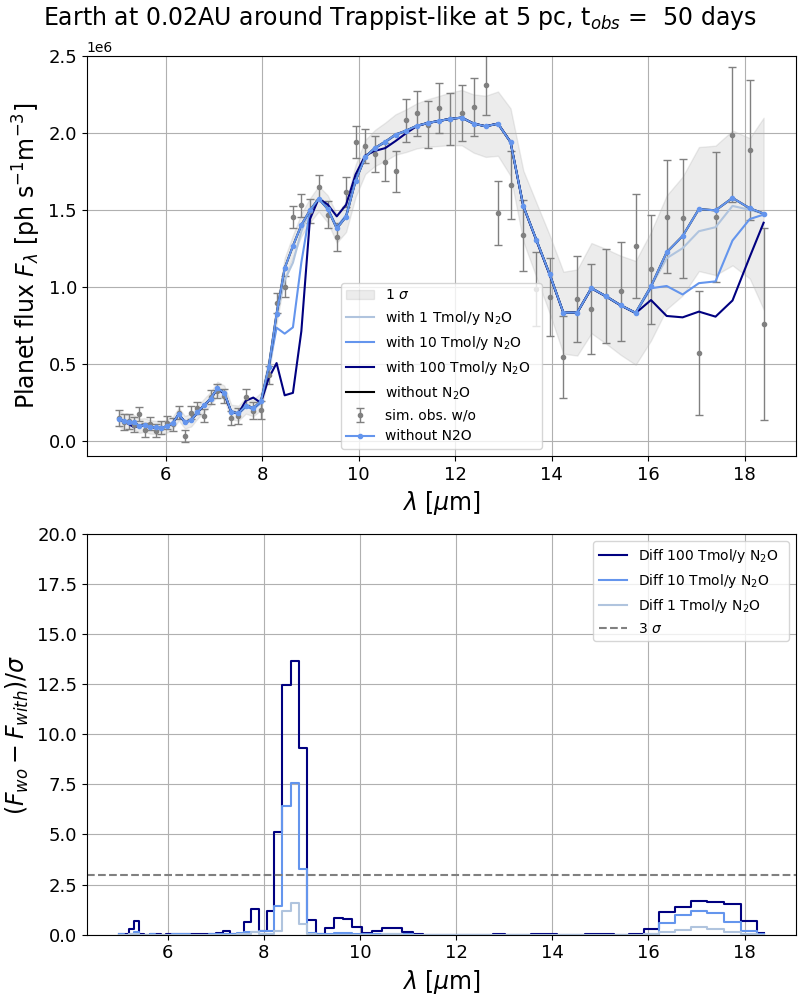}

\caption{Detectability of various fluxes of N$_2$O in the emission spectrum of an Earth-like planet around a (from top left to bottom right:) Sun-like, K6V, Proxima Centauri type and Trappist-like star at 5pc, after 50 days of observation with \textit{LIFE}.Top: planet flux for atmospheres with and without N$_2$O . The grey area represents
the 1-$\sigma$ sensitivity; the dark grey error bars show an individual simulated observation. Bottom: Statistical
significance of the detected differences between an atmospheric models with and  without N$_2$O. For the Sun-like case the main contribution to the detectability is in the 16-18 $\mu m$ band. For the later type star cases the differences especially in the 8-9 $\mu m$ (see also discussion in Section \ref{sec:wl_dep}.}
\label{fig:N20_50d}

\end{figure*}

\subsubsection{Sun-like G star}

The results for an Earth-like planet orbiting a Sun-like star at 5 pc distance are listed in Table \ref{table:N2Osun} and shown in Figure \ref{fig:N20_50d} for a 50 day observation (more cases for 10 and 100 days are shown in Appendix \ref{app:n2o_times} Figures \ref{fig:N20_10d} and \ref{fig:N20_100d} ). In these figures (also for the other N$_2$O cases in sections \ref{sec_proxn2o} to \ref{sec_trapn2o}) the top panel shows the planet flux for atmospheres with and without N$_2$O and the grey area represents the 1-$\sigma$ sensitivity calculated with LIFEsim while the dark grey error bars show an individual simulated observation. The bottom panel analyses the statistical significance of the detected differences between an atmospheric models with and  without N$_2$O. Our analysis shows that the 100 Tmol/yr case is likely observable already after 10 days (with SNR of 6.6/3.3 in our two metrics), while the 10 Tmol/yr model seems to become observable around or slightly after day 50. The 1 Tmol/yr is beyond our sensitivity , only reaching SNRs of 2.1/1.3 even for integration times of 100 days. Noteworthy here is that this detection is driven by the longer (16-18 $\mu$m) wavelength bands of N$_2$O.

\begin{table*}[ht]
\caption{Modelled SNR ratios for various intergration times and different fluxes of N$_2$O at 5 pc Sun like star}             
\label{table:N2Osun}      
\centering          
\begin{tabular}{ c r r r }     
\hline\hline       
 T$_{int}$ [d] & Feature & Band-int. SNR & Max. line sigma\\ 
\hline                    
10 & 1 $ \frac{Tmol}{yr}$ N$_2$O & 0.7 & 0.4 \\
10 & 10 $ \frac{Tmol}{yr}$ N$_2$O & 2.3 & 1.2 \\
10 &100 $ \frac{Tmol}{yr}$ N$_2$O & 6.6 & 3.3 \\

\hline
50 & 1 $ \frac{Tmol}{yr}$ N$_2$O & 1.5 & 0.9 \\
50 & 10 $ \frac{Tmol}{yr}$ N$_2$O & 5.1 & 2.7 \\
50 &100 $ \frac{Tmol}{yr}$ N$_2$O & 14.7 & 7.5 \\

\hline
100 & 1 $ \frac{Tmol}{yr}$ N$_2$O & 2.1 & 1.3 \\
100 & 10 $ \frac{Tmol}{yr}$ N$_2$O & 7.2 & 3.8 \\
100 &100 $ \frac{Tmol}{yr}$ N$_2$O & 20.7 & 10.6 \\

\hline                  
\end{tabular}
\end{table*}

This prototypical science case of an Earth twin scenario with these additional fluxes will be discussed in more detail in sections \ref{sec:retr_test} and \ref{sec:tec_req} where we use these for exemplary retrievals and derive consequences for the current mission design in the context of science and technology requirements.

\subsubsection{Proxima Centauri type star}\label{sec_proxn2o}

The results for an Earth-like planet orbiting a M5.5 Ve Proxima Centauri type star at 5 pc distance are listed in Table \ref{table:N2Oprox} and shown in Figure \ref{fig:N20_50d} for a 50 day observation (more cases simulating 10 and 100 days of integration time shown in Appendix \ref{app:n2o_times}, Figures \ref{fig:N20_10d} and \ref{fig:N20_100d}). In section \ref{sec:golden_t} we discuss the case for a planet orbiting Proxima Centauri at its actual distance of 1.3 pc as an example for a ``golden target" that allows very detailed, potentially even time resolved \textit{LIFE} observations. For this scenario the 100 as well as the 10 Tmol/yr case are detectable in the 10 day simulations with SNRs consistently above 5, while the 1 Tmol/yr seems to only barely touch the detection threshold after 100 days (with SNRs of 4.3/3.0). This case is one of the best observable ones in this study. In comparison to the previous case, here the shorter (8-10 $\mu$m) wavelength bands of N$_2$O contribute the most to the bandintegrated SNR (see Section \ref{sec:wl_dep}).

\begin{table*}[ht]
\caption{Modelled SNR ratios for various integration times and different fluxes of N$_2$O at 5 pc Proxima Centauri like star}             
\label{table:N2Oprox}      
\centering          
\begin{tabular}{ c r r r }     
\hline\hline       
 T$_{int}$ [d] & Feature & Band-int. SNR & Max. line sigma\\ 

\hline                    
10 & 1 $ \frac{Tmol}{yr}$ N$_2$O & 1.3 & 1.0 \\
10 & 10 $ \frac{Tmol}{yr}$ N$_2$O & 7.4 & 5.2 \\
10 &100 $ \frac{Tmol}{yr}$ N$_2$O & 18.3 & 11.1 \\

\hline
50 & 1 $ \frac{Tmol}{yr}$ N$_2$O & 3.0 & 2.2 \\
50 & 10 $ \frac{Tmol}{yr}$ N$_2$O & 16.5 & 11.6 \\
50 &100 $ \frac{Tmol}{yr}$ N$_2$O & 40.9 & 24.8 \\

\hline
100 & 1 $ \frac{Tmol}{yr}$ N$_2$O & 4.3 & 3.0 \\
100 & 10 $ \frac{Tmol}{yr}$ N$_2$O & 23.4 & 16.4 \\
100 &100 $ \frac{Tmol}{yr}$ N$_2$O & 57.9 & 35.0 \\
\hline                  
\end{tabular}
\end{table*}

\subsubsection{K6V}
The results for an Earth-like planet orbiting a K6V star at 5 pc distance are listed in Table \ref{table:N2OK6V} and shown in Figure \ref{fig:N20_50d} for a 50 day observation (more cases simulating 10 and 100 days of integration time are shown in Appendix \ref{app:n2o_times}, Figures \ref{fig:N20_10d} and \ref{fig:N20_100d}). The results are very similar to and even slightly more significant than the Proxima Centauri type case. Once again the 100 and 10 Tmol/yr cases are detectable in the 10 day simulations reaching SNR of over 10/5 in our two metrics. The 1 Tmol/yr case crosses the detection threshold with an SNR of 8.1/3.7 at around 50 days.

\begin{table*}[ht]
\caption{Modelled SNR ratios for various integration times and different fluxes of N$_2$O at 5 pc K6V star}             
\label{table:N2OK6V}      
\centering          
\begin{tabular}{ c r r r }     
\hline\hline       
 T$_{int}$ [d] & Feature & Band-int. SNR & Max. line sigma\\ 
\hline                    
10 & 1 $ \frac{Tmol}{yr}$ N$_2$O & 3.6 & 1.7 \\
10 & 10 $ \frac{Tmol}{yr}$ N$_2$O & 10.8 & 5.1 \\
10 &100 $ \frac{Tmol}{yr}$ N$_2$O & 15.0 & 6.4 \\

\hline
50 & 1 $ \frac{Tmol}{yr}$ N$_2$O & 8.1 & 3.7 \\
50 & 10 $ \frac{Tmol}{yr}$ N$_2$O & 24.1 & 11.4 \\
50 &100 $ \frac{Tmol}{yr}$ N$_2$O & 33.6 & 14.3 \\

\hline
100 & 1 $ \frac{Tmol}{yr}$ N$_2$O & 11.5 & 5.3 \\
100 & 10 $ \frac{Tmol}{yr}$ N$_2$O & 34.1 & 16.1 \\
100 &100 $ \frac{Tmol}{yr}$ N$_2$O & 47.5 & 20.3 \\

\hline                  
\end{tabular}
\end{table*}

\subsubsection{TRAPPIST-1 like star}\label{sec_trapn2o}

The results for an Earth-like planet orbiting a TRAPPIST-1 like star at 5 pc distance are listed in Table \ref{table:N2Otrap} and shown in Figure \ref{fig:N20_50d} for a 50 day observation (more cases simulating 10 and 100 days of integration time are shown in Appendix \ref{app:n2o_times},  Figures \ref{fig:N20_10d} and \ref{fig:N20_100d}). Similar to the above cases for the K6V and Proxima Centauri type host stars, the 100 and 10 Tmol/yr case cross the detection threshold in the 10 day simulations with the SNR metrics at 4.8/9.7 and 3.4/6.1 respectively.
The 1 Tmol/yr case only seems on the edge of detectability even after 100 days reaching bandintegrated SNR of 3.0.

\begin{table*}[ht]
\caption{Modelled SNR ratios for various integration times and different fluxes of N$_2$O at 5 pc TRAPPIST-1 like star}             
\label{table:N2Otrap}      
\centering          
\begin{tabular}{ c r r r }     
\hline\hline       
 T$_{int}$ [d] & Feature & Band-int. SNR & Max. line sigma\\ 
\hline                    
10 & 1 $ \frac{Tmol}{yr}$ N$_2$O & 0.9 & 0.7 \\
10 & 10 $ \frac{Tmol}{yr}$ N$_2$O & 4.8 & 3.4 \\
10 &100 $ \frac{Tmol}{yr}$ N$_2$O & 9.7 & 6.1 \\

\hline
50 & 1 $ \frac{Tmol}{yr}$ N$_2$O & 2.1 & 1.6 \\
50 & 10 $ \frac{Tmol}{yr}$ N$_2$O & 10.8 & 7.6 \\
50 &100 $ \frac{Tmol}{yr}$ N$_2$O & 21.8 & 13.7 \\

\hline
100 & 1 $ \frac{Tmol}{yr}$ N$_2$O & 3.0 & 2.2 \\
100 & 10 $ \frac{Tmol}{yr}$ N$_2$O & 15.2 & 10.7 \\
100 &100 $ \frac{Tmol}{yr}$ N$_2$O & 30.8 & 19.3 \\

\hline                  
\end{tabular}
\end{table*}

\subsection{Detectability of CH$_3$X}

Similarly to the N$_2$O cases discussed in the last section, we modelled the detectability of CH$_3$X for a set of observation times of 10 days, 50 days and 100 days. Again, we used the two metrics for the detectability introduced above: a maximum difference in single line in units of the respective sensitivity in that channel and a bandpass-integrated SNR. 

\subsubsection{Proxima Centauri type star}

The results for an Earth-like planet orbiting a Proxima Centauri (M5.5 Ve) type star at 5 pc distance from the Sun are listed in Table \ref{table:CH3Xprox} and shown in Figure \ref{fig:CH3x50_prox} for a 50 day observation (modelled observations of 10 and 100 days are shown in Appendix \ref{app:ch3x_times}, Figures \ref{fig:prox_10d} - \ref{fig:prox_100d}). After ten days the 100x and 10x CH$_3$X and 100x CH$_3$Cl cases are detectable with SNRs above 10/5 in the two metric used here. After 50 days the 100x CH$_3$Br case becomes detectable, reaching SNRs of 14.4/7.1. After 100 days all the 1x cases are still below our detectability limit, while all the 10x and 100x cases seem detectable.

\begin{figure}[ht]
\centering

\includegraphics[width=.3\textwidth]{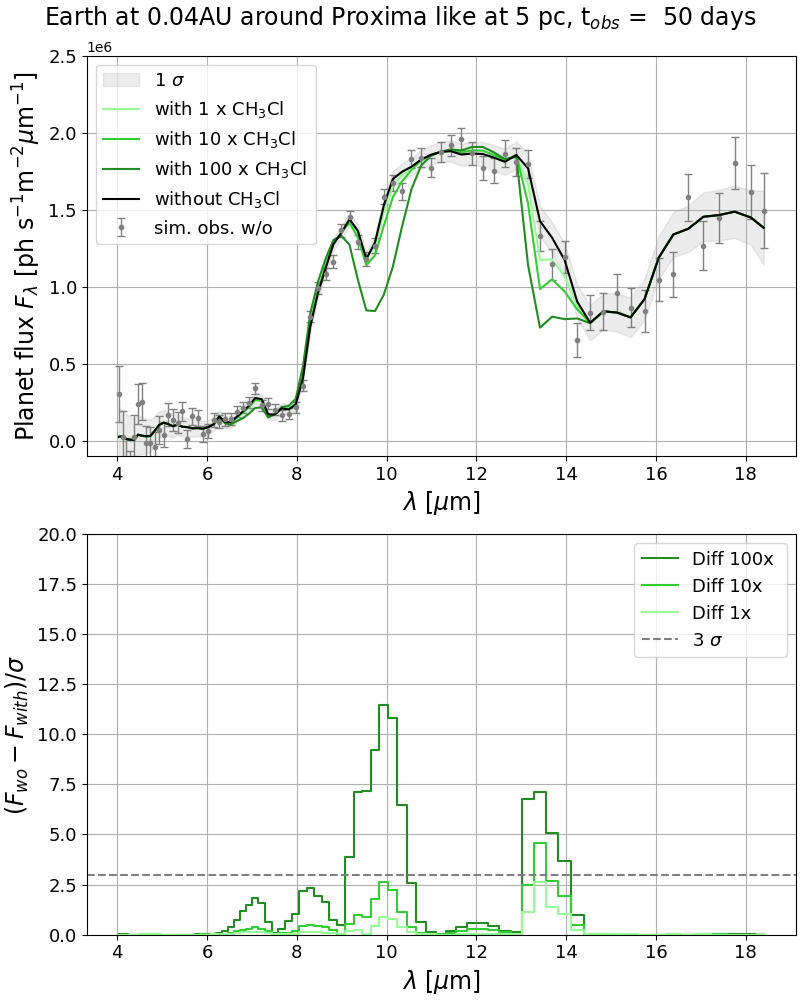}
\includegraphics[width=.3\textwidth]{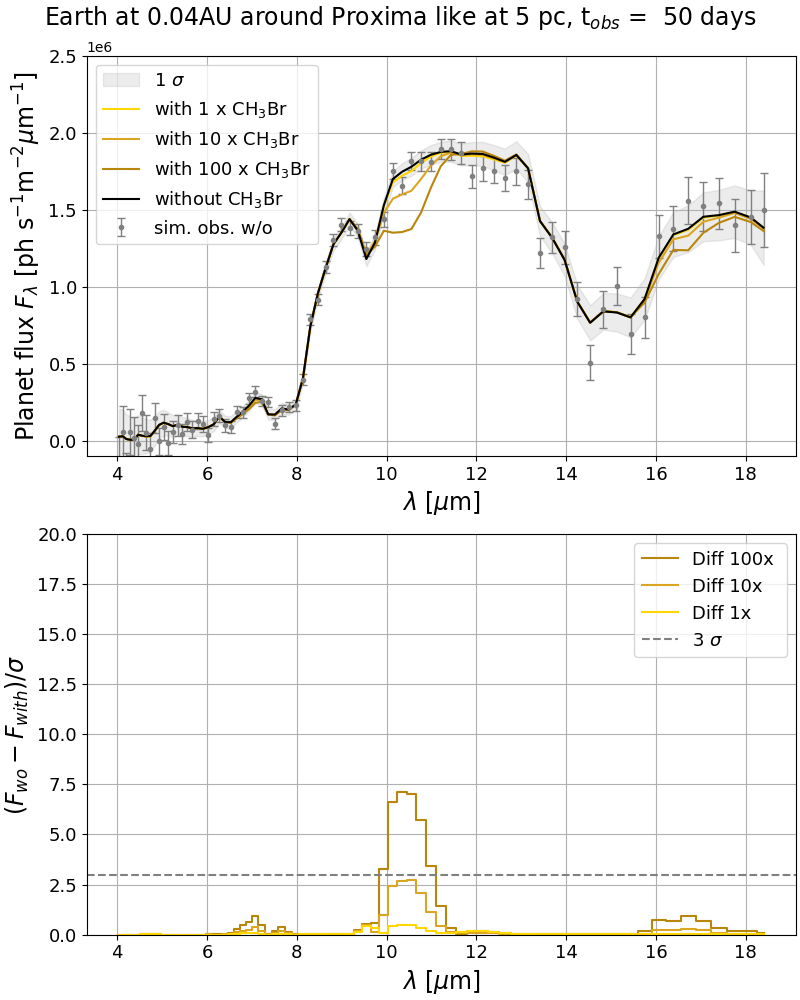}
\includegraphics[width=.3\textwidth]{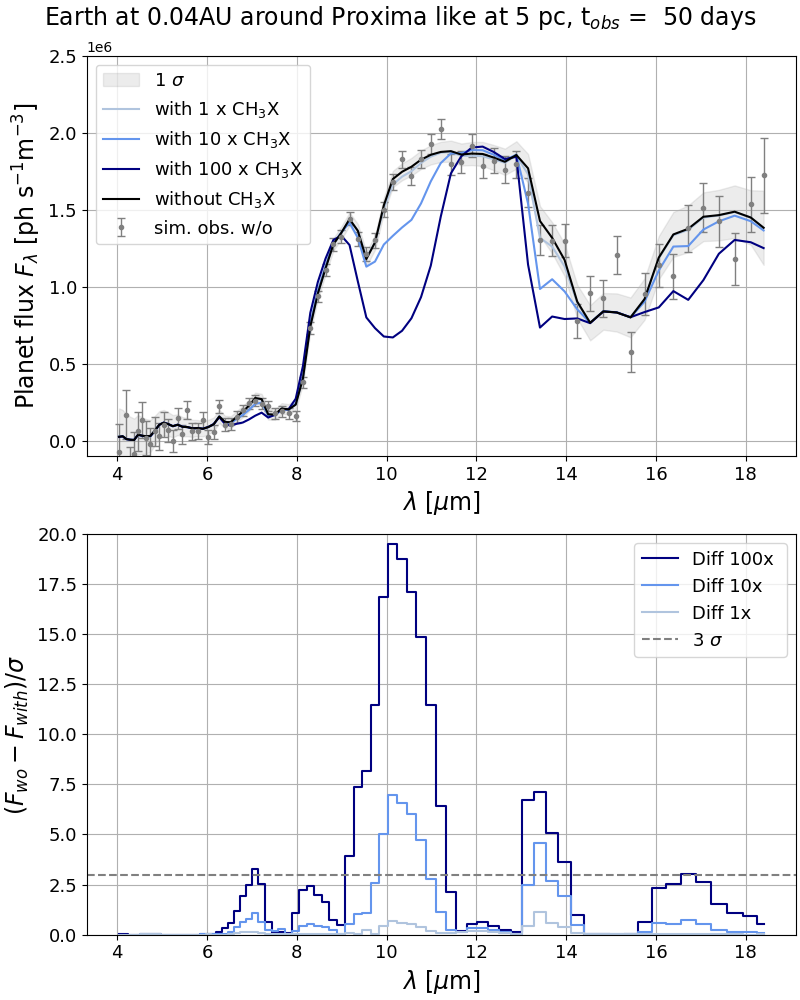}
\caption{Detectability of various levels of CH$_3$Cl and CH$_3$Br fluxes in the emission spectrum of an Earth-like planet around an Proxima Centauri type star, after 50 days of
observation with \textit{LIFE}. Top: planet flux for atmospheres with and without CH$_3$Cl and CH$_3$Br. The grey area represents
the 1-$\sigma$ sensitivity; the grey error bars show an individual simulated observation. Bottom: Statistical
significance of the detected differences between an atmospheric model with and without CH$_3$Cl and CH$_3$Br.}
\label{fig:CH3x50_prox}
\end{figure}

\begin{table*}[ht]
\caption{Modelled SNR ratios for various integration times and different fluxes of CH3X around Proxima Centauri type host at 5 pc}             
\label{table:CH3Xprox}      
\centering          
\begin{tabular}{ c r r r }     
\hline\hline       
 T$_{int}$ [d] & Feature & Band-int. SNR & Max. line sigma\\ 
\hline                    
10  & 1x CH$_3$X & 0.8 & 0.5\\

10  & 10x CH$_3$X & 6.9 & 3.1\\

10  & 100x CH$_3$X & 21.0 & 8.7\\

10  & 1x CH$_3$Cl & 1.6 & 1.2\\

10  & 10x CH$_3$Cl & 3.4 & 2.0\\

10  & 100x CH$_3$Cl & 11.5 & 5.1\\

10  & 1x CH$_3$Br & 0.5 & 0.2\\

10  & 10x CH$_3$Br & 2.4 & 1.2\\

10  & 100x CH$_3$Br & 6.4 & 3.2\\
\hline

50  & 1x CH$_3$X & 1.9 & 1.1\\

50  & 10x CH$_3$X & 15.4 & 7.0\\

50  & 100x CH$_3$X & 46.9 & 19.5\\

50  & 1x CH$_3$Cl & 3.6 & 2.6\\

50  & 10x CH$_3$Cl & 7.6 & 4.6\\

50  & 100x CH$_3$Cl & 25.8 & 11.5\\

50  & 1x CH$_3$Br & 1.1 & 0.5\\

50  & 10x CH$_3$Br & 5.3 & 2.7\\

50  & 100x CH$_3$Br & 14.4 & 7.1\\
\hline

100  & 1x CH$_3$X & 2.7 & 1.6\\

100  & 10x CH$_3$X & 21.8 & 9.9\\

100  & 100x CH$_3$X & 66.3 & 27.6\\

100  & 1x CH$_3$Cl & 5.1 & 3.7\\

100  & 10x CH$_3$Cl & 10.8 & 6.4\\

100  & 100x CH$_3$Cl & 36.5 & 16.2\\

100  & 1x CH$_3$Br & 1.6 & 0.7\\

100  & 10x CH$_3$Br & 7.5 & 3.8\\

100  & 100x CH$_3$Br & 20.3 & 10.1\\
\hline                  
\end{tabular}
\end{table*}

\subsubsection{AD Leonis type star}

The results for an Earth-like planet orbiting a AD Leonis type  M3.5V star at 5 pc distance from the Sun are listed in Table \ref{table:CH3xADL} and shown in Figure \ref{fig:CH3x50_ADL} for a 50 day observation (more cases for 10 and 100 day observations are shown in Appendix \ref{app:ch3x_times}, Figures \ref{fig:ADL_10d} - \ref{fig:ADL_100d}). In this scenario only the 100x and (since this case is dominated by CH$_3$Cl) 100x cases are detectable after 10 days with SNRs of 9.8/4.6 and 8.7/4.6. It would, however take more than 50 and up to 100 days in the \textit{LIFE} baseline setup to make the various 10x cases observable. All of the cases with 1x and the 10x CH$_3$Br case do not cross the threshold if detectability even after 100 days.

\begin{figure}[ht]
\centering
\includegraphics[width=.3\textwidth]{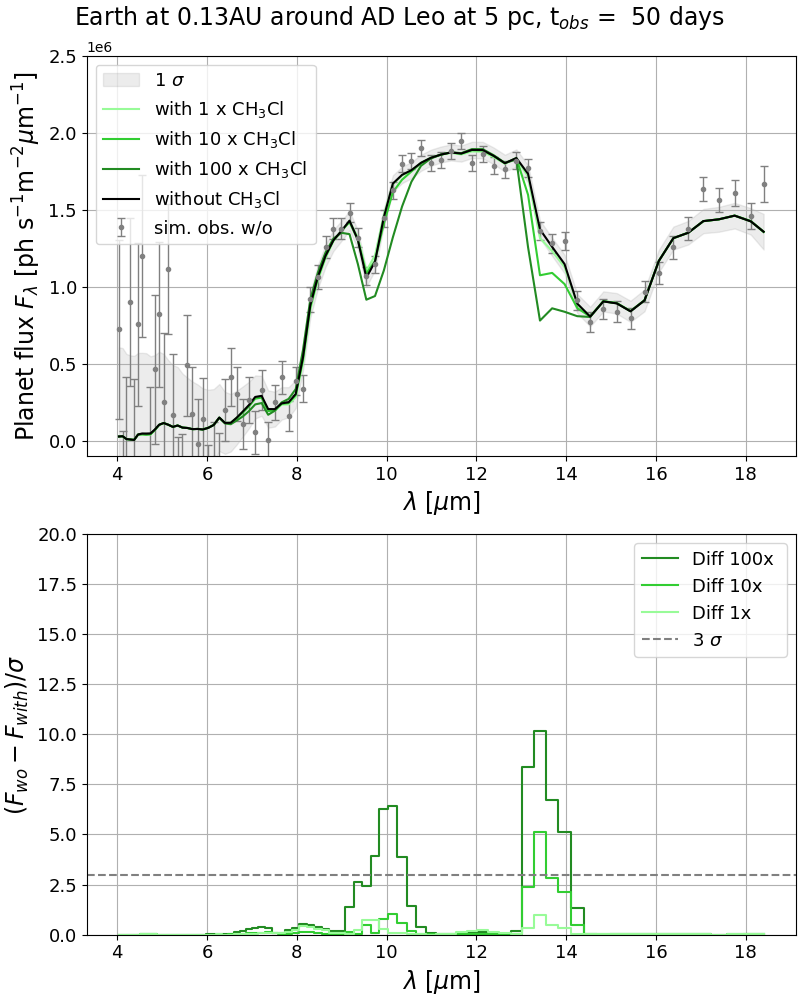}
\includegraphics[width=.3\textwidth]{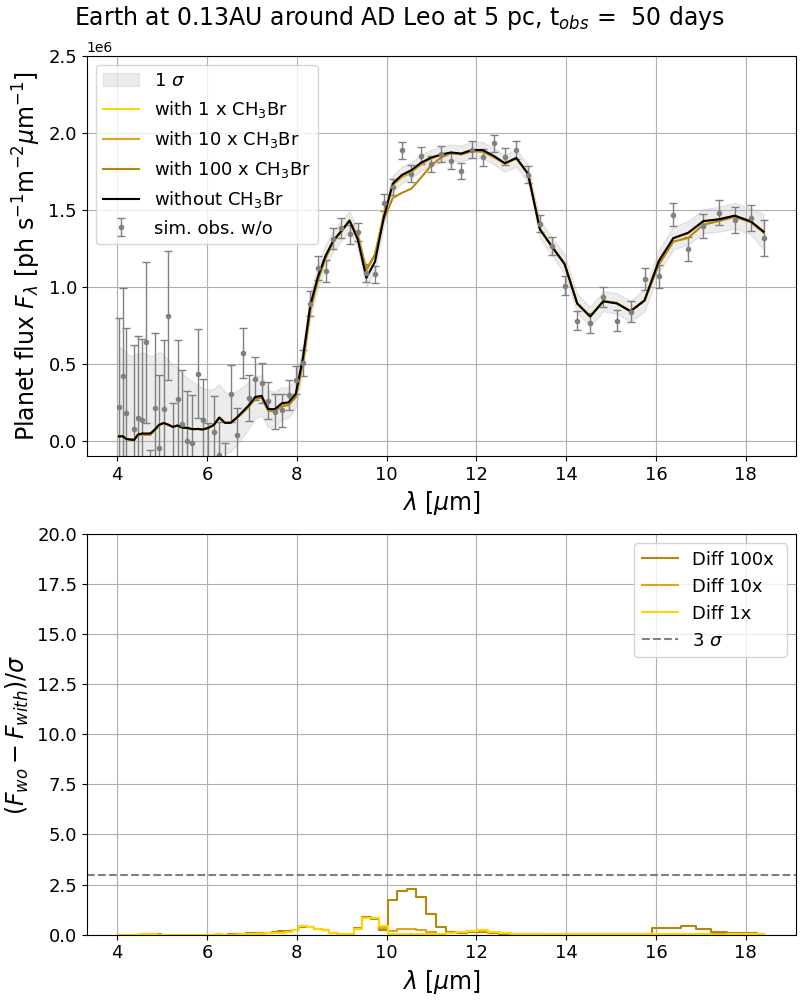}
\includegraphics[width=.3\textwidth]{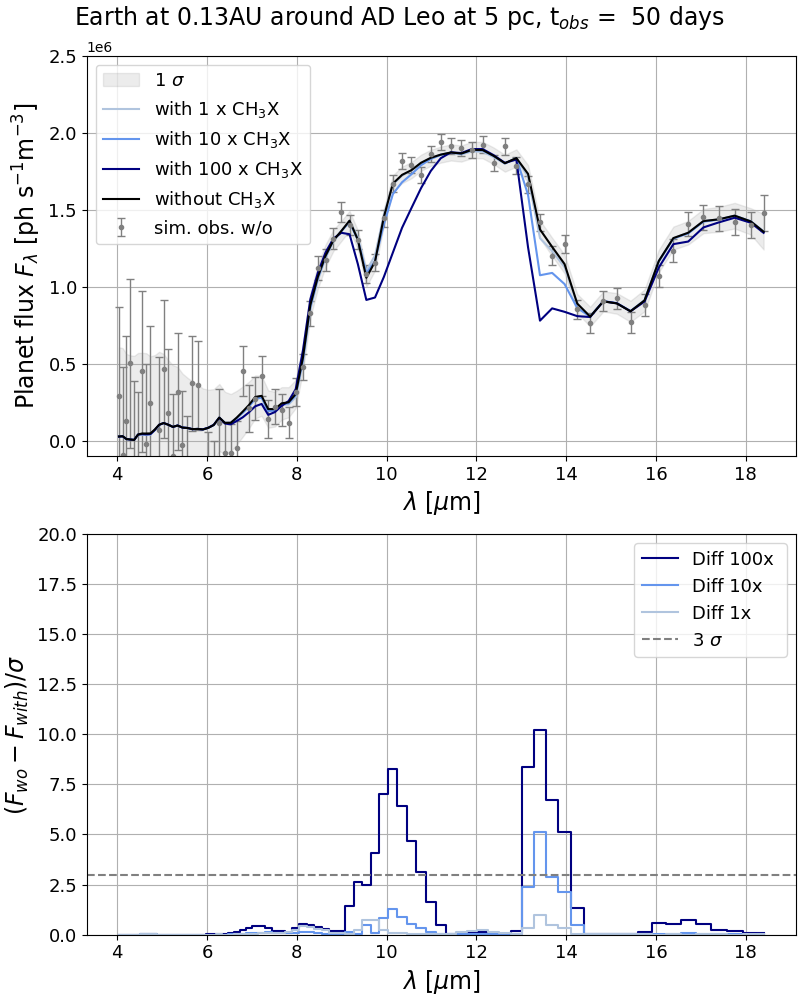}

\caption{Detectability of various levels of CH$_3$Cl and CH$_3$Br fluxes in the emission spectrum of an Earth-like planet around an AD Leonis type star, after 50 days of
observation with\textit{LIFE}.Top: planet flux for atmospheres with and without CH$_3$Cl and CH$_3$Br. The grey area represents
the 1-$\sigma$ sensitivity; the grey error bars show an individual simulated observation. Bottom: Statistical significance of the detected differences between an atmospheric model with and without CH$_3$Cl and CH$_3$Br.}
\label{fig:CH3x50_ADL}
\end{figure}

\begin{table*}[ht]
\caption{Modelled SNR ratios for various integration times and different fluxes of CH3X around AD Leonis type host at 5 pc}             
\label{table:CH3xADL}      
\centering          
\begin{tabular}{ c r r r }     
\hline\hline       
 T$_{int}$ [d] & Feature & Band-int. SNR & Max. line sigma\\ 
\hline                    
10  & 1x CH$_3$X & 0.8 & 0.4\\

10  & 10x CH$_3$X & 3.1 & 2.3\\

10  & 100x CH$_3$X & 9.8 & 4.6\\

10  & 1x CH$_3$Cl & 0.8 & 0.4\\

10  & 10x CH$_3$Cl & 3.1 & 2.3\\

10  & 100x CH$_3$Cl & 8.7 & 4.6\\

10  & 1x CH$_3$Br & 0.7 & 0.4\\

10  & 10x CH$_3$Br & 0.7 & 0.4\\

10  & 100x CH$_3$Br & 2.0 & 1.0\\

\hline
50  & 1x CH$_3$X & 1.8 & 1.0\\

50  & 10x CH$_3$X & 7.0 & 5.1\\

50  & 100x CH$_3$X & 21.8 & 10.2\\

50  & 1x CH$_3$Cl & 1.8 & 1.0\\

50  & 10x CH$_3$Cl & 6.9 & 5.1\\

50  & 100x CH$_3$Cl & 19.4 & 10.2\\

50  & 1x CH$_3$Br & 1.5 & 0.8\\

50  & 10x CH$_3$Br & 1.6 & 0.8\\

50  & 100x CH$_3$Br & 4.5 & 2.3\\

\hline
100  & 1x CH$_3$X & 2.5 & 1.4\\

100  & 10x CH$_3$X & 9.9 & 7.2\\

100  & 100x CH$_3$X & 30.9 & 14.4\\

100  & 1x CH$_3$Cl & 2.5 & 1.4\\

100  & 10x CH$_3$Cl & 9.7 & 7.2\\

100  & 100x CH$_3$Cl & 27.4 & 14.4\\

100  & 1x CH$_3$Br & 2.1 & 1.2\\

100  & 10x CH$_3$Br & 2.2 & 1.2\\

100  & 100x CH$_3$Br & 6.4 & 3.2\\
\hline                  
\end{tabular}
\end{table*}

\subsubsection{K6V type star}
The results for an Earth-like planet orbiting a K6V type star at 5 pc distance from the Sun are listed in Table \ref{table:CH3xK6V} and shown in Figure \ref{fig:CH3x50_K6V} for a 50 day observation (more cases shown in Appendix \ref{app:ch3x_times}, Figures \ref{fig:K6V_10d} - \ref{fig:K6V_100d}). This is the least favorable case discussed in this paper and almost none of the cases seem to be observable. Only for 100x CH$_3$X and 100x CH$_3$Cl after 100 days of observation do we barely cross the detection threshold with our SNR metrics at 7.8 and 4.9 repspectively.

\begin{figure}[ht]
\centering
\includegraphics[width=.3\textwidth]{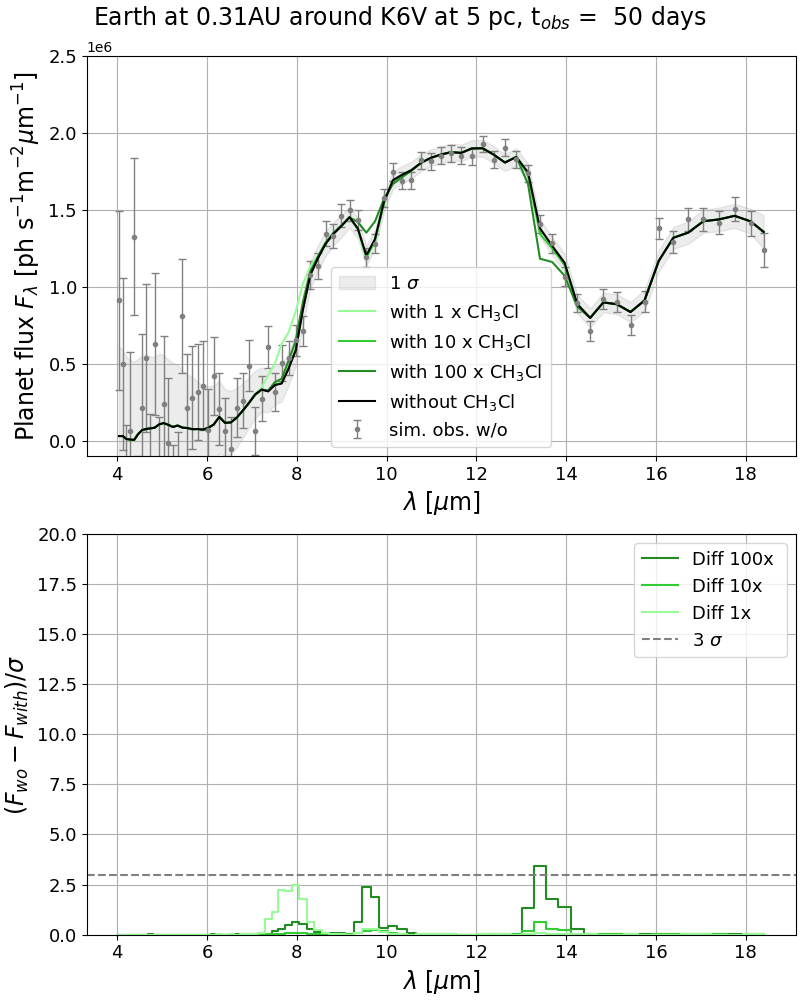}
\includegraphics[width=.3\textwidth]{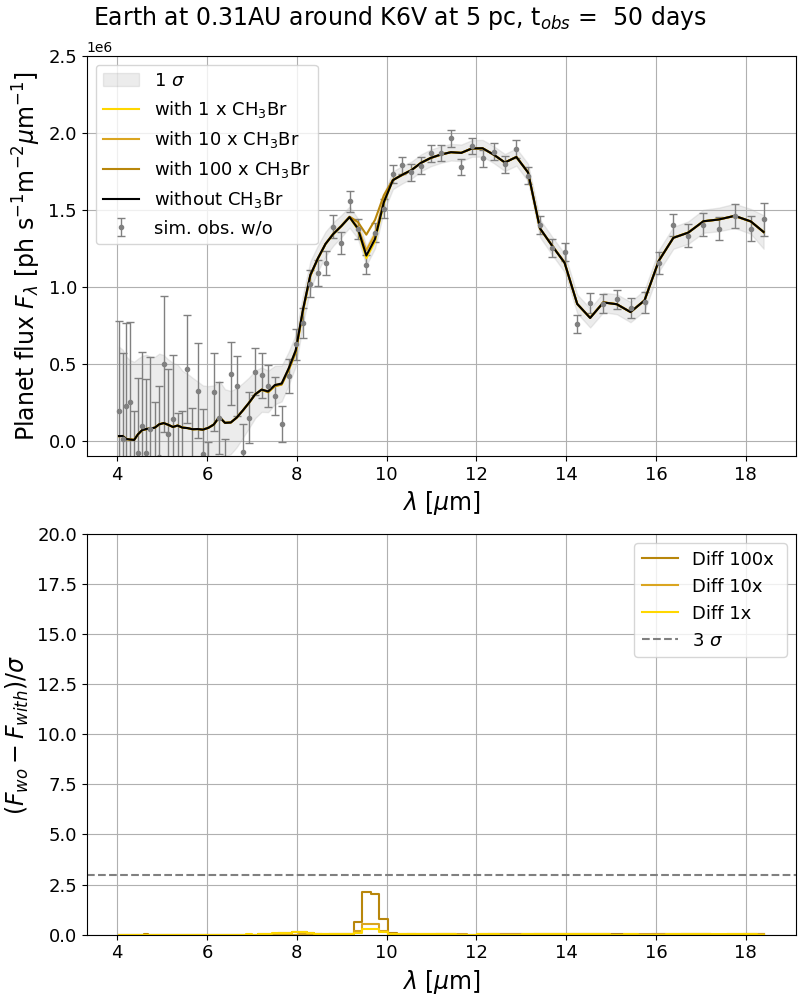}
\includegraphics[width=.3\textwidth]{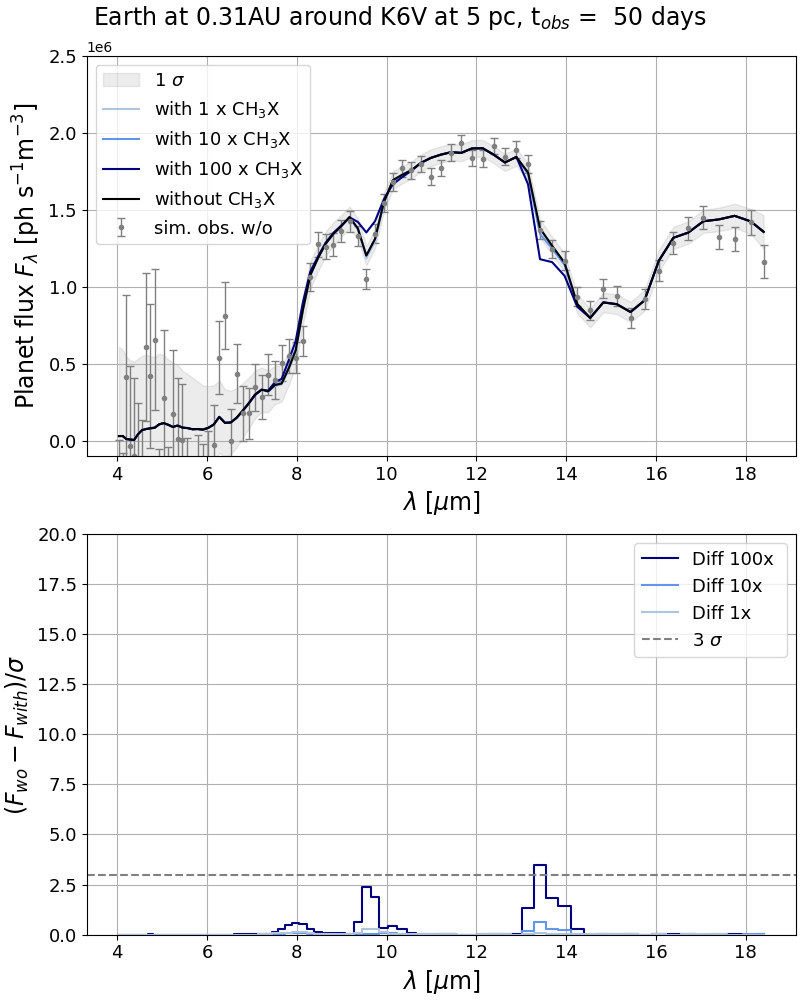}

\caption{Detectability of various levels of CH$_3$Cl and CH$_3$Br fluxes in the emission spectrum of an Earth-like planet around a K6V type star, after 50 days of
observation with\textit{LIFE}.Top: planet flux for atmospheres with and without CH$_3$Cl and CH$_3$Br. The grey area represents
the 1-$\sigma$ sensitivity; the grey error bars show an individual simulated observation. Bottom: Statistical
significance of the detected differences between an atmospheric model with and without CH$_3$Cl and CH$_3$Br. None of these cases seem to be detectable.}
\label{fig:CH3x50_K6V}
\end{figure}

\begin{table*}[ht]
\caption{Modelled SNR ratios for various integration times and different fluxes of CH3X around K6V star at 5 pc}             
\label{table:CH3xK6V}      
\centering          
\begin{tabular}{ c r r r }     
\hline\hline       
 T$_{int}$ [d] & Feature & Band-int. SNR & Max. line sigma\\ 
\hline                    
10  & 1x CH$_3$X & 0.2 & 0.1\\

10  & 10x CH$_3$X & 0.4 & 0.3\\

10  & 100x CH$_3$X & 2.5 & 1.6\\

10  & 1x CH$_3$Cl & 2.1 & 1.1\\

10  & 10x CH$_3$Cl & 0.4 & 0.3\\

10  & 100x CH$_3$Cl & 2.4 & 1.5\\

10  & 1x CH$_3$Br & 0.2 & 0.1\\

10  & 10x CH$_3$Br & 0.3 & 0.2\\

10  & 100x CH$_3$Br & 1.4 & 1.0\\
\hline
50  & 1x CH$_3$X & 0.5 & 0.3\\

50  & 10x CH$_3$X & 0.8 & 0.6\\

50  & 100x CH$_3$X & 5.5 & 3.5\\

50  & 1x CH$_3$Cl & 4.6 & 2.5\\

50  & 10x CH$_3$Cl & 0.9 & 0.6\\

50  & 100x CH$_3$Cl & 5.4 & 3.4\\

50  & 1x CH$_3$Br & 0.5 & 0.3\\

50  & 10x CH$_3$Br & 0.8 & 0.5\\

50  & 100x CH$_3$Br & 3.1 & 2.1\\
\hline
100  & 1x CH$_3$X & 0.7 & 0.4\\

100  & 10x CH$_3$X & 1.1 & 0.9\\

100  & 100x CH$_3$X & 7.8 & 4.9\\

100  & 1x CH$_3$Cl & 6.5 & 3.5\\

100  & 10x CH$_3$Cl & 1.2 & 0.9\\

100  & 100x CH$_3$Cl & 7.7 & 4.8\\

100  & 1x CH$_3$Br & 0.7 & 0.4\\

100  & 10x CH$_3$Br & 1.1 & 0.8\\

100  & 100x CH$_3$Br & 4.4 & 3.0\\
  \hline             
\end{tabular}
\end{table*}

\subsubsection{TRAPPIST-1 type star}

The results for an Earth-like planet orbiting a TRAPPIST-1 type star at 5 pc distance from the Sun are listed in Table \ref{table:CH3xtrap} and shown in Figure \ref{fig:CH3x50_trap} for a 50 day observation (more cases for 10 and 100 days of observation are shown in Appendix \ref{app:ch3x_times}, Figures \ref{fig:trap_10d} - \ref{fig:trap_100d}). 
For this case the 100x CH$_3$X case is detectable wit SNRs of 9.3/4.0 already after 10 days. The 10x CH$_3$X (10.3/4.7) as well as the 100x CH$_3$Cl (11.8/5.2) become detectable after 50 days. Even after 100 days of simulated observation do all 1x cases and the 10x CH$_3$Cl and CH$_3$Br stay undetectable.

\begin{figure}[ht]
\centering
\includegraphics[width=.3\textwidth]{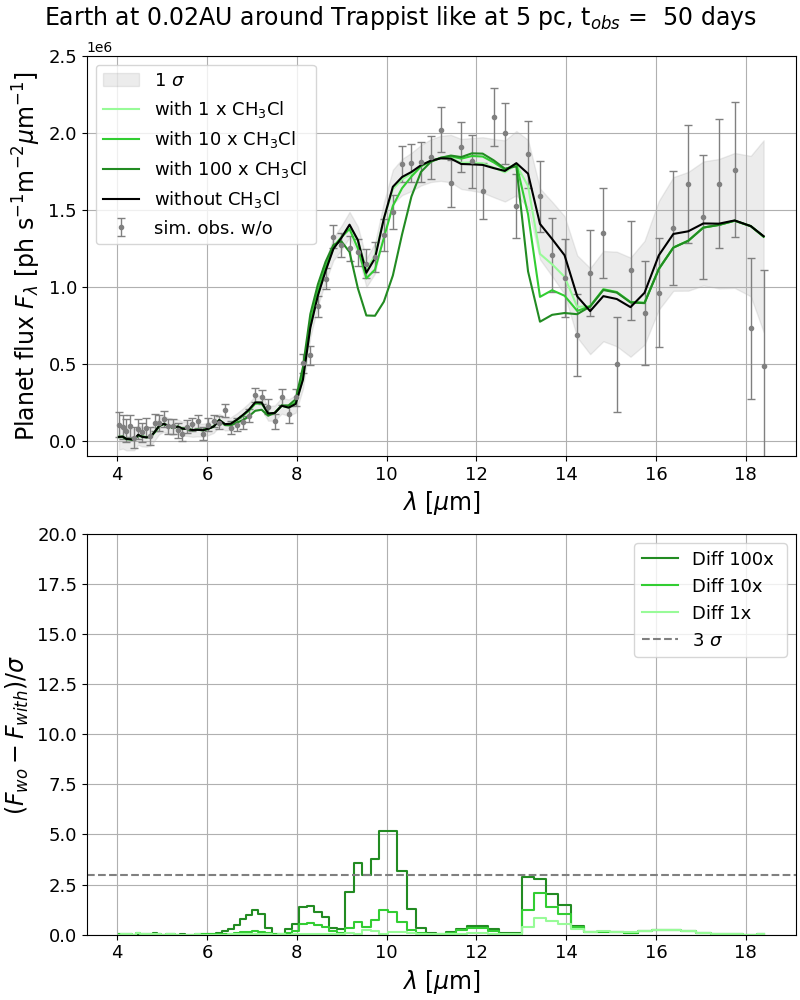}
\includegraphics[width=.3\textwidth]{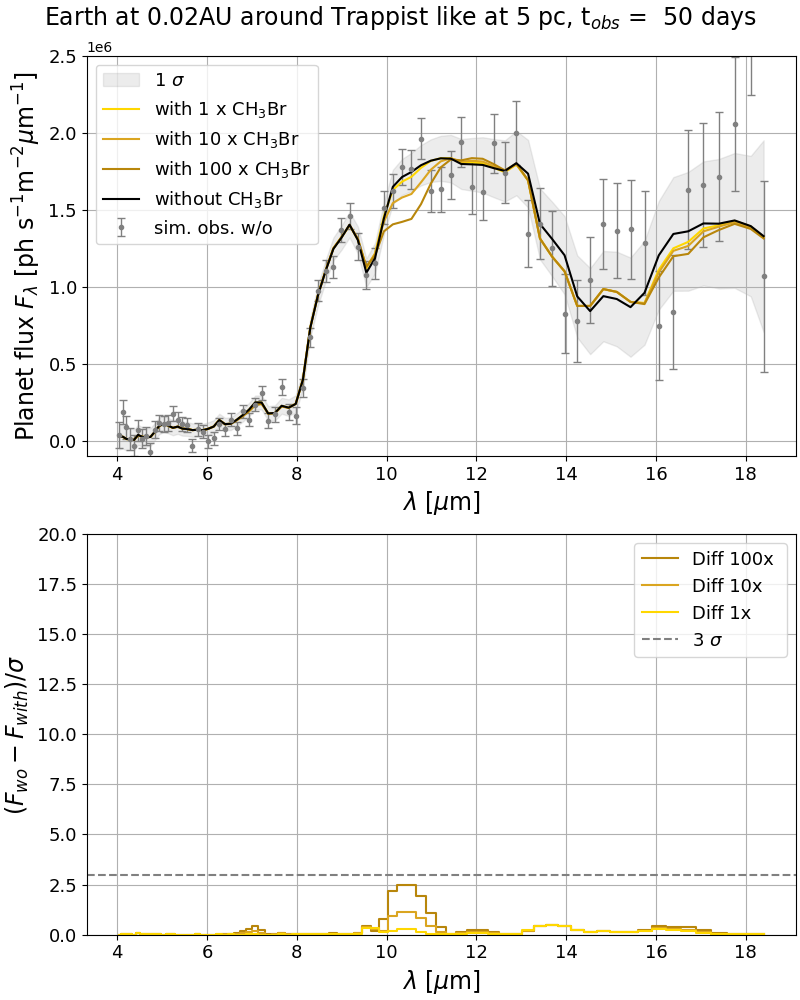}
\includegraphics[width=.3\textwidth]{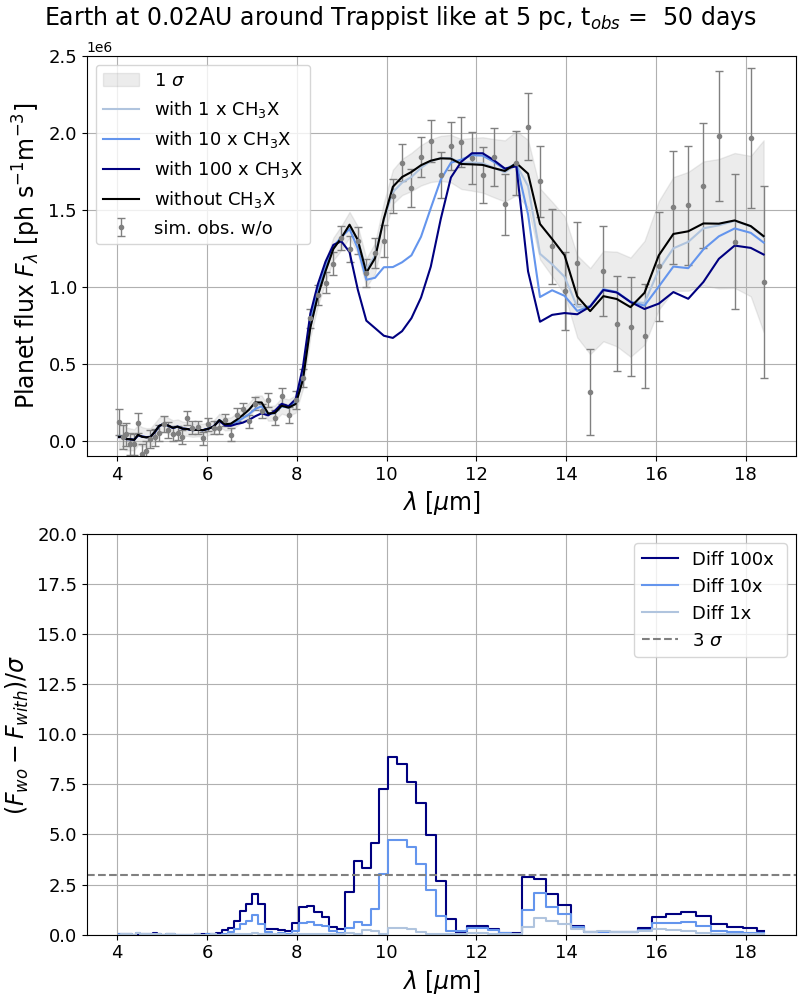}

\caption{Detectability of various levels of CH$_3$Cl and CH$_3$Br fluxes in the emission spectrum of an Earth-like planet around a TRAPPIST-1 type star, after 50 days of
observation with\textit{LIFE}.Top: planet flux for atmospheres with and without CH$_3$Cl and CH$_3$Br. The grey area represents
the 1-$\sigma$ sensitivity; the grey error bars show an individual simulated observation. Bottom: Statistical
significance of the detected differences between an atmospheric model with and without CH$_3$Cl and CH$_3$Br.}
\label{fig:CH3x50_trap}
\end{figure}

 \begin{table*}[ht]
\caption{Modelled SNR ratios for various integration times and different fluxes of CH3X around TRAPPIST-like star at 5 pc}             
\label{table:CH3xtrap}      
\centering          
\begin{tabular}{ c r r r }     
\hline\hline       
 T$_{int}$ [d] & Feature & Band-int. SNR & Max. line sigma\\ 
\hline  
10  & 1x CH$_3$X & 0.7 & 0.4\\

10  & 10x CH$_3$X & 4.6 & 2.1\\

10  & 100x CH$_3$X & 9.3 & 4.0\\

10  & 1x CH$_3$Cl & 0.6 & 0.4\\

10  & 10x CH$_3$Cl & 1.7 & 0.9\\

10  & 100x CH$_3$Cl & 5.3 & 2.3\\

10  & 1x CH$_3$Br & 0.5 & 0.2\\

10  & 10x CH$_3$Br & 1.1 & 0.5\\

10  & 100x CH$_3$Br & 2.2 & 1.1\\
\hline
50  & 1x CH$_3$X & 1.5 & 0.8\\

50  & 10x CH$_3$X & 10.3 & 4.7\\

50  & 100x CH$_3$X & 20.8 & 8.9\\

50  & 1x CH$_3$Cl & 1.5 & 0.8\\

50  & 10x CH$_3$Cl & 3.9 & 2.1\\

50  & 100x CH$_3$Cl & 11.8 & 5.2\\

50  & 1x CH$_3$Br & 1.2 & 0.5\\

50  & 10x CH$_3$Br & 2.4 & 1.2\\

50  & 100x CH$_3$Br & 5.0 & 2.5\\
\hline
100  & 1x CH$_3$X & 2.2 & 1.2\\

100  & 10x CH$_3$X & 14.6 & 6.7\\

100  & 100x CH$_3$X & 29.5 & 12.5\\

100  & 1x CH$_3$Cl & 2.1 & 1.2\\

100  & 10x CH$_3$Cl & 5.5 & 2.9\\

100  & 100x CH$_3$Cl & 16.7 & 7.3\\

100  & 1x CH$_3$Br & 1.7 & 0.7\\

100  & 10x CH$_3$Br & 3.4 & 1.6\\

100  & 100x CH$_3$Br & 7.0 & 3.5\\

\hline                  
\end{tabular}
\end{table*}

\subsection{Comparison to full spectral retrievals}\label{sec:retr_test}

In order to benchmark the results of the previously presented exercise of comparing spectra with and without the species present in the atmosphere, we performed a few exemplary retrievals. This was done in the following manner: the input spectra were produced as described in the sections above while our retrievals were performed with the retrieval suite described in \citet{LIFE3} and \citet{LIFE5}. The retrieval routine couples the forward model \textit{petitRADTRANS} \citep{2019A&A...627A..67M} with the parameter estimation module \textit{pyMultiNest} \citep{2014A&A...564A.125B}. The retrieval routine - in constrast to the forward models used for the input spetra - neglected clouds and assumed pressure-constant abundance profiles for every species included in the calculation. The spectral calculation within the retrieval was performed considering HITRAN 2020 opacities, assuming air broadening and a line wing cutoff at 25 cm$^{-1}$ from the line core. 

As shown in \cite{LIFE5}, neglecting clouds in retrievals when considering spectra of partly cloudy planets is overall not impacting a correct retrieval of the chemical composition of the atmosphere. However, the opacity linelists used in the retrievals are generally not consistent with the linelists used by PSG and SMART that produced the input spectra. As discussed in \citet{2022SPIE12180E..3LA}, differences in opacities might cause considerable differences in retrievals. Other differences in the implementation of scattering and raytracing methods might also generate biases. For this reason, an intercomparison among these three forward models (as well as other ones) has been ongoing within the community (Villanueva et al., in prep.) within the Climates Using Interactive Suites of Intercomparisons Nested for Exoplanet Studies (CUISINES) sponsored by the Nexus for Exoplanet System Science (NExSS).

Besides relaying the most basic system parameters (such as the distance of the system and stellar type of the host) this was done without any other prior knowledge of the atmosphere or about the process of generating the model spectra. This experimental setup reproduces the circumstances of an actual observation very well, in which case we also would not know any of the true values a priori. 

The main question we asked was for which cases (i.e. which levels of band-integrated/max line SNR) we would get a detection and/or upper limits for the presence of the species of interest.

\begin{figure*}[ht]
\centering
\includegraphics[width=.9\textwidth]{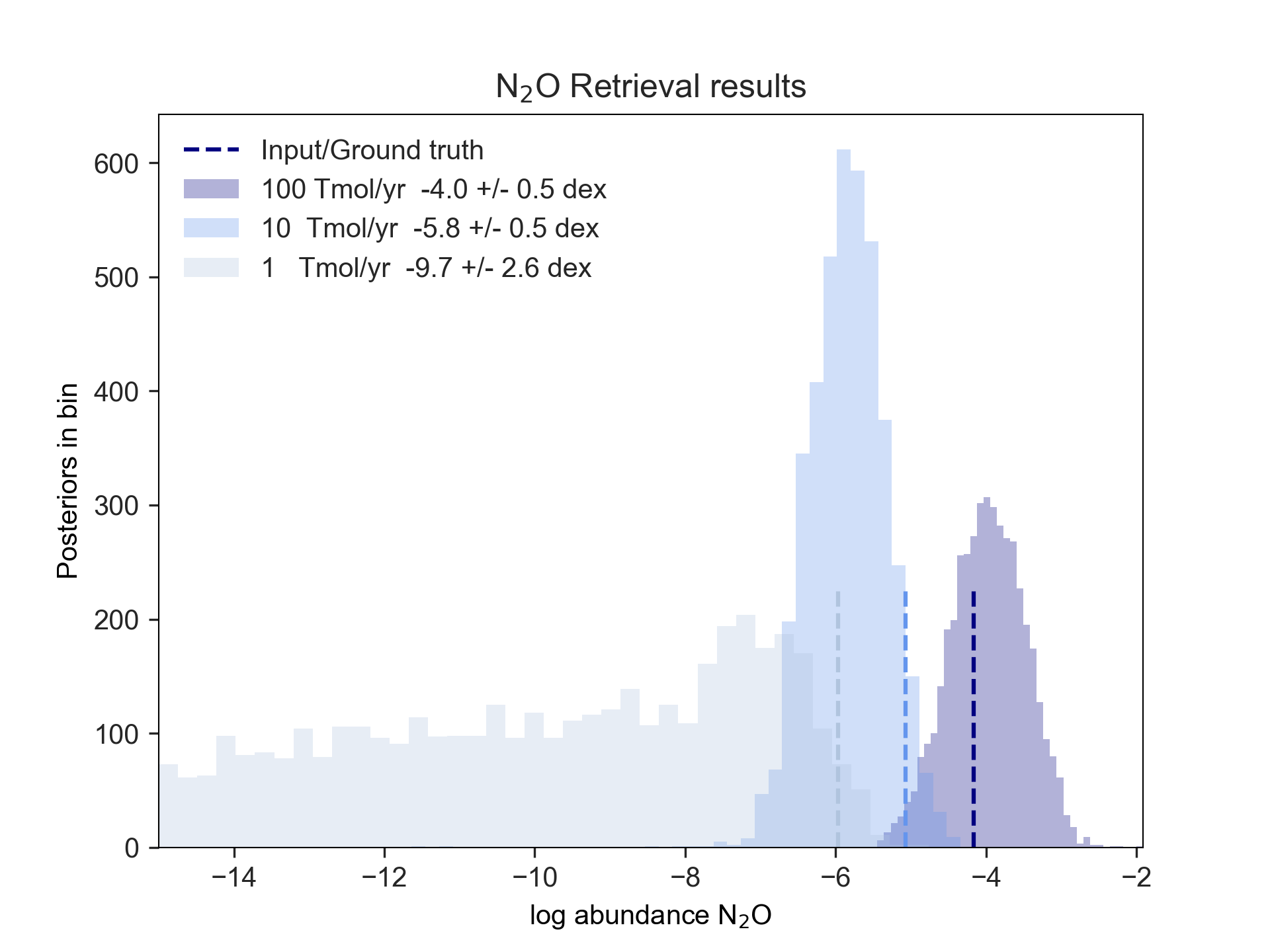}

\caption{Comparison - Retrieval. Histograms of the posterior distributions for N$_2$O abundances in the 3 retrieval cases compared to the ``ground truth" model values. While the 100 and 10 Tmol/yr case are retrieved within  $\sim$ 1.5 sigma, the 1 Tmol/yr case only produces an upper limit. Biases seen here can be largely attributed to a degeneracy between abundances (in particular H$_2$O and N$_2$O) as shown in the cornerplots in Appendix Section \ref{sec:app_retr}.  } \label{fig:comp_ret}
\end{figure*}

As example we chose the \textit{N$_2$O Earth at 5 pc around Sun-like star 50d} cases and ran retrievals on the the spectra for 1, 10 and 100 Tmol/y fluxes of N$_2$O. In the direct comparison analysis above these slightly more challenging cases cases were distiguishable from a no N$_2$O case at a band integrated SNR of about 1.5,5.1 and 14.7 (max. channel SNR of 0.9, 2.7, 7.5). This gives us the opportunity to compare our ability to retrieve the N$_2$O abundances to one order of magnitude in our SNR metric, giving as a first order estimate of the detection threshold in that metric. Furthermore it connects this study with \citet{LIFE3} and \citet{LIFE5} who also focused on Sun-like hosts in their studies of Earth-twins.

The results are shown in Figure \ref{fig:comp_ret}, where we show the histograms of the posteriors for the N$_2$O abundances for the different cases, in comparison to the ``ground truth" of the input spectra. All our retrievals derive N$_2$O abundances within 1.5 sigma of the input value of the original models run through the \textit{LIFE} observation simulator. For the 1 Tmol/yr case we can only retrieve an upper limit. More details for the retrievals, such as corner plots and values for all retrieved parameters are presented in Appendix \ref{sec:app_retr}.

This analysis confirms our first order estimate that a band-integrated SNR of about 5-10 should be sufficient to detect the feature in question and justifies our choice of this line as a rational, rather pessimistic detection limit in our reasoning.

\section{Discussion and conclusions}\label{sec:disc}

\subsection{Wavelength dependent sensitivity between different cases}\label{sec:wl_dep}
In our simulations, the sensitivity of the interferometric measurement is fundamentally limited by astrophysical noise sources. In the longer wavelength regime of the simulated measurement, this astrophysical noise is dominated by the thermal emission of the local-zodical dust in the solar system. 
The diffuse nature of the emission makes the amount of local-zodi induced noise that is present after the nulling independent from the length of the chosen interferometric baseline. Therefore, towards longer wavelengths the instrument sensitivity is independent from the observed target. \footnote{The local-zodi noise does strongly depend on the position of the target in relation to the ecliptic plane \citet{LIFE2}, which is kept equal ($lat_s=0.78$) for all cases simulated in the ‘characterization phase’ in this study.} 
At short wavelengths, the noise is dominated by the host star. As the flux ratio between the exoplanet and its host star becomes more favourable towards later stellar types, the sensitivity of the observations increase for such star types. Coupled with more prominent spectra features, this results in the demonstrated higher significance detection in the shorter wavelength bands for the M-star targets compared to the G- and K-star targets.

\subsection{Golden Targets and dynamic atmospheric signatures}\label{sec:golden_t}
\begin{figure*}[ht]
\centering
\includegraphics[width=.9\textwidth]{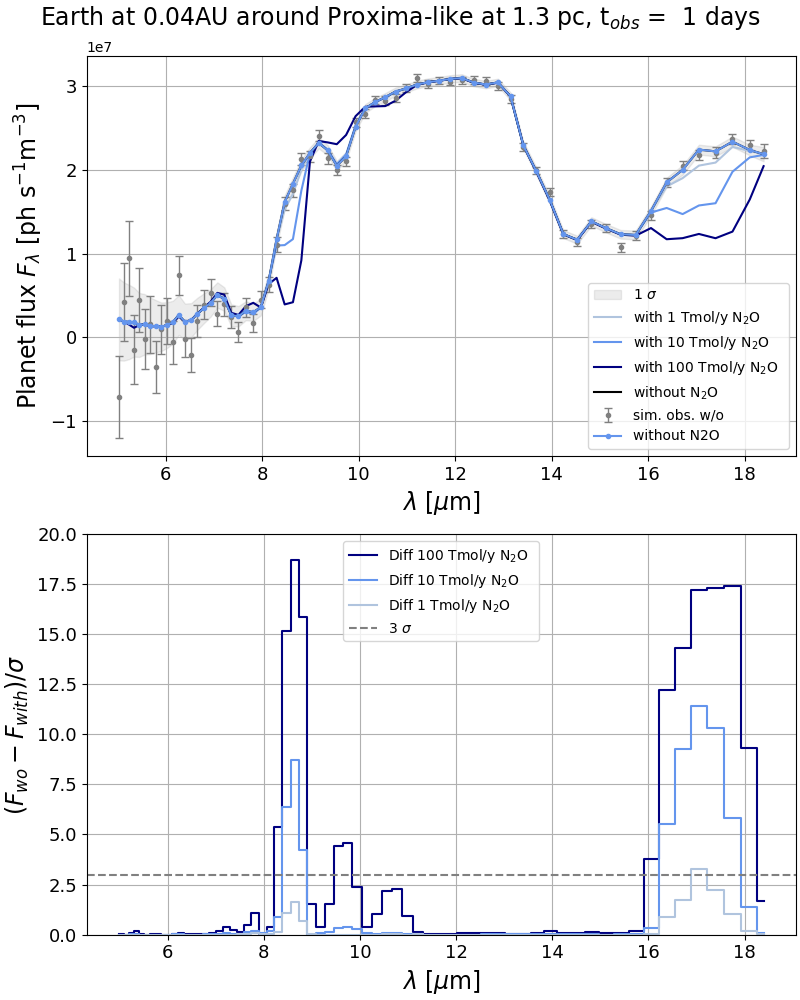}

\caption{Detectability of various fluxes of N$_2$O in the emission spectrum of an Earth-like planet around Proxima Centauri at its original distance, after only 1 day of
observation with \textit{LIFE}. Top: planet flux for atmospheres with and without N$_2$O . The grey area represents
the 1-$\sigma$ sensitivity; the dark grey error bars show an individual simulated observation. Bottom: Statistical
significance of the detected differences between an atmospheric model with and without N$_2$O.}
\label{fig:n2o_prox_orig}
\end{figure*}

One of the most interesting outcomes of this analysis is the expected quality of data obtainable for the closest target planets.
For example for the Proxima Centauri case at original distance (see Fig. \ref{fig:n2o_prox_orig}) or similar cases within $<$ 3pc distance from the Sun , ``spectral temporal resolution" of a few days is possible, meaning that changes caused by fluxes of the same magnitude assumed in our models could be detected on timescales of days or weeks. This could be used to characterize exoplanetary variability and dynamic processes in the atmospheres.

\subsection{Consequences for Science and Technology Requirements}\label{sec:tec_req}

\begin{figure}[ht]
\centering
\includegraphics[width=.53\textwidth]{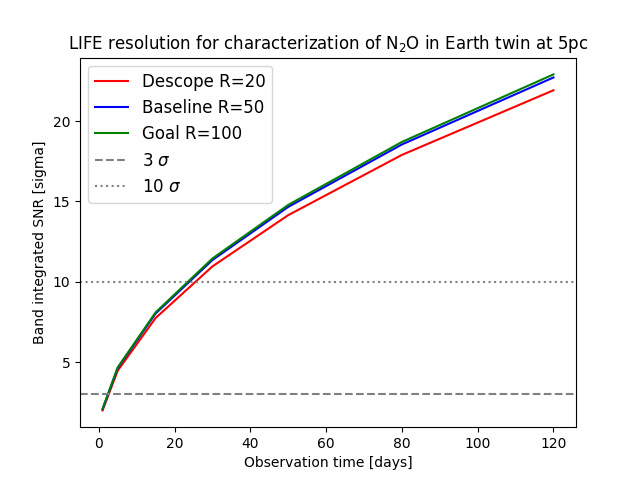}\quad
\includegraphics[width=.53\textwidth]{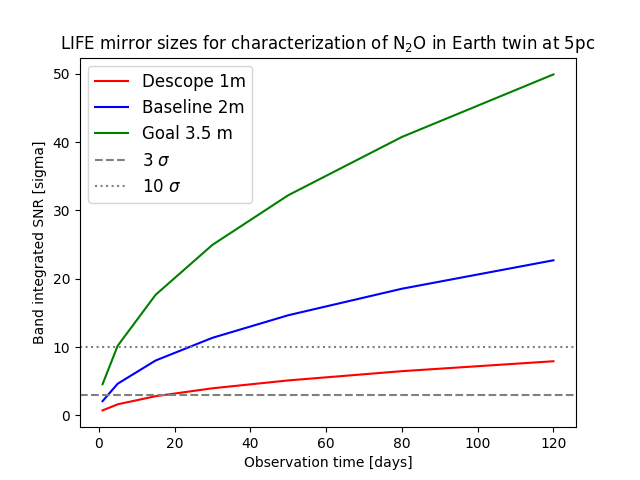}\quad
\includegraphics[width=.53\textwidth]{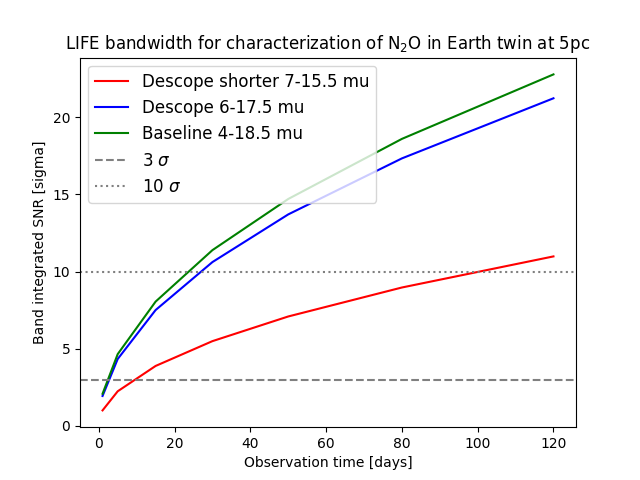}\quad

\caption{Dependence of observation time needed to detect 10 Tmol/y N$_2$O in the Earth twin around Sun like star at 5 pc example case on various mission parameters.}
\label{fig:tec_req}
\end{figure}

Fig. \ref{fig:tec_req} shows the dependence of the observation time needed to constrain N$_2$O for the 10 Tmol/y Earth twin case with various design parameters of the \textit{LIFE} mission. 
The bottom plot of Figure \ref{fig:tec_req} shows a relatively weak dependence on the shorter wavelength cut off but a very strong dependence on the longer wavelength which is explained by the location of the N$_2$O features in the MIR for the G star case.\footnote{It is important to note that in full retrievals, the contextual information provided by other wavelength bands is very important to constrain the abundances. This fact is not included in this analysis and should hence be taken with a grain of salt.} The Figure \ref{fig:tec_req} center panel shows how for the 2~m mirror size at 5\% throughput (or an equivalent size/throughput combination) it would take about 15 days to reach the 10 sigma band integrated threshold while it would only take about 5 days in the 3.5m configuration. For a descoped 1m setup we would not reach that in a reasonable time ($>$ 100 days). Fig. \ref{fig:tec_req} top panel shows almost no dependence on spectral resolution, due to the relatively wide bands of N$_2$O.

\subsection{Synergies and comparison with other observational platforms}\label{sec:syn_dis}

\begin{figure*}[h]
\centering
\includegraphics[width=.9\textwidth]{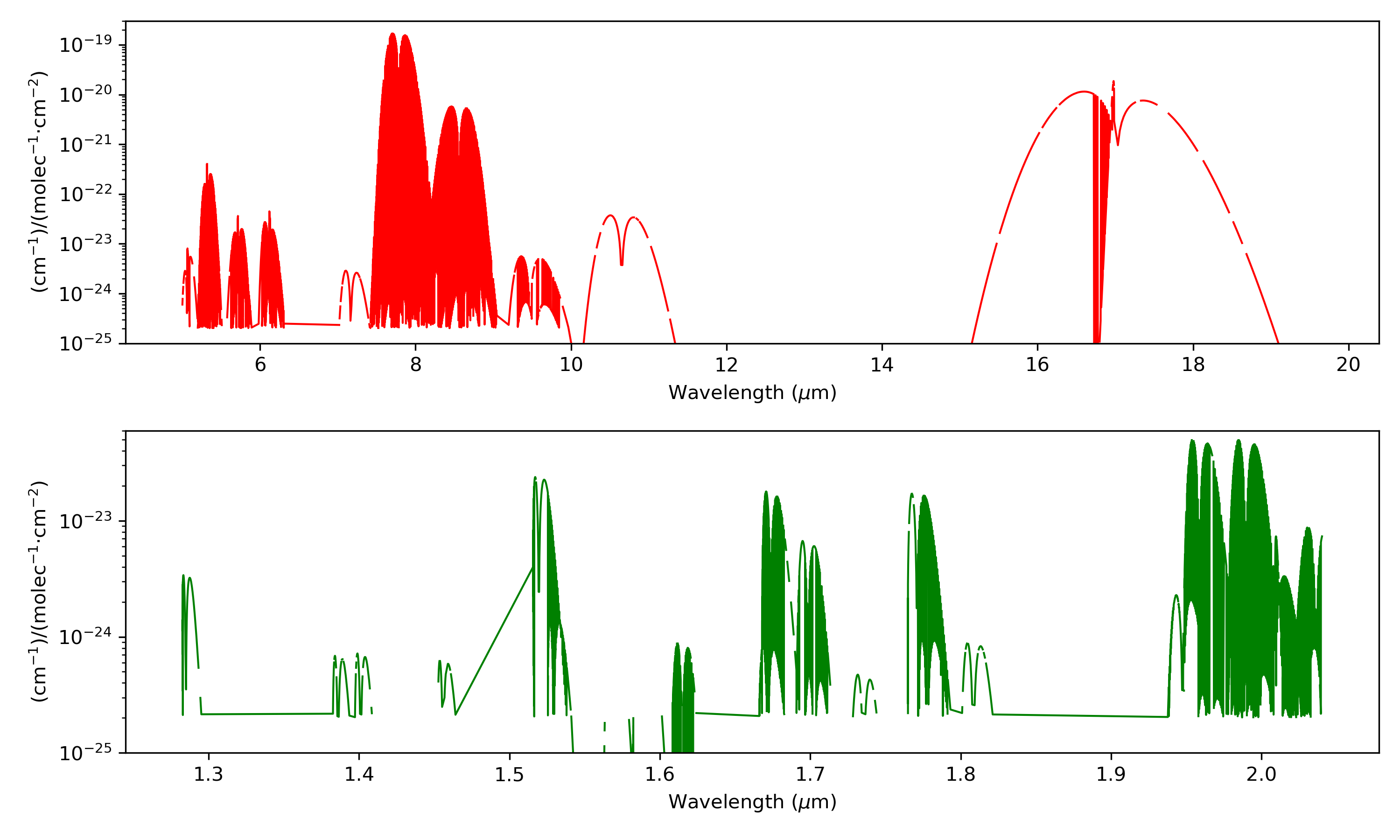}

\caption{HITRAN line intensities for N$_2$O showing orders of magnitude stronger features in the MIR vs. the NIR. This comparison is not possible for CH$_3$Br/CH$_3$Cl, because their absolute cross-sections have not been systematically measured in the 0.2-2 micron region relevant to reflected light observations of terrestrial planets (e.g., HWO). }
\label{fig:n2o_hitran}
\end{figure*}

\citet{2022ApJ...937..109S} and \citet{2022ApJ...938....6L} showed that some of the scenarios discussed here may be observable with JWST in transiting systems such as TRAPPIST-1. However these observations would need to add the signal from 10-100s of individual transits and will therefore only be feasible for very few, selected, transiting planets around very late type stars. On the other hand, it was shown that the long baselines of \textit{LIFE}, which result in extremely high spatial resolution will be particularly well suited for systems around late type stars with very close in habitable zones, that are also inaccessible to direct imaging techniques at shorter wavelengths \citep[][ see also Fig.\ref{fig:HZplanet_distro}]{LIFE2, LIFE10}. Furthermore N$_2$O shows orders of magnitude difference in the line intensities between the strongest MIR regions and the strongest NIR regions (see Fig.\ref{fig:n2o_hitran}).
The ability to characterize terrestrial temperate exoplanets around late type stars and the access to molecules that only show strong features in the MIR are two of the main synergies of \textit{LIFE} with HWO, which will mostly focus on FGK systems and observe in the UV to NIR range.

If there are late type star exoplanetary systems in the solar neighborhood with planets that exhibit global biospheres producing N$_2$O and CH$_3$X signals, \textit{LIFE} will be the best suited future mission to systematically search for and eventually detect them.

\subsection{Summary of results}

   \begin{enumerate}
      \item Our first order analysis of the detectability of exoplanetary spectra with the modelled fluxes of potentially biogenic gases shows that most of the discussed cases are indeed observable with a space telescope like LIFE with observation times between 10 and 100 days for habitable planetary systems within 5 pc.
      
      \item We compare some of these results to selected retrievals, confirm the detectability and generally show consistency with the direct comparison of forward models in the detectability metric used here.
      
        \item Using simulations of potential yields of the LIFE mission we can show that dozens of terrestrial temperate exoplanets can be tested for these molecules in the way discussed here. In terms of distance the cases at 5 pc are typical for M star host and likely the best cases for FGK hosts.
      
      \item For some ``golden targets" (e.g. a very nearby system like Proxima Centauri) \textit{LIFE} could detect levels of the discussed species on very short timescales, delivering high ``spectral-temporal" resolution of days to week to observe 10s of Tmol/yr fluxes, potentially opening up future avenues for variability studies.
      

      \item Comparison of the \textit{LIFE} baseline with other potential setups of the space telescope confirm observability also for different instrumental configuration and informs the mission design about trade offs driven by the detectability of these features.

   \end{enumerate}

\section*{Acknowledgements}
We thank the anonymous reviewer for their careful reading of our manuscript and their many insightful comments and suggestions.
This work has been carried out within the framework of the National Centre of Competence in Research PlanetS supported by the Swiss National Science Foundation under grants 51NF40\_182901 and 51NF40\_205606. EA acknowledges the financial support of the SNSF.
DP, ML, and ES gratefully acknowledge support from the NASA Exobiology program through grant No. 80NSSC20K1437, the NASA Interdisciplinary Consortia for Astrobiology Research (ICAR) Program via grant Nos. 80NSSC21K0905 and 80NSSC21K0905, and the Virtual Planetary Laboratory funded by NASA Astrobiology Program grant No. 80NSSC18K0829. DP acknowledges additional support from the NASA FINESST program via grant 80NSSC22K1319. We thank Philipp Huber, Adrian Glauser and the ETH \textit{LIFE} team for very informative and instructive discussions about LIFE\textsc{sim} and flux units.

%



\software{LIFE\textsc{sim} \citep{LIFE2}, petitRT \citep{2019A&A...627A..67M}, 
 Atmos (https://github.com/ VirtualPlanetaryLaboratory/atmos; Arney et al. 2016), SMART (Meadows \& Crisp 1996), Planetary Spectrum Generator (https://psg.gsfc.nasa. gov/; Villanueva et al. 2018)         }



\clearpage
\appendix

\section{Retrieval Details}\label{sec:app_retr}

In this appendix section we show the cornerplots with all posteriors, the derived pressure-temperature profiles and the fitted spectra for the three cases discussed in Section \ref{sec:retr_test}.

\begin{figure}[ht]
  \centering   
  \begin{overpic}[width=0.96\textwidth]{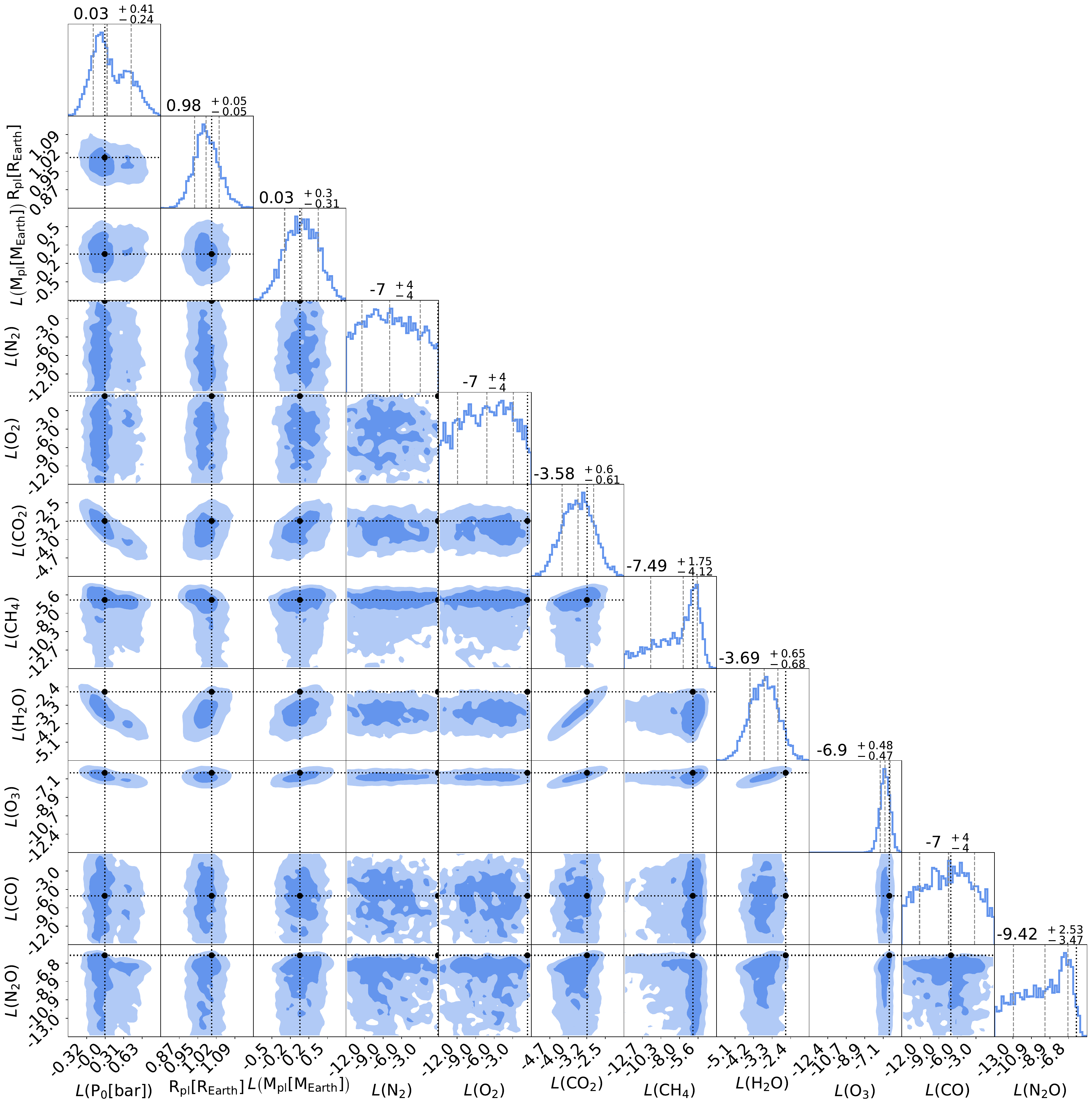}
     \put(50,60){\includegraphics[scale=0.6]{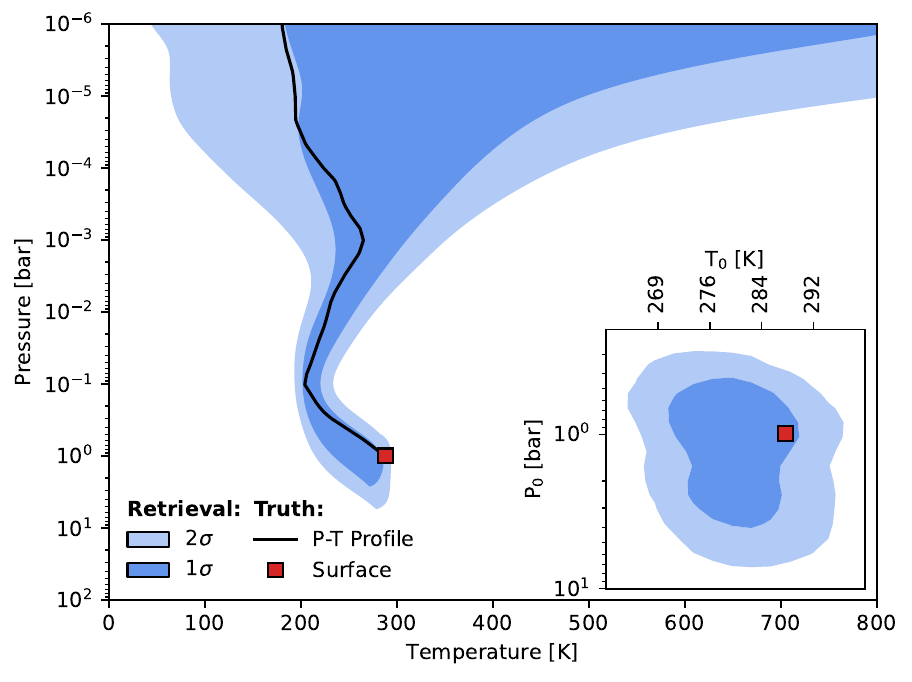}}  
  \end{overpic}
  
  \quad\includegraphics[width=0.96\textwidth]{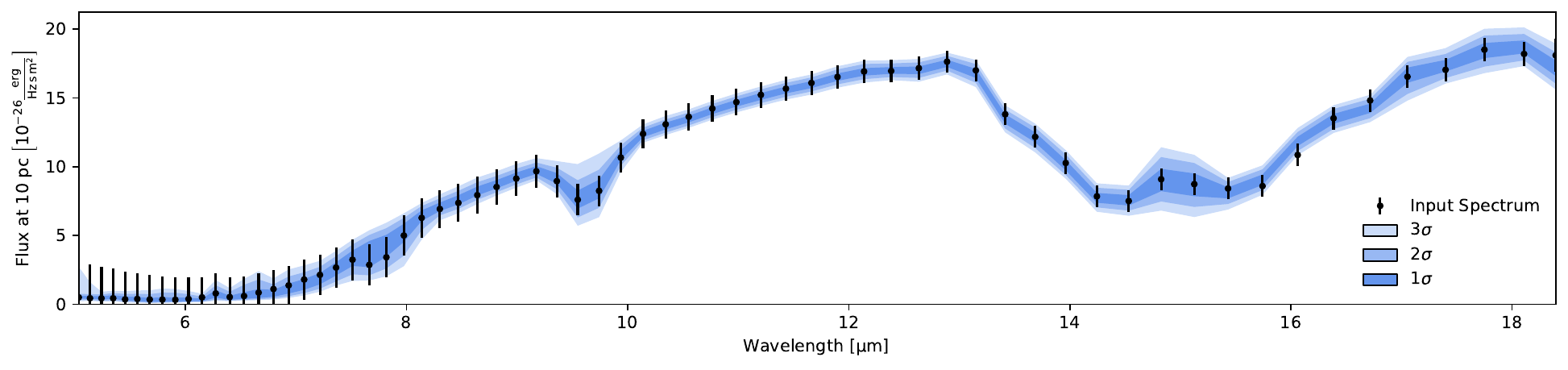}
\caption{Retrieval results for 1 Tmol/y N$_2$O case.}
\label{fig:retr_1tmol}
\end{figure}

\clearpage
\begin{deluxetable*}{cc}[ht]
\tablecaption{1 Tmol/yr retrieval results. Here, $L(\cdot)$ stands for $log_{10}(\cdot)$. }
\tablewidth{0pt}
\tablehead{
\colhead{Value} & \colhead{Retrieved result}
}
\startdata
$L\left(\mathrm{P_0}\left[\mathrm{bar}\right]\right)$ & 0.03 $_{\,-0.24}^{\,+0.41}$ \\ 
$\mathrm{R_{pl}}\left[\mathrm{R_{Earth}}\right]$ & 0.98 $_{\,-0.05}^{\,+0.05}$ \\ 
$L\left(\mathrm{M_{pl}}\left[\mathrm{M_{Earth}}\right]\right)$ & 0.03 $_{\,-0.31}^{\,+0.30}$ \\ 
$L\left(\mathrm{N_2}\right)$ & --\\ 
$L\left(\mathrm{O_2}\right)$ & -- \\ 
$L\left(\mathrm{CO_2}\right)$ & -3.58 $_{\,-0.61}^{\,+0.60}$ \\ 
$L\left(\mathrm{CH_4}\right)$ & -7.49 $_{\,-4.12}^{\,+1.75}$ \\ 
$L\left(\mathrm{H_2O}\right)$ & -3.69 $_{\,-0.68}^{\,+0.65}$ \\ 
$L\left(\mathrm{O_3}\right)$ & -6.90 $_{\,-0.47}^{\,+0.48}$ \\ 
$L\left(\mathrm{CO}\right)$ & -- \\ 
$L\left(\mathrm{N_2O}\right)$ & -9.42 $_{\,-3.47}^{\,+2.53}$ \\ 
\enddata
\end{deluxetable*}

\begin{figure}[ht]
  \centering   
  \begin{overpic}[width=0.96\textwidth]{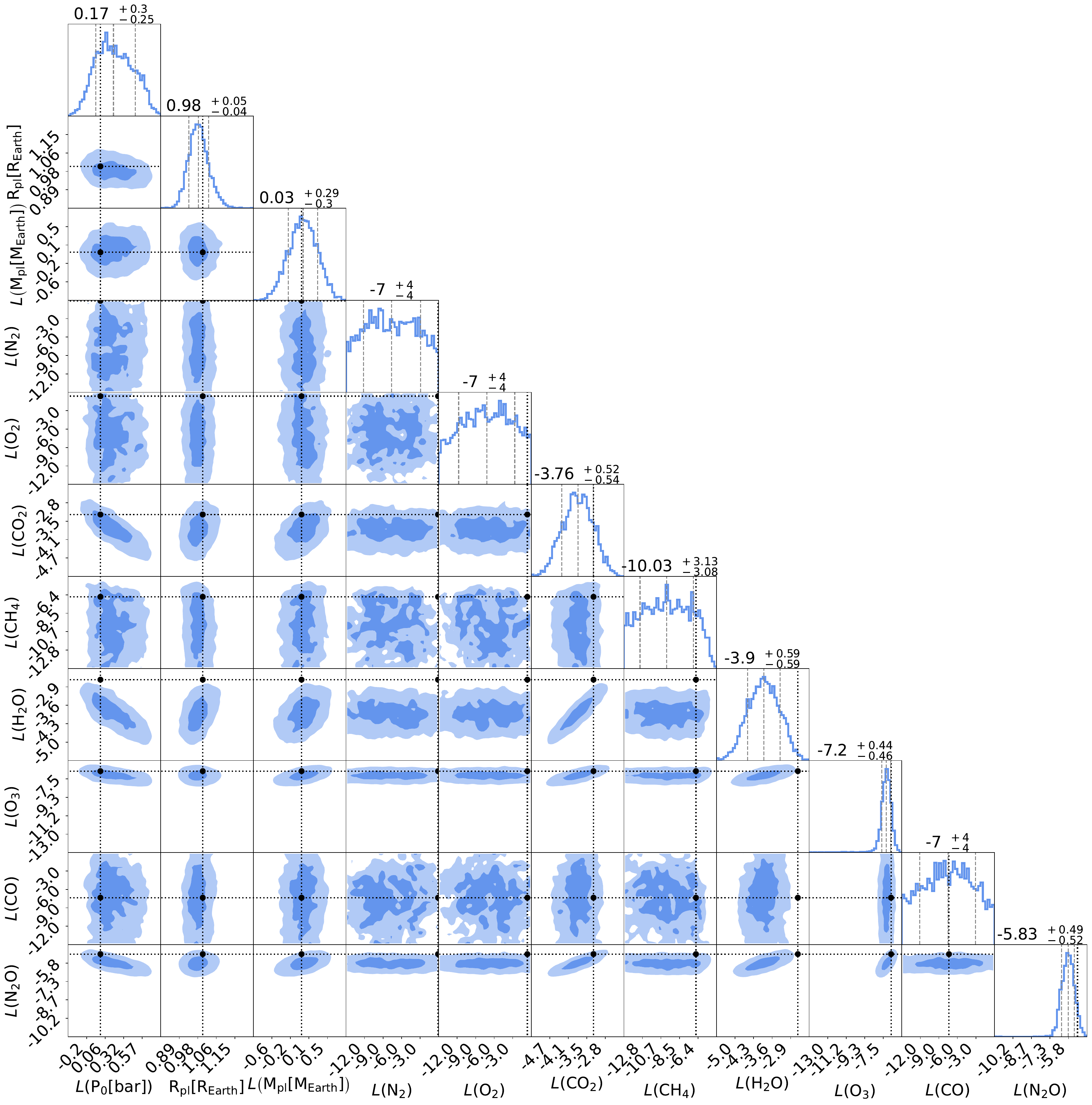}
     \put(50,60){\includegraphics[scale=0.6]{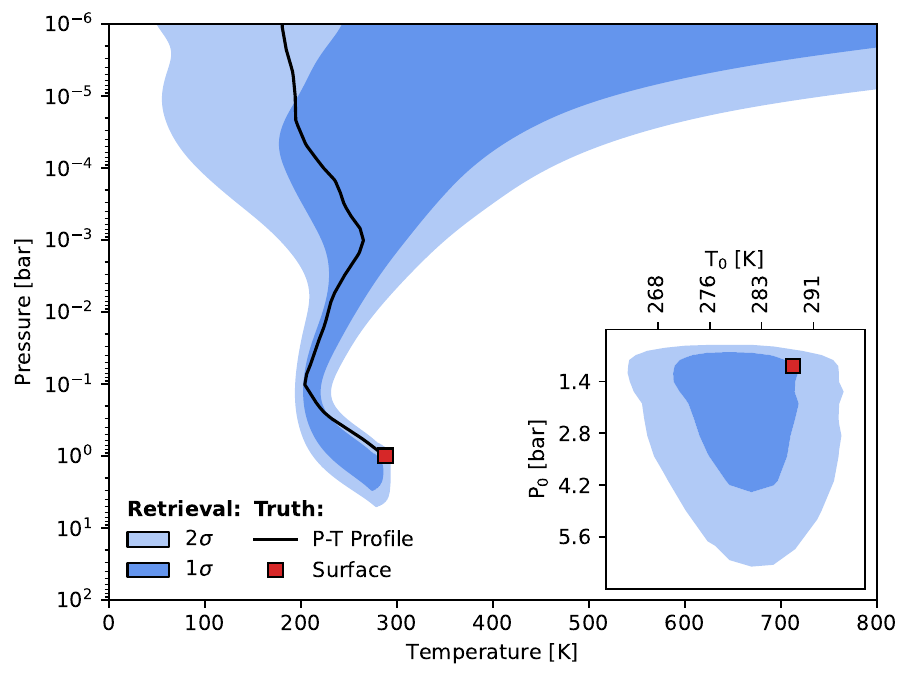}}  
  \end{overpic}
  
  \quad\includegraphics[width=0.96\textwidth]{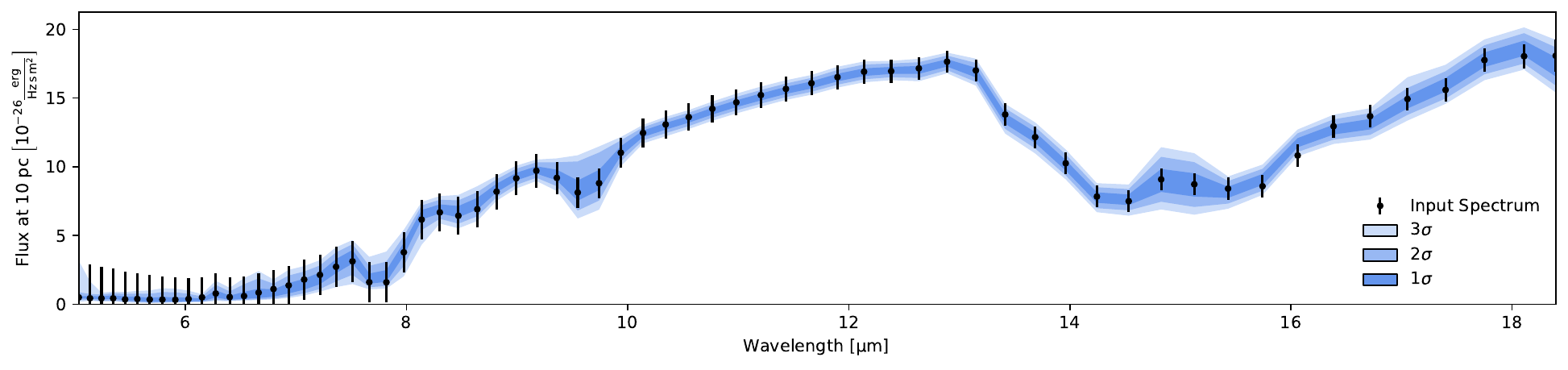}
\caption{Retrieval results for 10 Tmol/y N$_2$O case.}
\label{fig:retr_10tmol}
\end{figure}

\clearpage

\begin{deluxetable*}{cc}[ht]
\tablecaption{10 Tmol/yr retrieval results. Here, $L(\cdot)$ stands for $log_{10}(\cdot)$. }
\tablewidth{0pt}
\tablehead{
\colhead{Value} & \colhead{Retrieved result}
}
\startdata
$L\left(\mathrm{P_0}\left[\mathrm{bar}\right]\right)$ & 0.17 $_{\,-0.25}^{\,+0.30}$ \\ 
$\mathrm{R_{pl}}\left[\mathrm{R_{Earth}}\right]$ & 0.98 $_{\,-0.04}^{\,+0.05}$ \\ 
$L\left(\mathrm{M_{pl}}\left[\mathrm{M_{Earth}}\right]\right)$ & 0.03 $_{\,-0.30}^{\,+0.29}$ \\ 
$L\left(\mathrm{N_2}\right)$ & -- \\ 
$L\left(\mathrm{O_2}\right)$ & -- \\ 
$L\left(\mathrm{CO_2}\right)$ & -3.76  $_{\,-0.54}^{\,+0.52}$ \\ 
$L\left(\mathrm{CH_4}\right)$ & -10 $_{\,-3}^{\,+3}$ \\ 
$L\left(\mathrm{H_2O}\right)$ & -3.90 $_{\,-0.59}^{\,+0.59}$ \\ 
$L\left(\mathrm{O_3}\right)$ & -7.20 $_{\,-0.46}^{\,+0.44}$ \\ 
$L\left(\mathrm{CO}\right)$ &-- \\ 
$L\left(\mathrm{N_2O}\right)$ & -5.83$_{\,-0.52}^{\,+0.49}$ \\ 
\enddata
\end{deluxetable*}

\begin{figure}[ht]
  \centering   
  \begin{overpic}[width=0.96\textwidth]{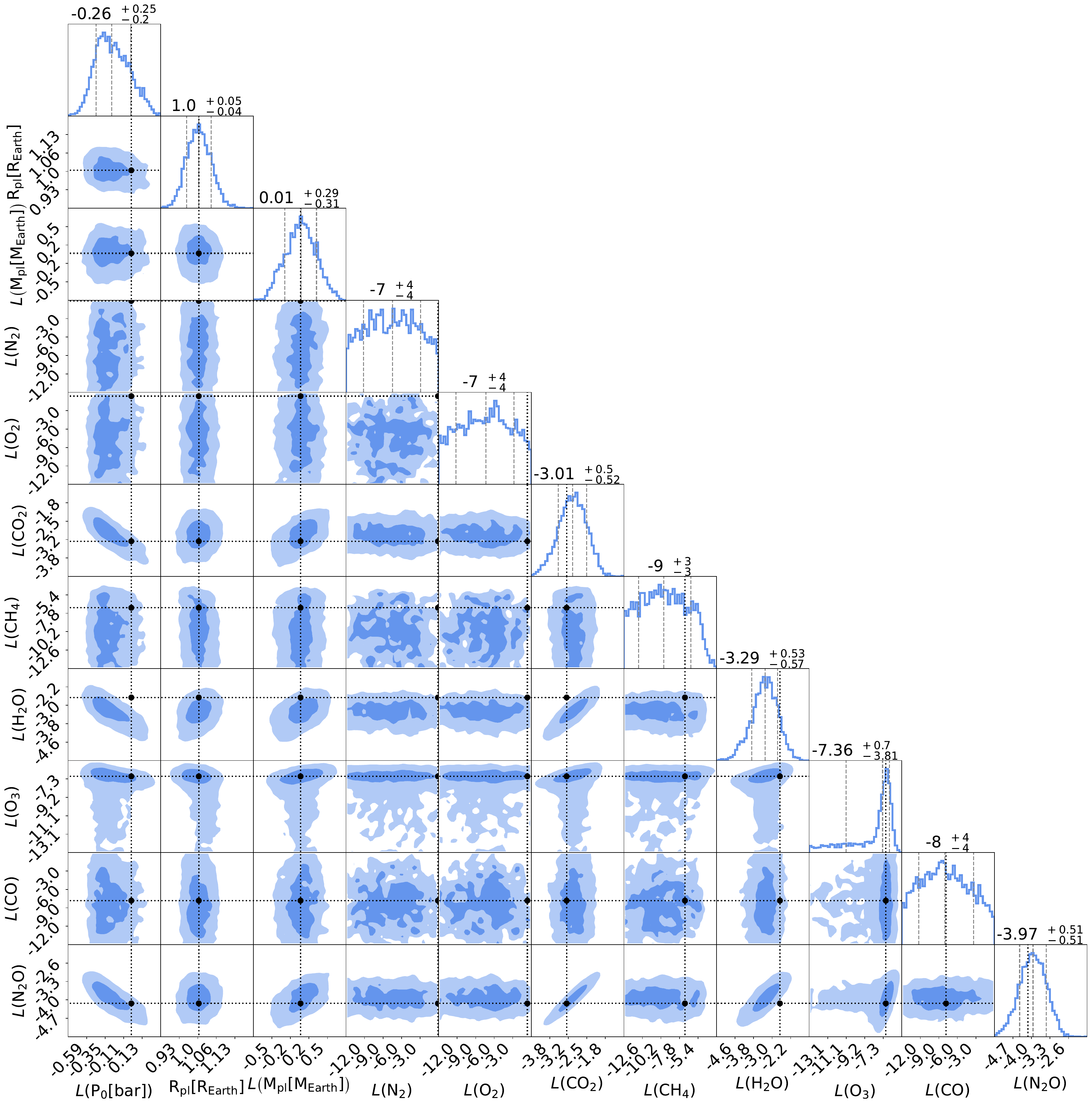}
     \put(50,60){\includegraphics[scale=0.6]{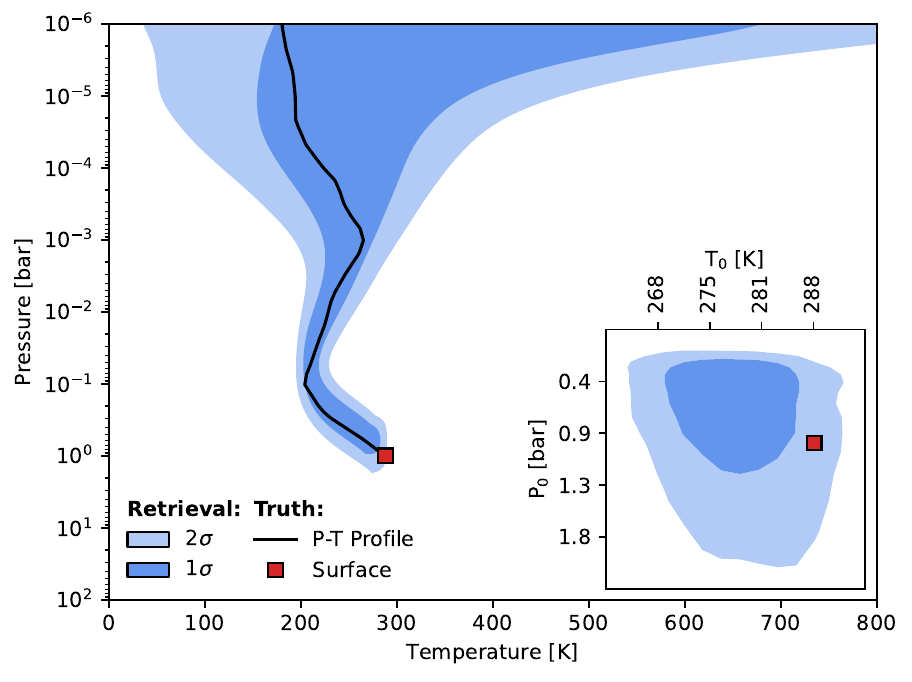}}  
  \end{overpic}
  
  \quad\includegraphics[width=0.96\textwidth]{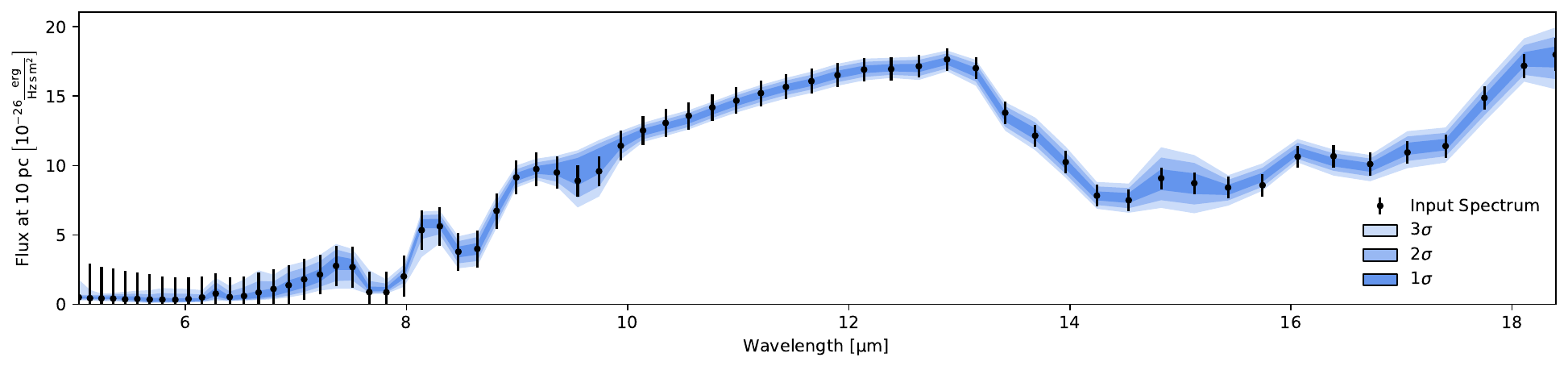}
\caption{Retrieval results for 100 Tmol/y N$_2$O case.}
\label{fig:retr_100tmol}
\end{figure}

\clearpage

\begin{deluxetable*}{cc}[ht]
\tablecaption{100 Tmol/yr retrieval results. Here, $L(\cdot)$ stands for $log_{10}(\cdot)$. }
\tablewidth{0pt}
\tablehead{
\colhead{Value} & \colhead{Retrieved result}
}
\startdata
$L\left(\mathrm{P_0}\left[\mathrm{bar}\right]\right)$ & -0.26 $_{\,-0.20}^{\,+0.25}$ \\ 
$\mathrm{R_{pl}}\left[\mathrm{R_{Earth}}\right]$ & 1.00$_{\,-0.04}^{\,+0.05}$ \\ 
$L\left(\mathrm{M_{pl}}\left[\mathrm{M_{Earth}}\right]\right)$ & 0.01 $_{\,-0.29}^{\,+0.31}$ \\ 
$L\left(\mathrm{N_2}\right)$ & -- \\ 
$L\left(\mathrm{O_2}\right)$ & -- \\ 
$L\left(\mathrm{CO_2}\right)$ & -3.01 $_{\,-0.52}^{\,+0.50}$ \\ 
$L\left(\mathrm{CH_4}\right)$ & -9 $_{\,-3}^{\,+3}$ \\ 
$L\left(\mathrm{H_2O}\right)$ & -3.29 $_{\,-0.57}^{\,+0.53}$ \\ 
$L\left(\mathrm{O_3}\right)$ & -7.36 $_{\,-3.81}^{\,+0.70}$ \\ 
$L\left(\mathrm{CO}\right)$ & -- \\ 
$L\left(\mathrm{N_2O}\right)$ & -3.97 $_{\,-0.51}^{\,+0.51}$ \\ 
\enddata
\end{deluxetable*}

\section{Modelled Observations for different observation times}

In the following two appendix sections we show the simulated observations for 10 and 100 days of observation in addition to the 50 day observations already shown in the main text.

\subsection{Supplementary Results N2O}\label{app:n2o_times}

\begin{figure*}[htp]

\centering
\includegraphics[width=.45\textwidth]{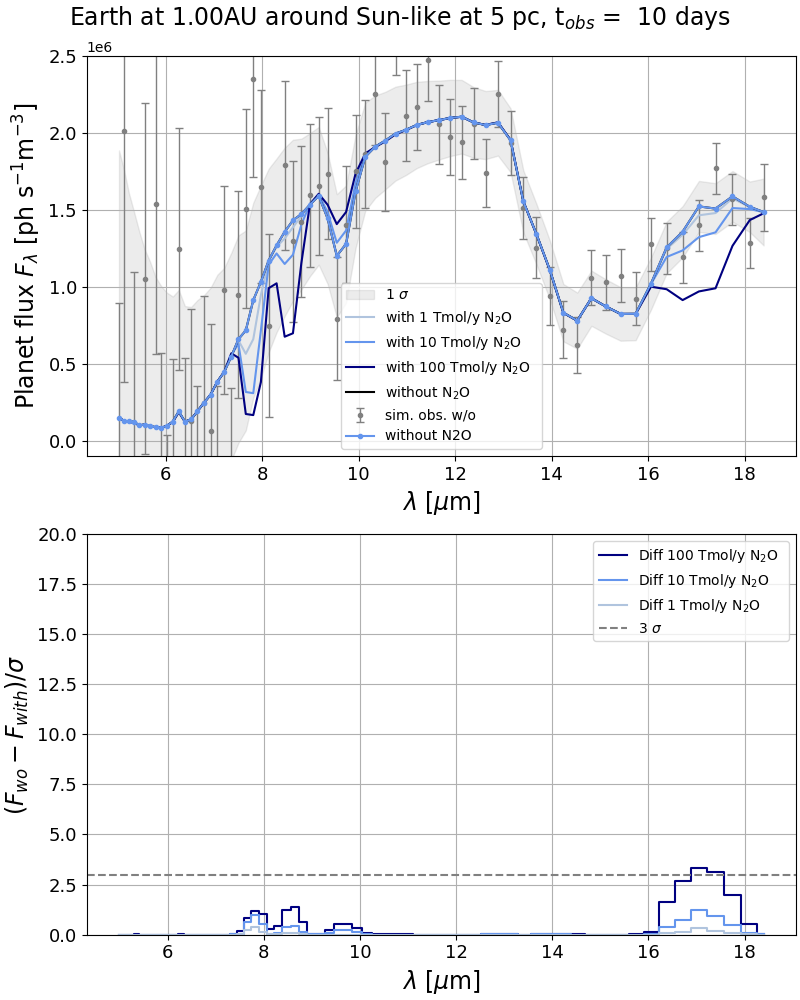}\hfill
\includegraphics[width=.45\textwidth]{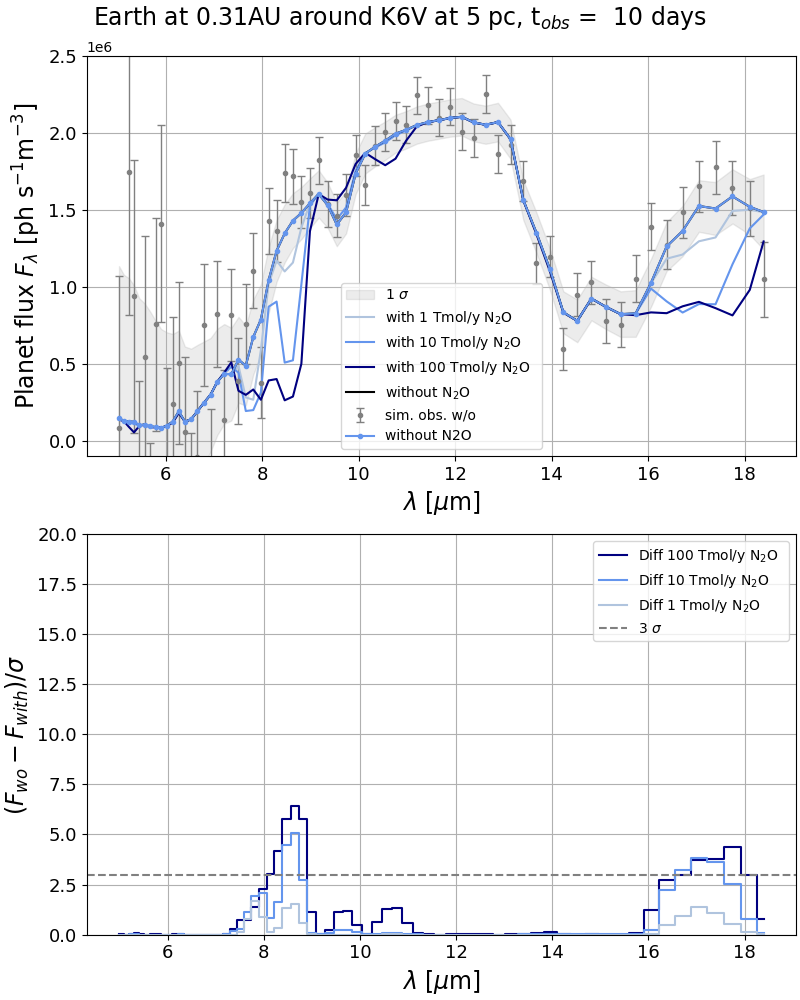}\hfill

\includegraphics[width=.45\textwidth]
{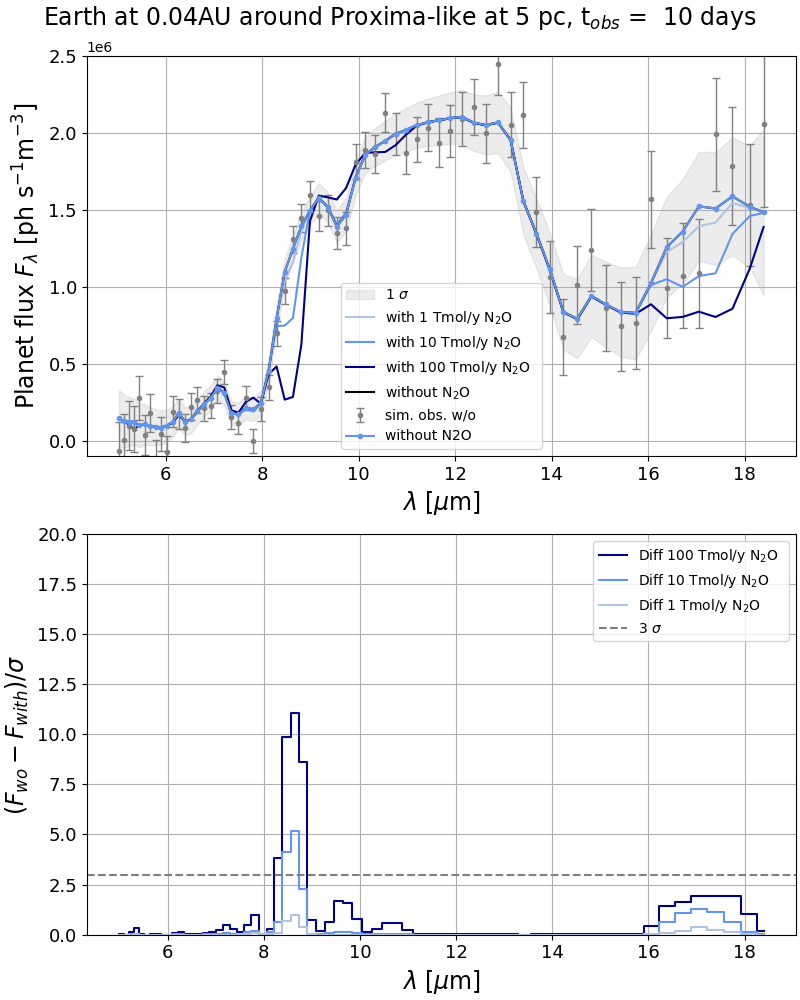}\hfill
\includegraphics[width=.45\textwidth]{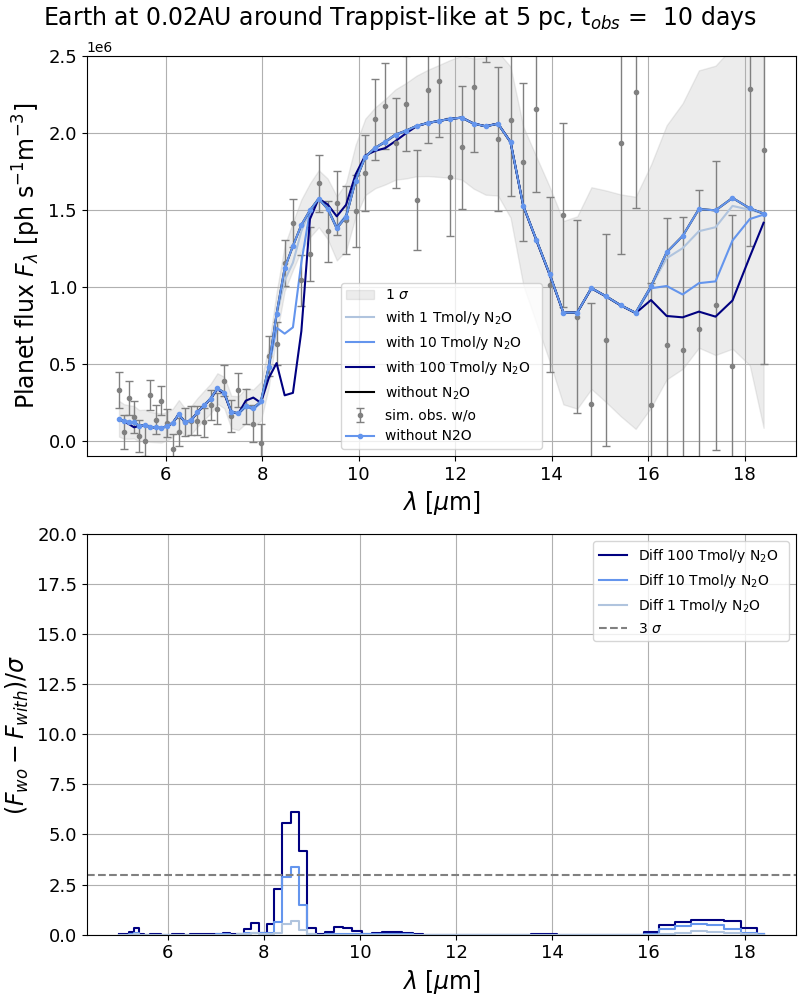}

\caption{Same as Figure \ref{fig:N20_50d} for 10 days of observation time.}
\label{fig:N20_10d}

\end{figure*}

\begin{figure*}[htp]

\centering
\includegraphics[width=.45\textwidth]{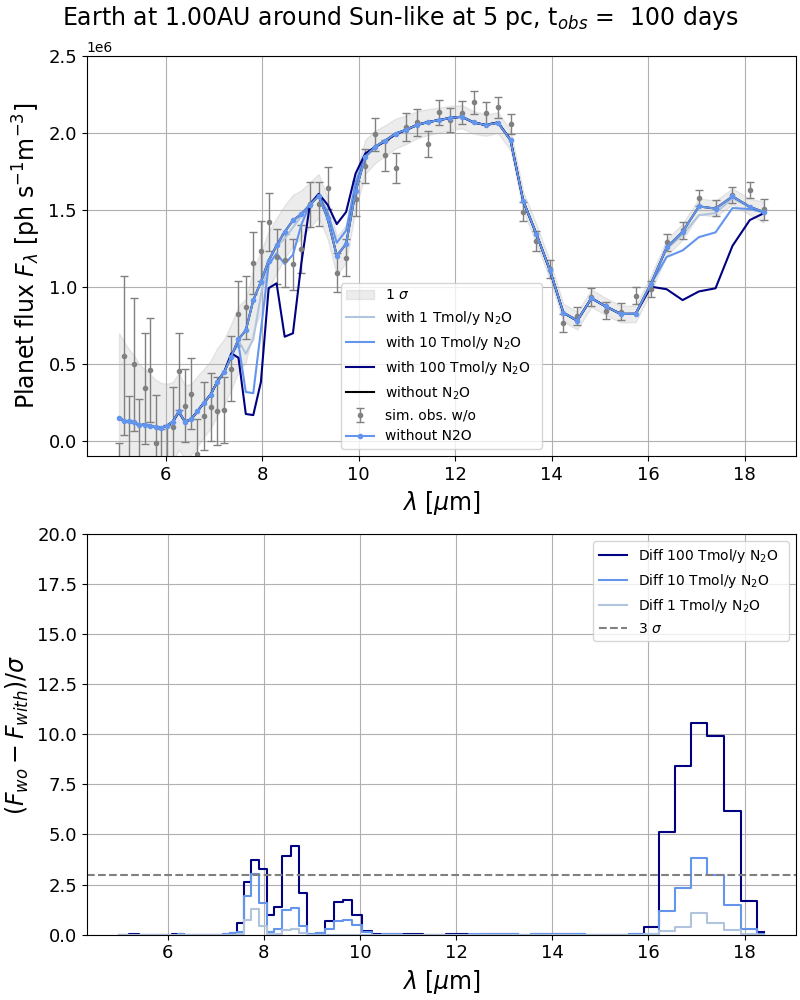}\hfill
\includegraphics[width=.45\textwidth]{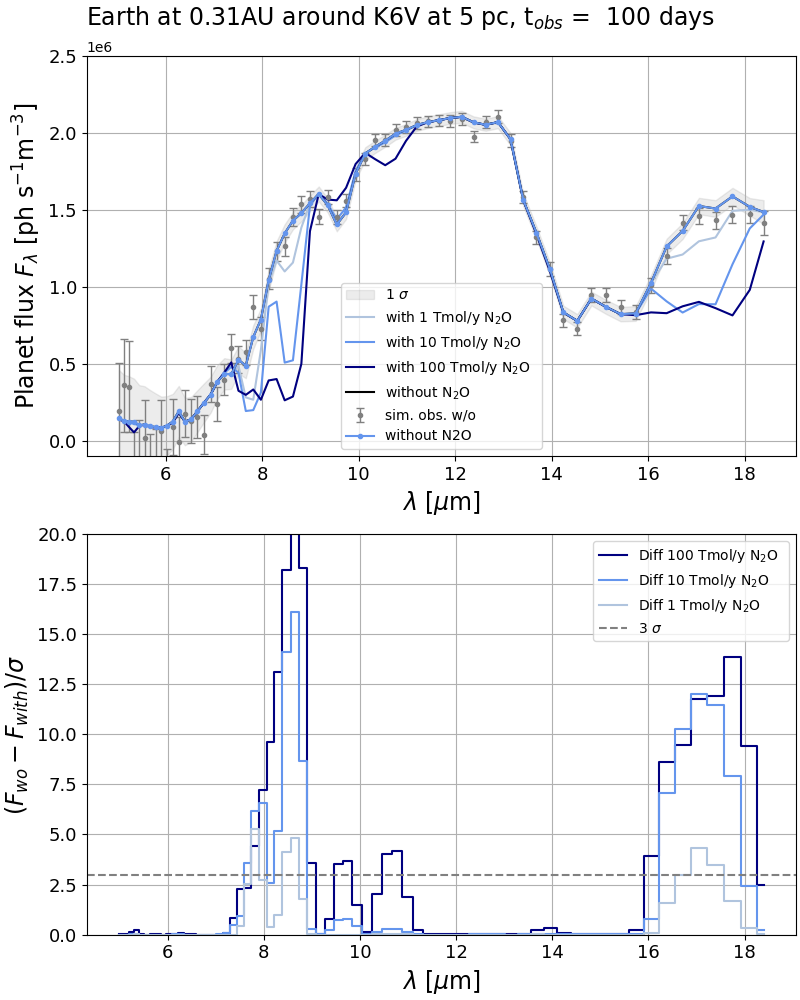}\hfill

\includegraphics[width=.45\textwidth]
{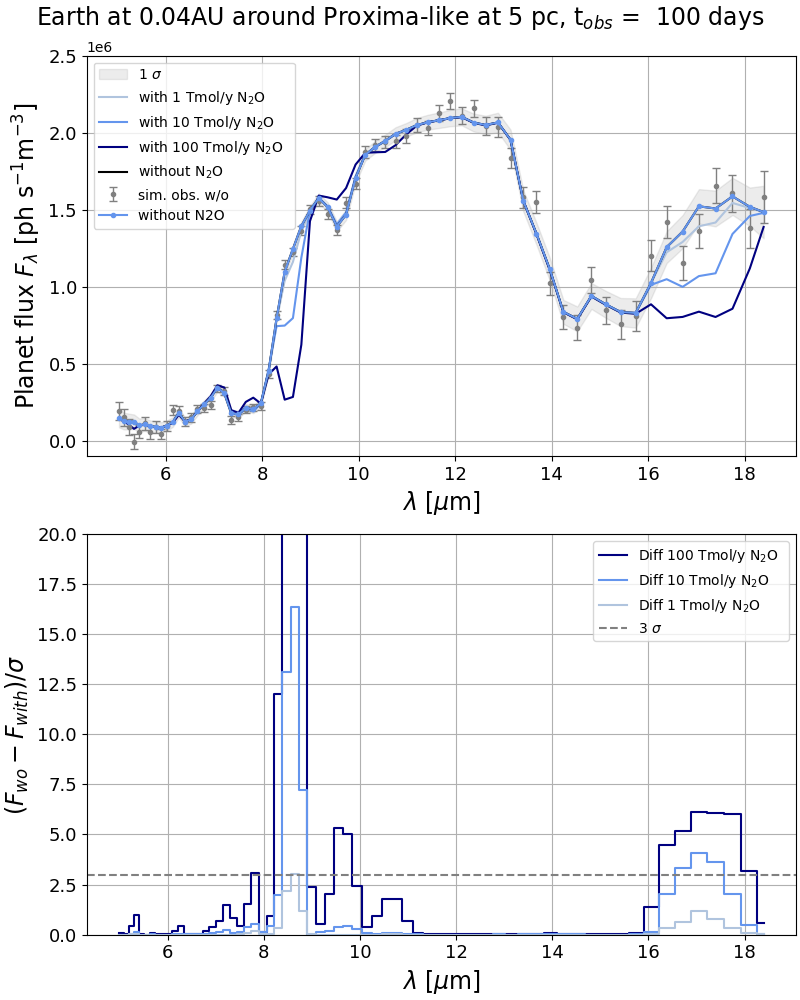}\hfill
\includegraphics[width=.45\textwidth]{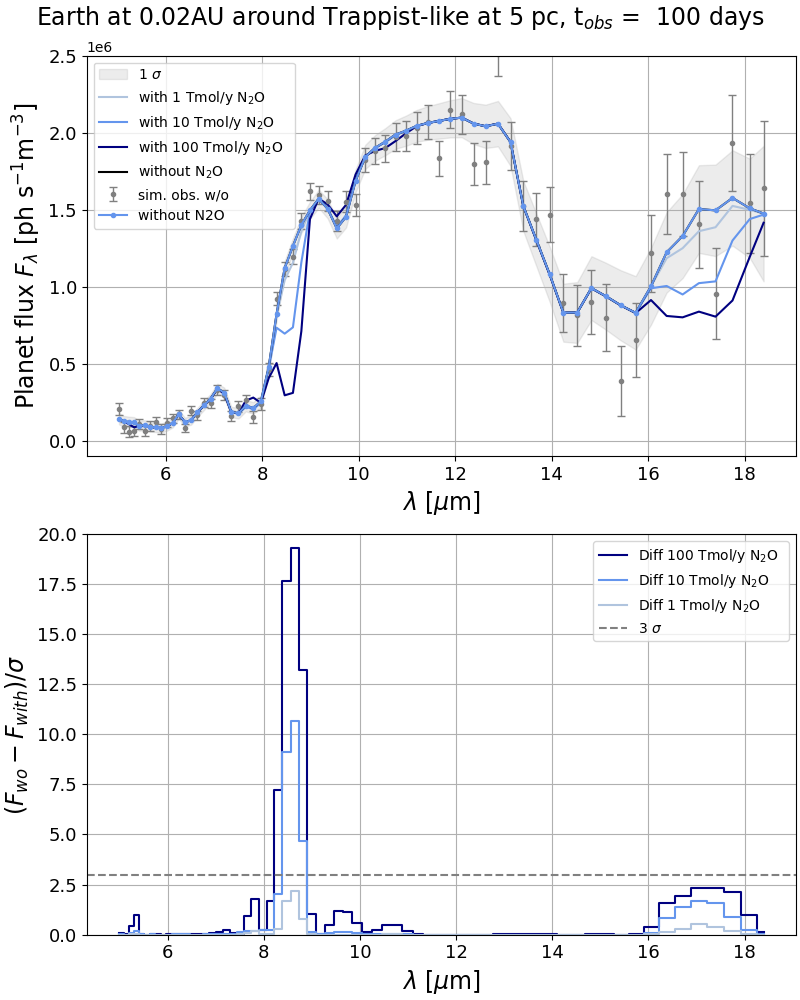}

\caption{Same as Figure \ref{fig:N20_50d} for 100 days of observation time.}
\label{fig:N20_100d}

\end{figure*}

\clearpage

\subsection{Supplementary Results CH3x}\label{app:ch3x_times}
\subsubsection{AD Leonis type host}
\begin{figure}[ht]
\centering
\includegraphics[width=.3\textwidth]{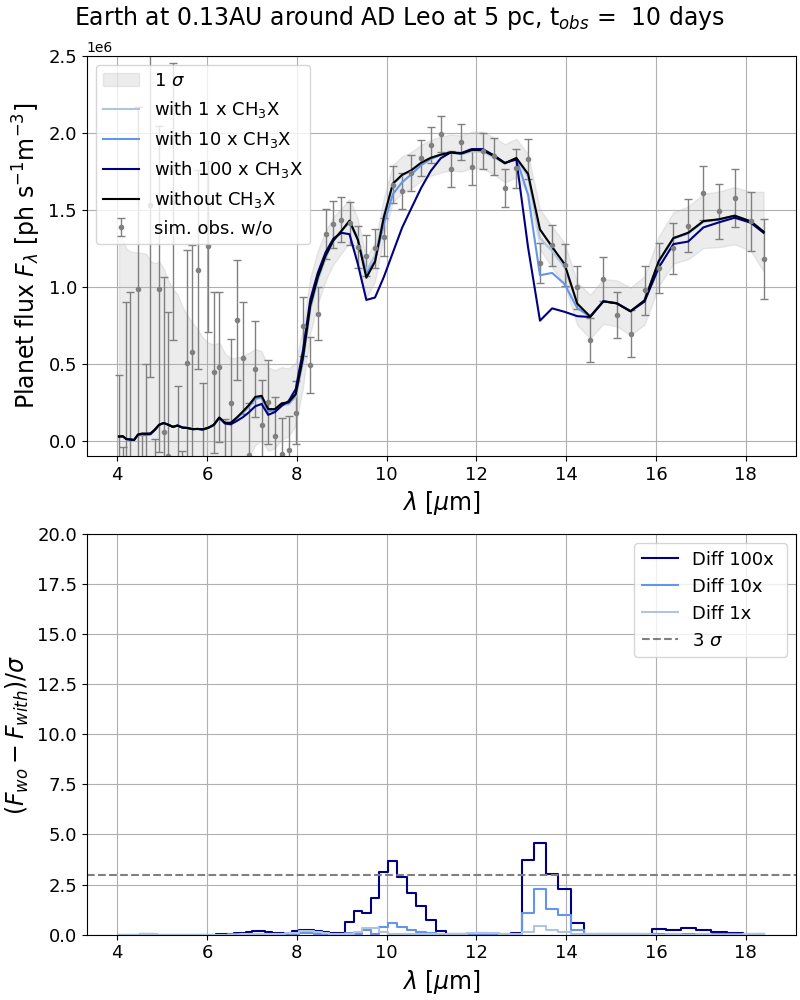}\quad
\includegraphics[width=.3\textwidth]{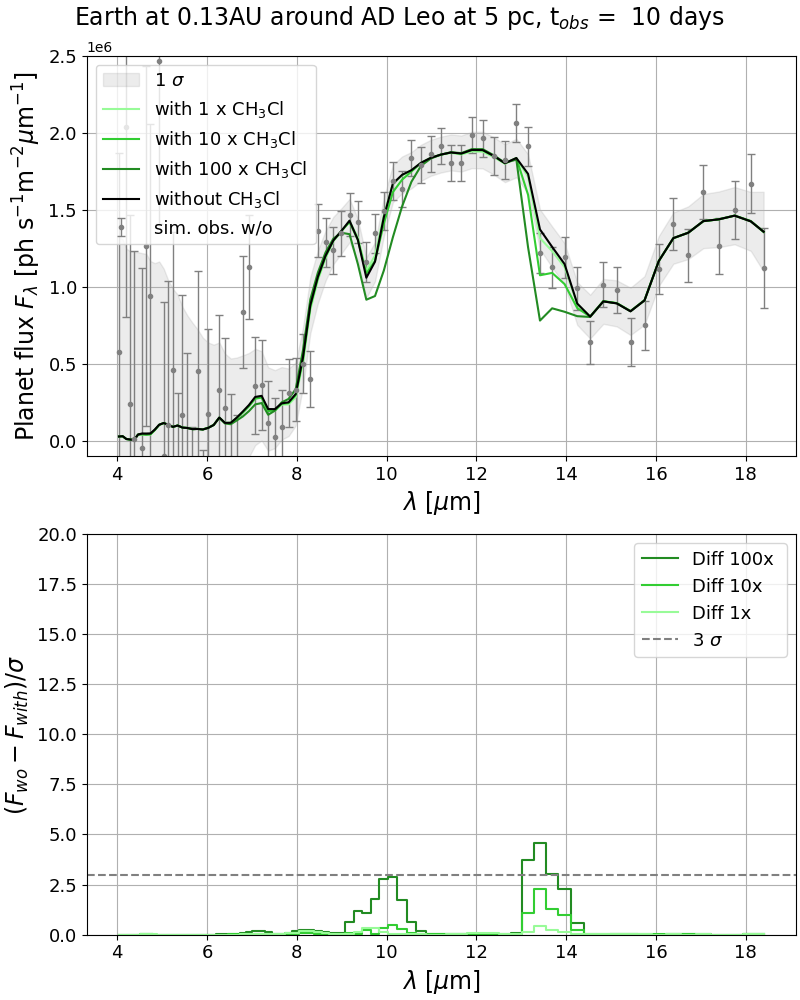}\quad
\includegraphics[width=.3\textwidth]{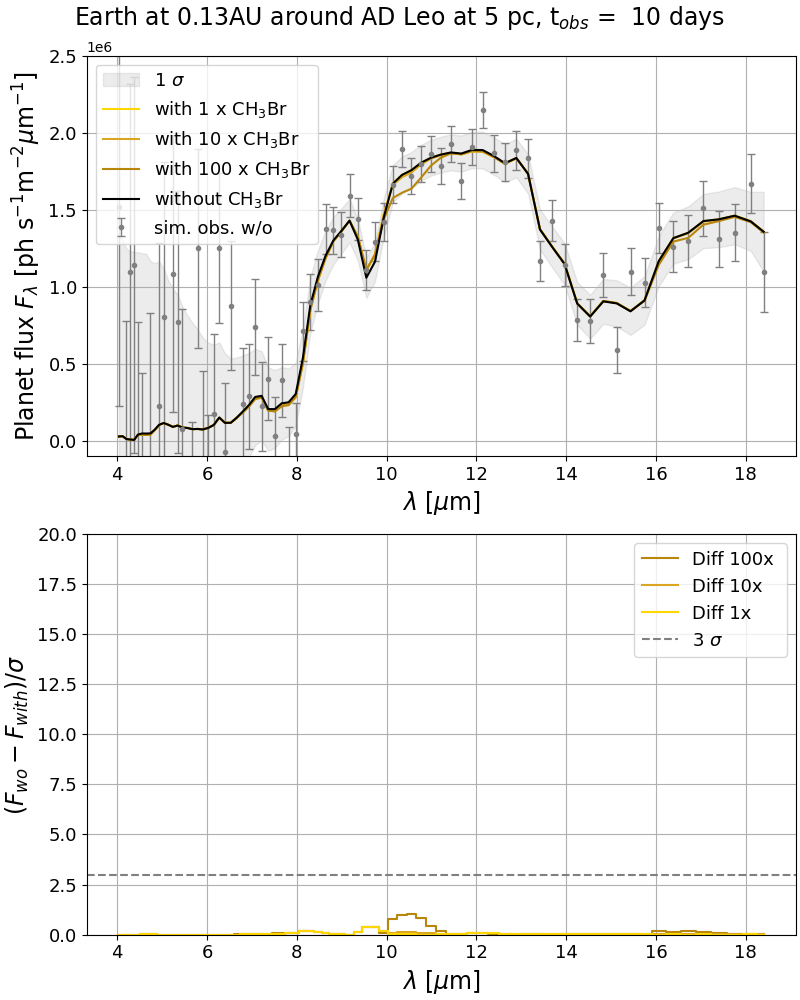}

\caption{Results for AD Leonis Type star in 10 days.}
\label{fig:ADL_10d}
\end{figure}

\begin{figure}[ht]
\centering
\includegraphics[width=.3\textwidth]{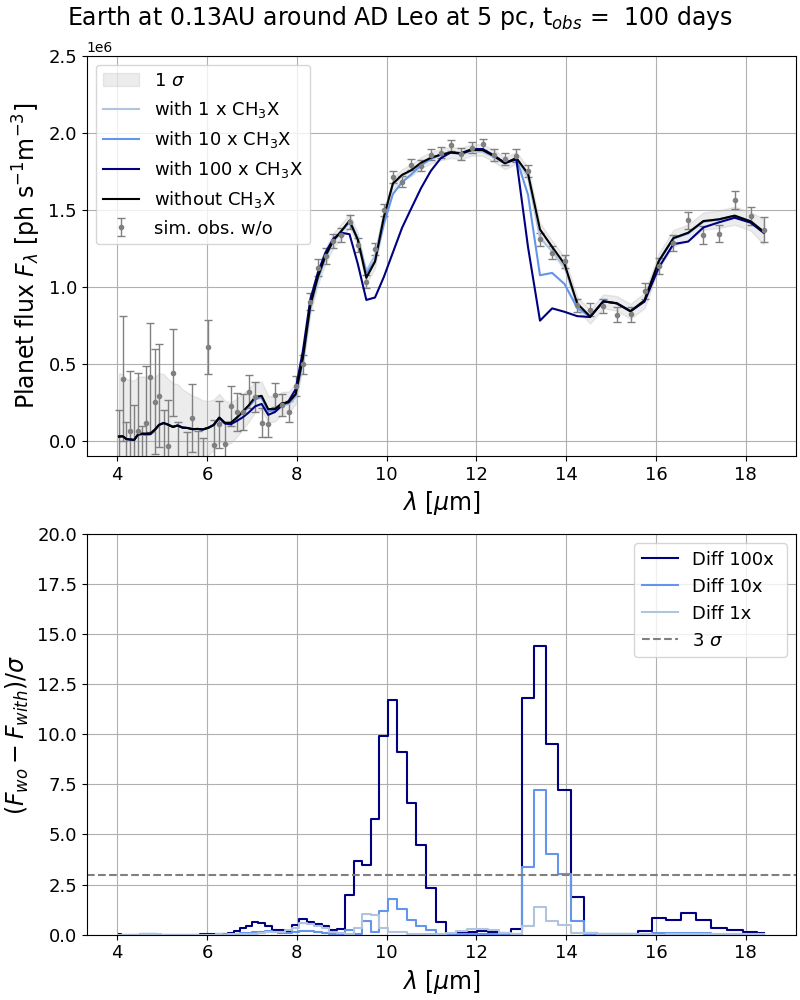}\quad
\includegraphics[width=.3\textwidth]{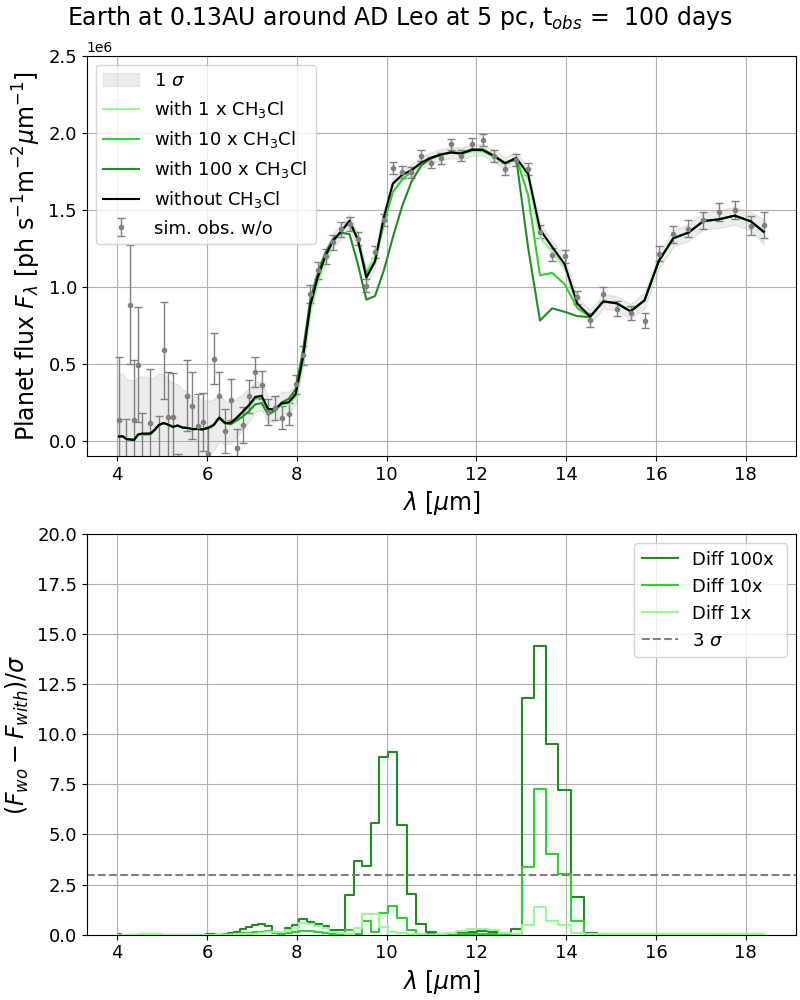}\quad
\includegraphics[width=.3\textwidth]{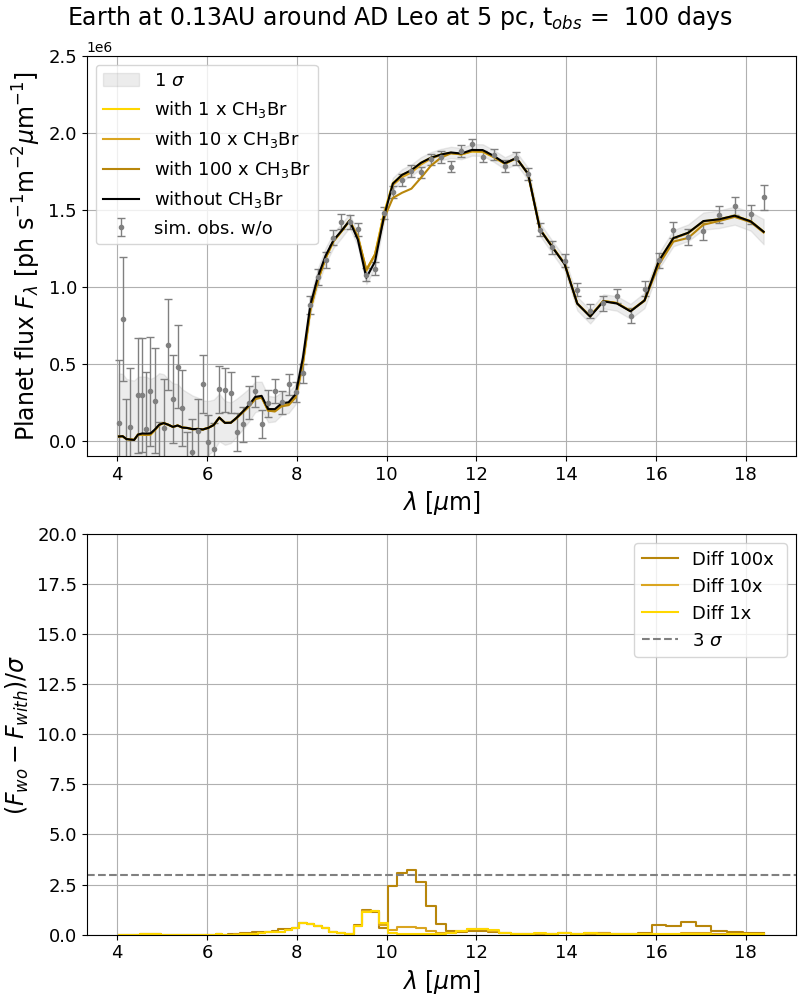}

\caption{Results for AD Leonis Type star in 100 days.}
\label{fig:ADL_100d}
\end{figure}

\clearpage

\subsubsection{TRAPPIST-1 type host}

\begin{figure}[ht]
\centering
\includegraphics[width=.3\textwidth]{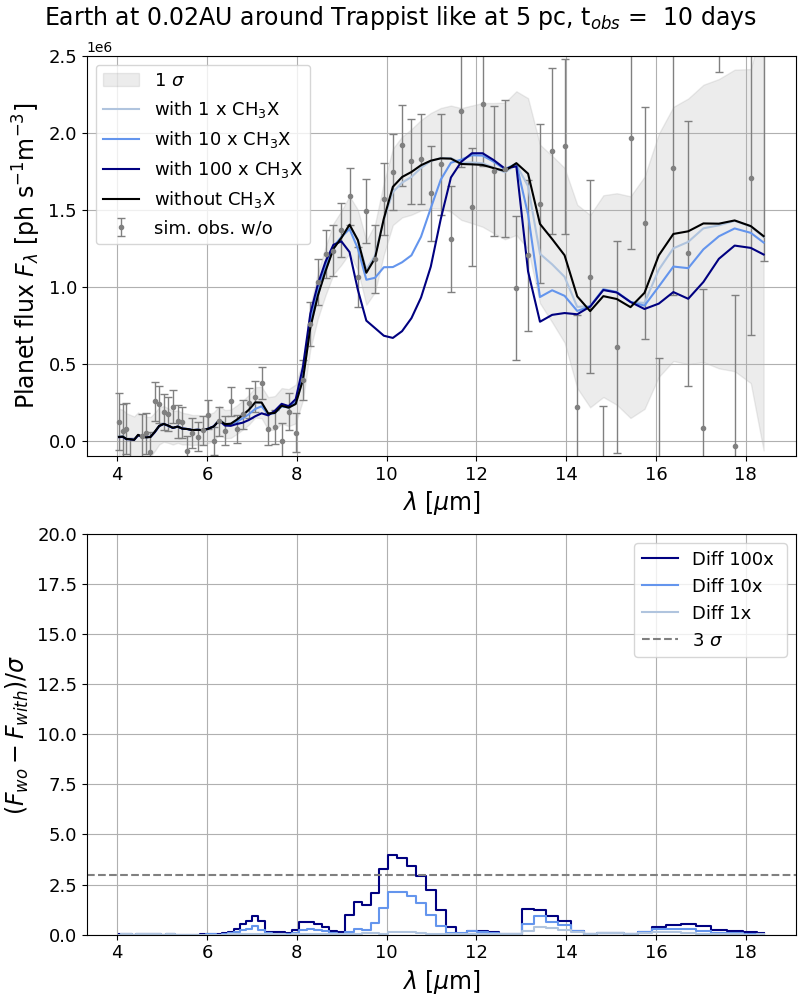}\quad
\includegraphics[width=.3\textwidth]{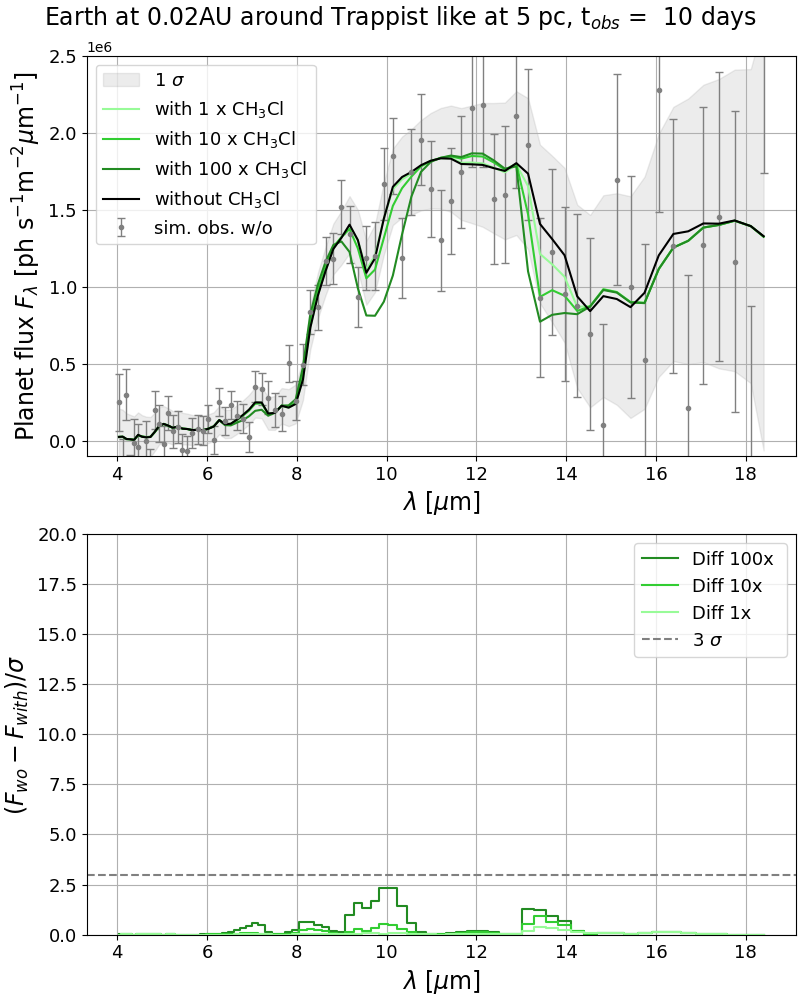}\quad
\includegraphics[width=.3\textwidth]{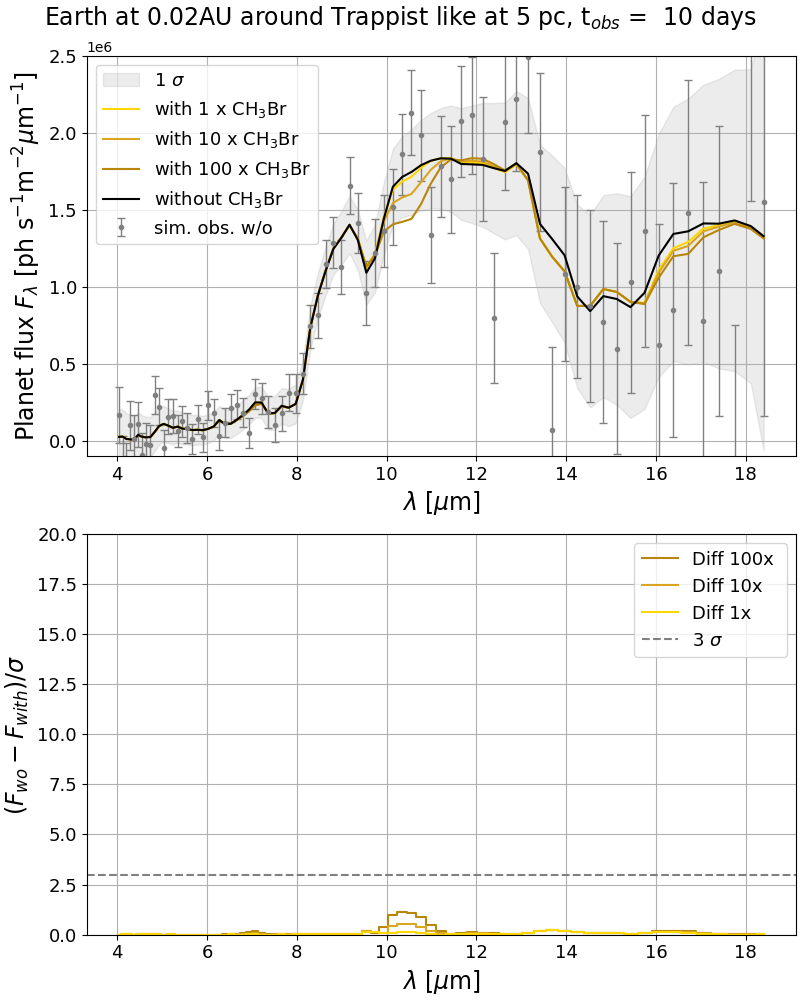}

\caption{Results for TRAPPIST-1 Type star in 10 days.}
\label{fig:trap_10d}
\end{figure}

\begin{figure}[ht]
\centering
\includegraphics[width=.3\textwidth]{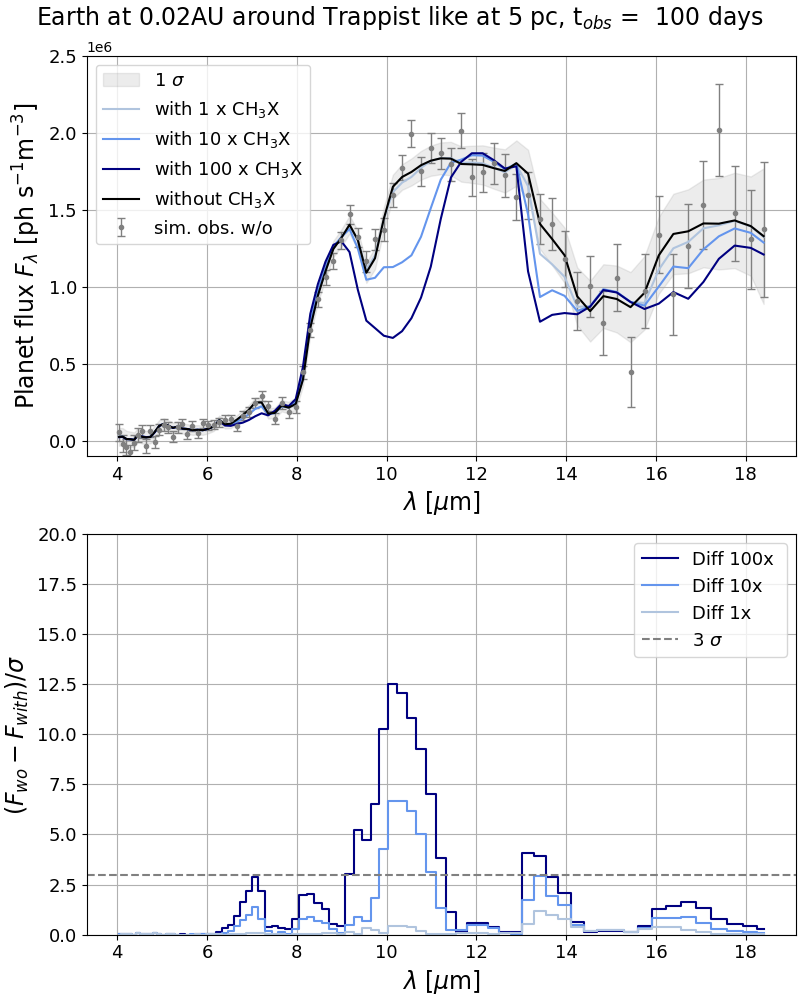}\quad
\includegraphics[width=.3\textwidth]{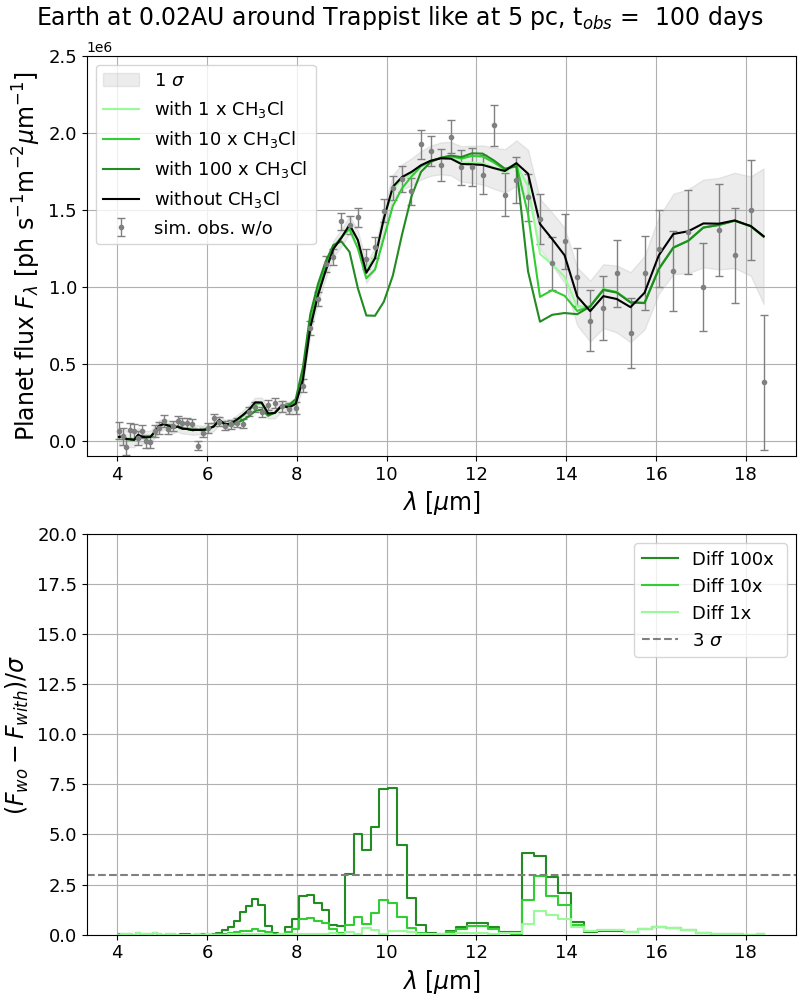}\quad
\includegraphics[width=.3\textwidth]{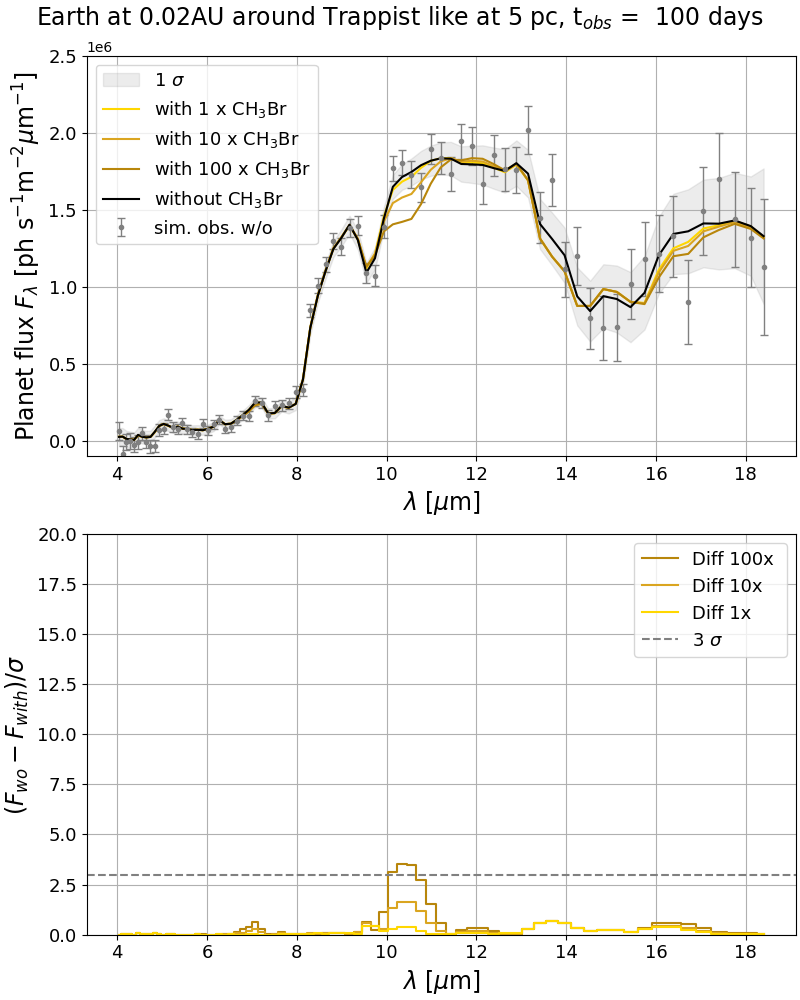}

\caption{Results for TRAPPIST-1 Type star in 100 days.}
\label{fig:trap_100d}
\end{figure}

\clearpage
\subsubsection{Proxima type host}
\begin{figure}[ht]
\centering
\includegraphics[width=.3\textwidth]{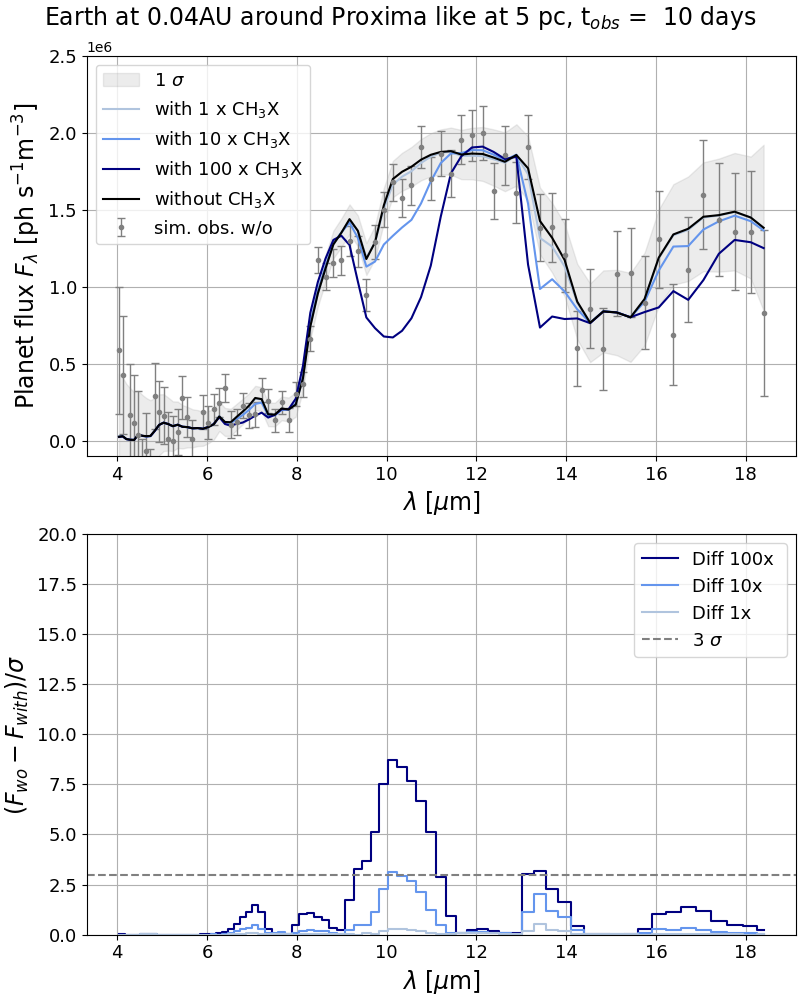}\quad
\includegraphics[width=.3\textwidth]{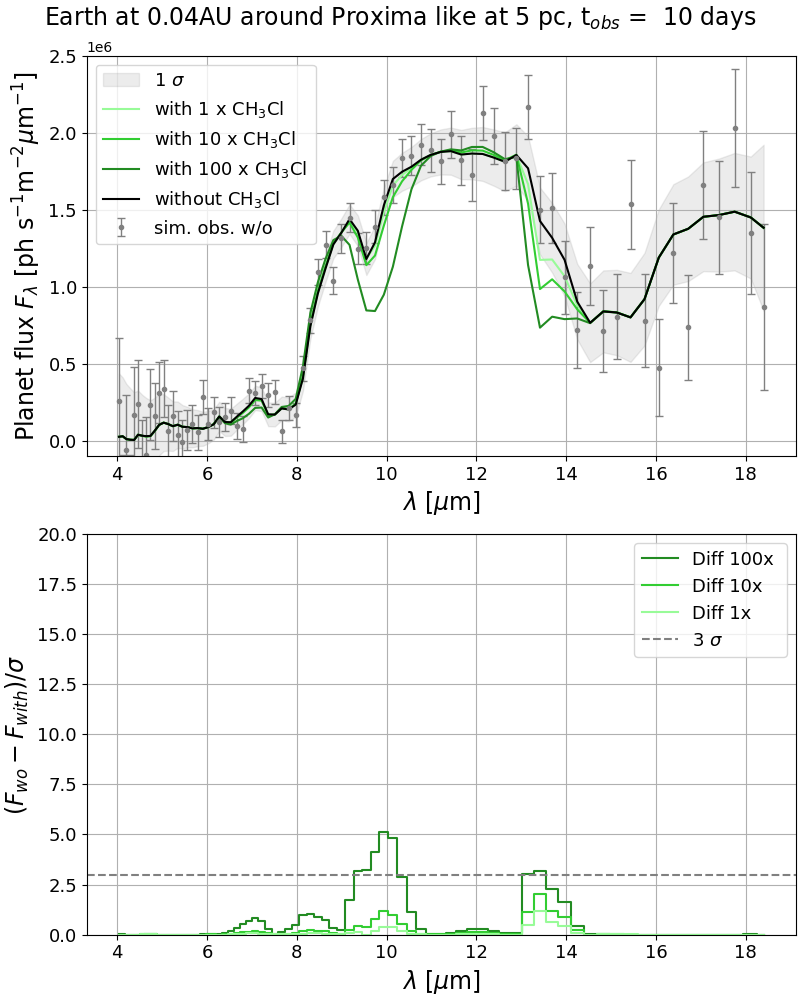}\quad
\includegraphics[width=.3\textwidth]{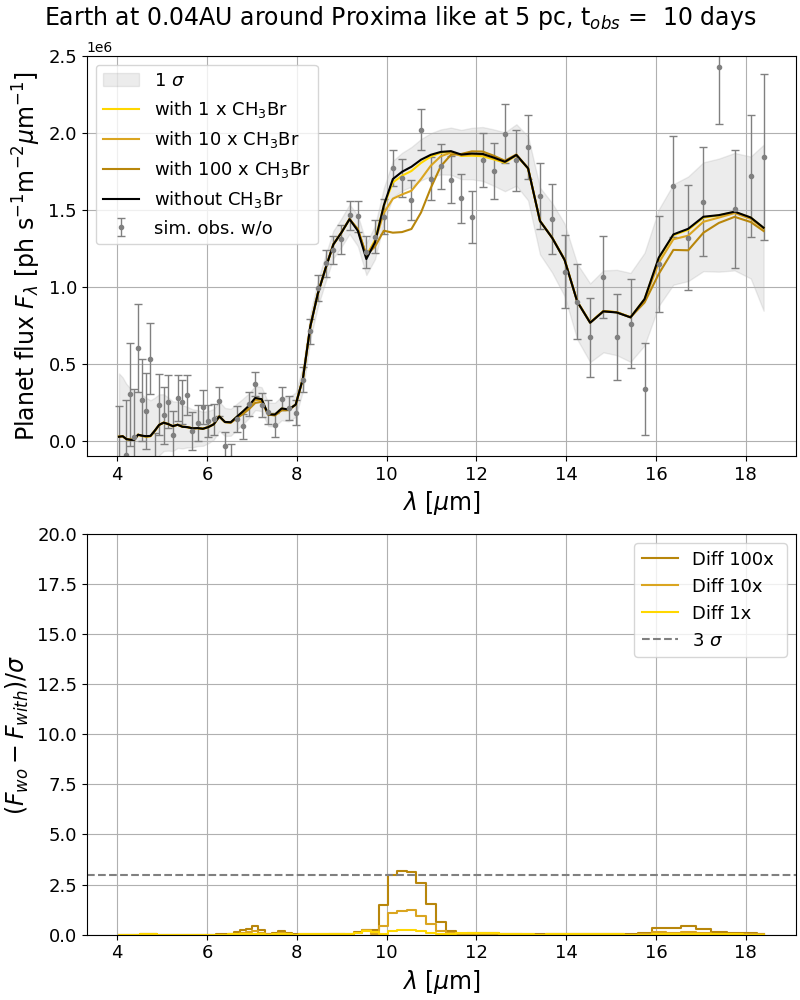}

\caption{Results for Proxima Type star in 10 days.}
\label{fig:prox_10d}
\end{figure}

\begin{figure}[ht]
\centering
\includegraphics[width=.3\textwidth]{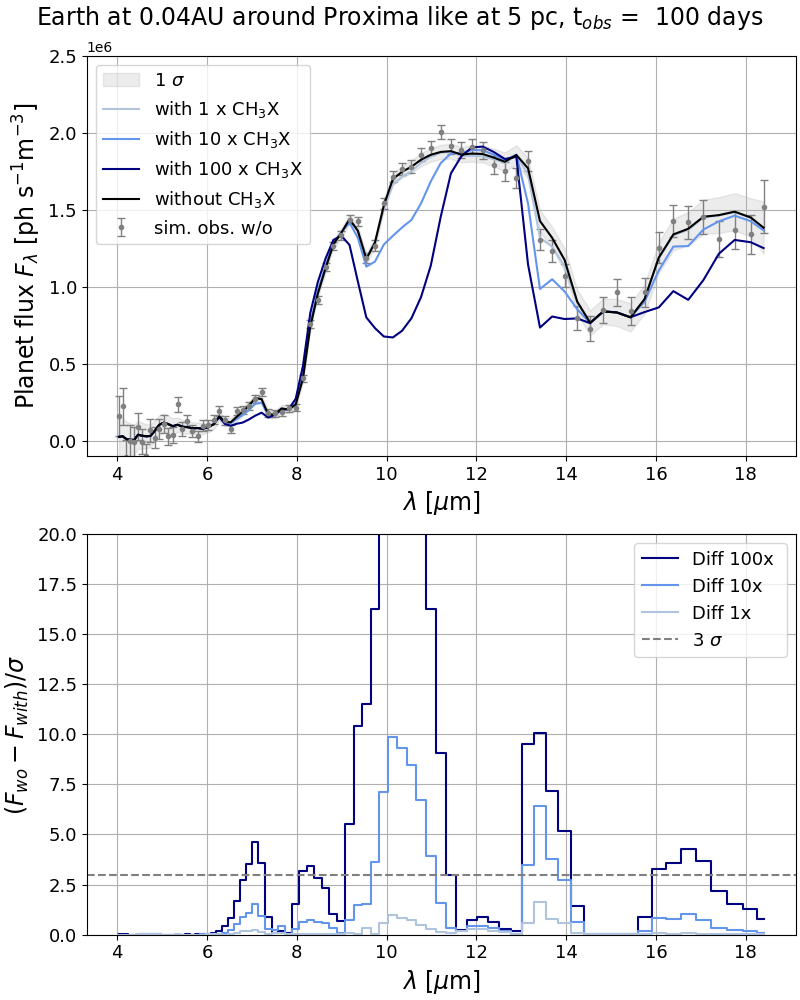}\quad
\includegraphics[width=.3\textwidth]{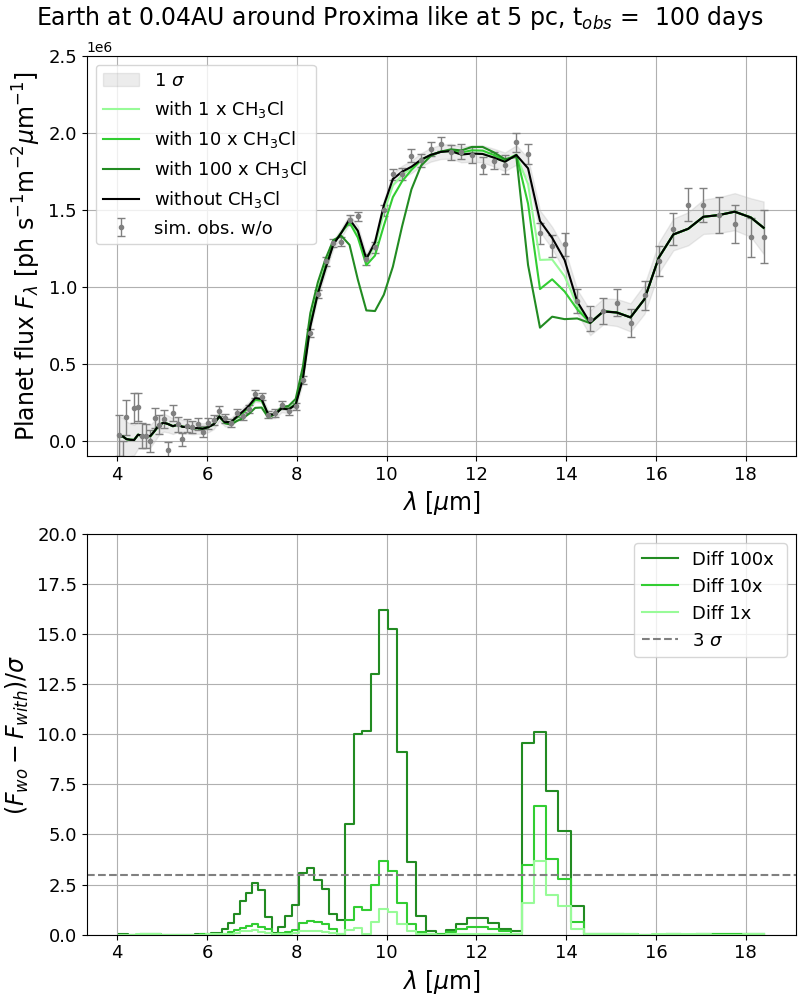}\quad
\includegraphics[width=.3\textwidth]{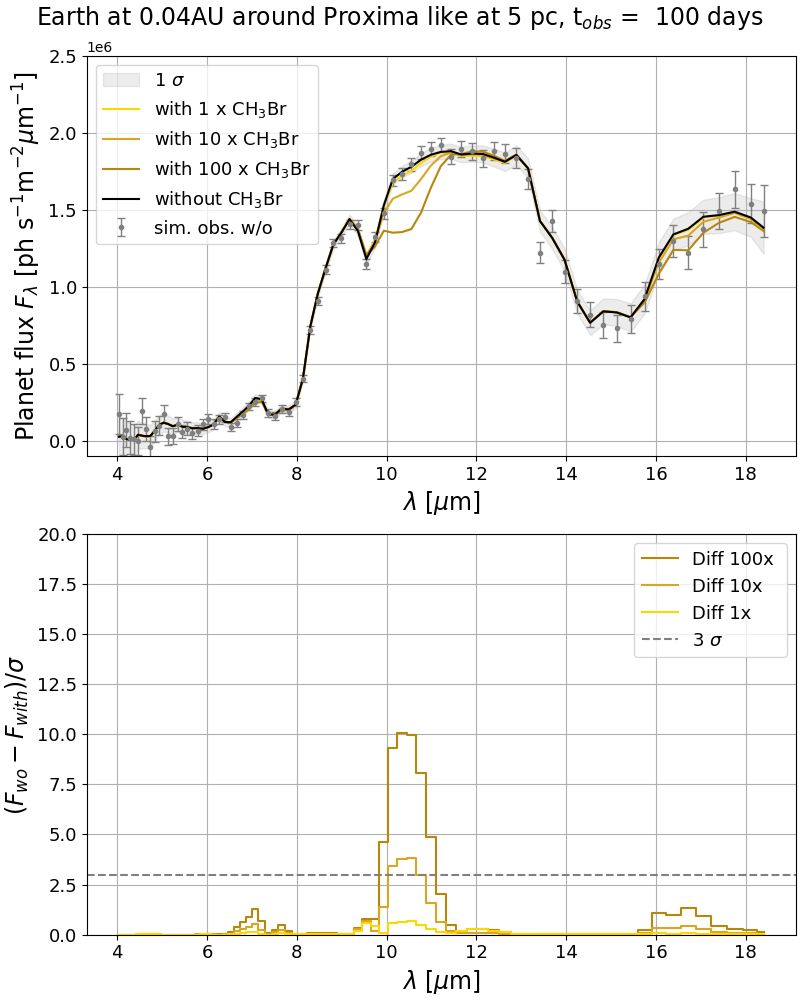}

\caption{Results for Proxima Type star in 100 days.}
\label{fig:prox_100d}
\end{figure}
\clearpage

\subsubsection{K6V type host}
\begin{figure}[ht]
\centering
\includegraphics[width=.3\textwidth]{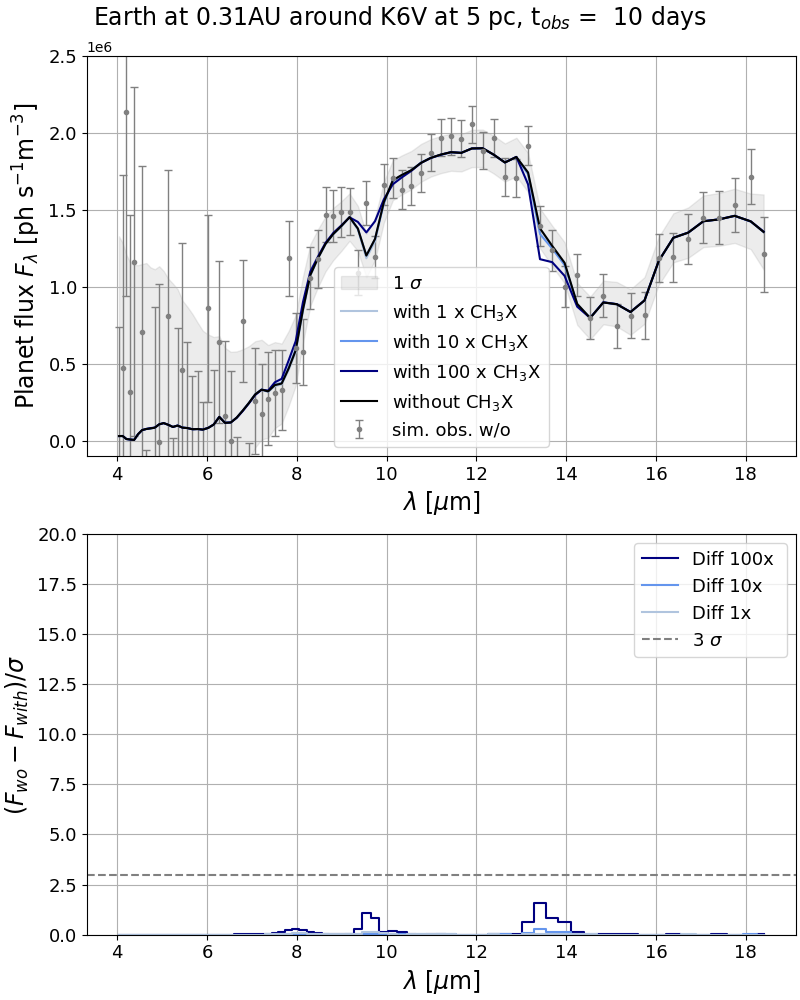}\quad
\includegraphics[width=.3\textwidth]{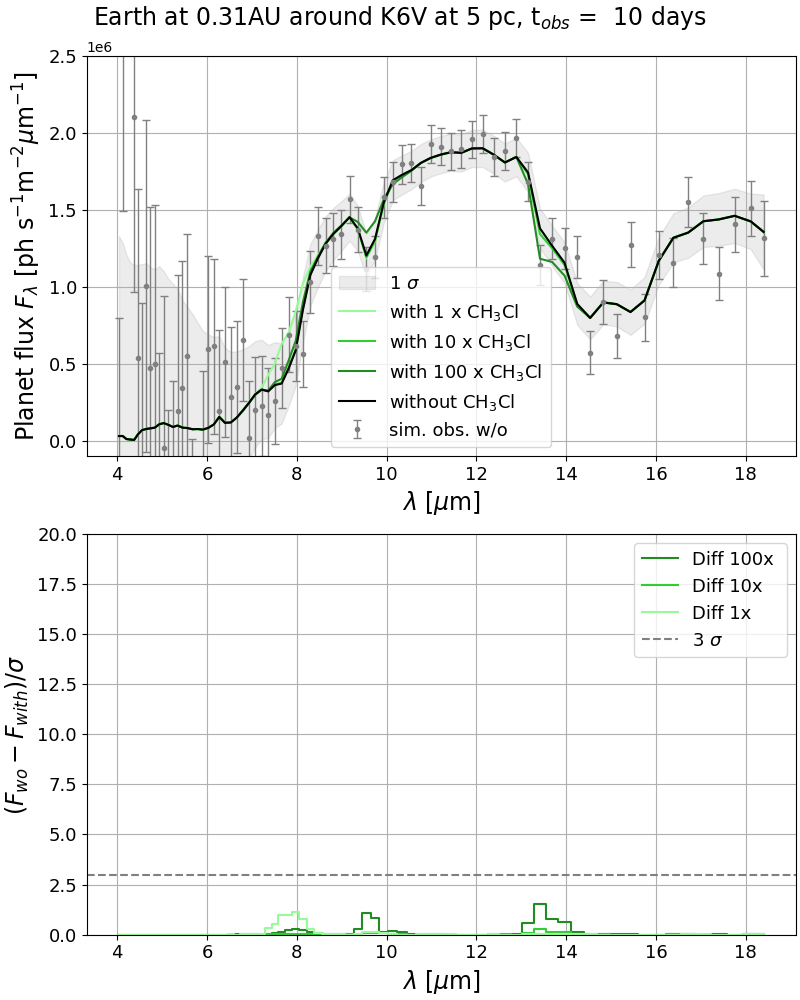}\quad
\includegraphics[width=.3\textwidth]{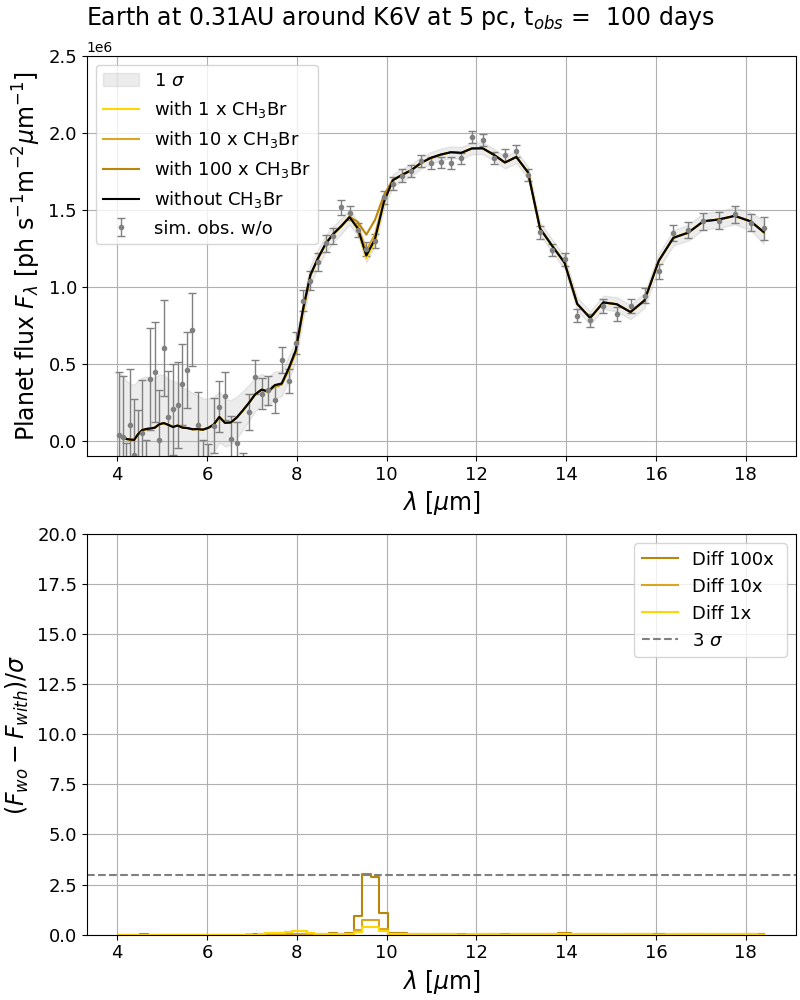}

\caption{Results for K6V Type star in 10 days.}
\label{fig:K6V_10d}
\end{figure}

\begin{figure}[ht]
\centering
\includegraphics[width=.3\textwidth]{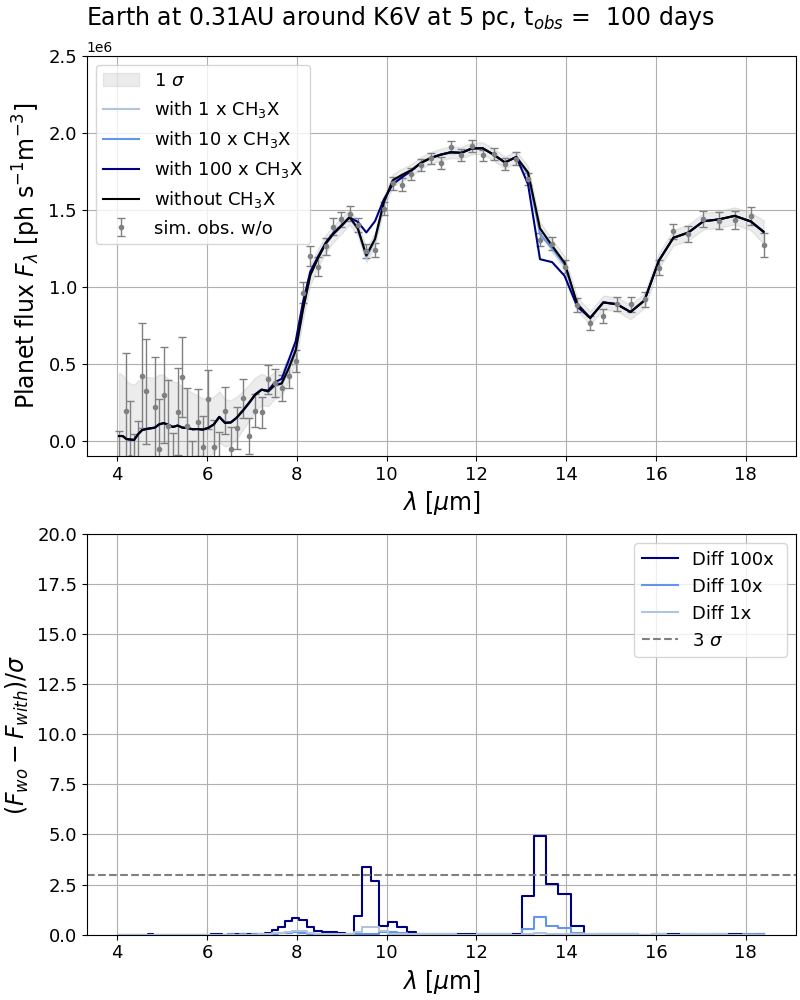}\quad
\includegraphics[width=.3\textwidth]{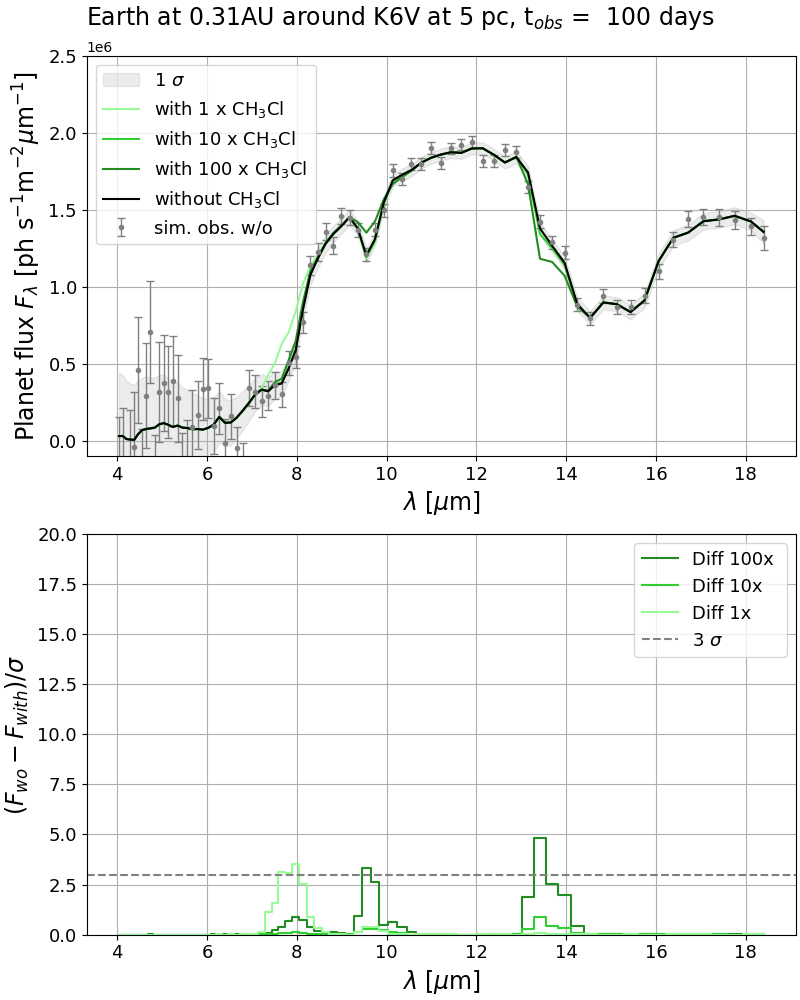}\quad
\includegraphics[width=.3\textwidth]{Figures/CH3X/CH3BR_K6V_5pc_0.31AU_100d.png}

\caption{Results for K6V Type star in 100 days.}
\label{fig:K6V_100d}
\end{figure}

\section{N$_{2}$O and CH$_{3}$X Flux-Abundance Details}\label{sec:app_fluxabund}

The predicted abundance of a trace gas, while highly dependent on its production flux, can also vary as a function of the background atmosphere and other assumptions embedded in the photochemical model used to calculate it including fundamental inputs such as stellar spectra, reaction rates, and molecular cross-sections \citep[e.g.,][]{2020ApJ...896..148R}. To facilitate future reproduction of our results, in  Tables \ref{table:N2OFLUXABUND} and \ref{table:CH3XFLUXABUND} we provide a cross-referencing of the N$_{2}$O and CH$_{3}$X fluxes and their respective predicted ground-level mixing ratios used in this study. As stated above, these calculations are sourced from \citet{2022ApJ...937..109S} for N$_{2}$O and \citet{2022ApJ...938....6L}  for CH$_{3}$Cl, CH$_{3}$Br, and CH$_{3}$X. 

\begin{deluxetable*}{ccccc}[ht]
\tablecaption{Flux vs. Ground-level Abundance for N$_{2}$O}
\label{table:N2OFLUXABUND} 
\tablewidth{0pt}
\tablehead{
\colhead{Scenario} & \colhead{Earth-like flux} & \colhead{1 $\frac{Tmol}{yr}$} & \colhead{10 $\frac{Tmol}{yr}$} & \colhead{100 $\frac{Tmol}{yr}$}
}
\startdata
Sun & 3.39$\times$10$^{-7}$ & 7.34$\times$10$^{-7}$ & 5.44$\times$10$^{-6}$ & 4.54$\times$10$^{-5}$ \\
K6V & 2.96$\times$10$^{-6}$ & 7.20$\times$10$^{-6}$ & 9.14$\times$10$^{-5}$ & 1.63$\times$10$^{-3}$ \\
Proxima Centauri & 1.14$\times$10$^{-6}$ & 2.75$\times$10$^{-6}$ & 2.82$\times$10$^{-5}$ & 6.54$\times$10$^{-4}$ \\
TRAPPIST-1 & 1.27$\times$10$^{-6}$ & 3.08$\times$10$^{-6}$ & 3.09$\times$10$^{-5}$ & 3.26$\times$10$^{-4}$ \\
\enddata
\end{deluxetable*}
\

\begin{deluxetable*}{ccccc}[ht]
\tablecaption{Flux vs. Ground-level Abundance for CH$_{3}$Cl, CH$_{3}$Br, and CH$_{3}$X}
\label{table:CH3XFLUXABUND} 
\tablewidth{0pt}
\tablehead{
\colhead{Scenario} & \colhead{Earth-like flux} & \colhead{10x Earth-like flux} & \colhead{100x Earth-like flux} & \colhead{1000x Earth-like flux}
}
\startdata
K6V–CH$_{3}$Cl & 4.82$\times$10$^{-9}$ & 4.82$\times$10$^{-8}$ & 3.64$\times$10$^{-7}$ & 2.09$\times$10$^{-6}$ \\ 
K6V–CH$_{3}$Br & 1.28$\times$10$^{-10}$ & 1.23$\times$10$^{-9}$ & 1.05$\times$10$^{-8}$ & 1.34$\times$10$^{-7}$ \\ 
K6V–CH$_{3}$X & 4.94$\times$10$^{-9}$ & 4.94$\times$10$^{-8}$ & 3.73$\times$10$^{-7}$ & 2.11$\times$10$^{-6}$ \\ \hline
AD Leo–CH$_{3}$Cl & 5.92$\times$10$^{-8}$ & 5.02$\times$10$^{-7}$ & 3.43$\times$10$^{-6}$ & 3.69$\times$10$^{-5}$ \\ 
AD Leo–CH$_{3}$Br & 3.47$\times$10$^{-9}$ & 3.42$\times$10$^{-8}$ & 4.61$\times$10$^{-7}$ & 1.49$\times$10$^{-6}$ \\
AD Leo–CH$_{3}$X & 6.27$\times$10$^{-8}$ & 5.39$\times$10$^{-7}$ & 3.57$\times$10$^{-6}$ & 4.77$\times$10$^{-5}$ \\ \hline
Proxima Centauri–CH$_{3}$Cl & 5.38$\times$10$^{-7}$ & 1.64$\times$10$^{-6}$ & 1.34$\times$10$^{-5}$ & 1.40$\times$10$^{-4}$ \\
Proxima Centauri–CH$_{3}$Br & 1.15$\times$10$^{-7}$ & 7.77$\times$10$^{-7}$ & 2.86$\times$10$^{-6}$ & 9.39$\times$10$^{-6}$ \\
Proxima Centauri–CH$_{3}$X & 2.87$\times$10$^{-7}$ & 3.80$\times$10$^{-6}$ & 3.21$\times$10$^{-5}$ & 3.94$\times$10$^{-4}$ \\ \hline
TRAPPIST-1-CH$_{3}$Cl & 1.26$\times$10$^{-7}$ & 1.18$\times$10$^{-6}$ & 9.65$\times$10$^{-6}$ & 7.45$\times$10$^{-5}$ \\
TRAPPIST-1–CH$_{3}$Br & 9.81$\times$10$^{-8}$ & 4.47$\times$10$^{-7}$ & 1.24$\times$10$^{-6}$ & 4.60$\times$10$^{-6}$ \\
TRAPPIST-1–CH$_{3}$X & 2.24$\times$10$^{-7}$ & 4.30$\times$10$^{-6}$ & 2.14$\times$10$^{-5}$ & 6.16$\times$10$^{-4}$ \\ 
\enddata
\end{deluxetable*}

    \clearpage

\bibliography{references_LIFE}{}
\bibliographystyle{aasjournal}



\end{document}